# Symmetry-Resolved Second Harmonic Generation in Quantum and Functional Materials

Xiaoyu Guo[a*], Chang Jae Roh[b*], Youngjun Ahn[c†]

[a]William H. Miller III Department of Physics and Astronomy, Johns Hopkins University, Baltimore, MD, 21218, USA

[b]Department of Quantum Matter Physics, University of Geneva, 24 Quai Ernest-Ansermet, Geneva 4 CH-1211, Switzerland

[c]Materials Science Division, Argonne National Laboratory, Lemont, IL, 60439, USA

*These authors contributed equally

†Correspondence to: youngjun.ahn@anl.gov

**Abstract**

Second harmonic generation (SHG) has evolved from a probe of noncentrosymmetric crystals into a symmetry-resolved optical method for identifying order parameters in quantum and functional materials. In particular, polarization-resolved rotational anisotropy (RA) measurements of SHG can connect nonlinear susceptibility tensors to the crystallographic and magnetic point groups of the underlying materials. This capability is especially powerful when the ordered state is weak, spatially confined, multipolar, magnetic, or hidden from conventional linear probe techniques. In this review article, we established a systematic symmetry-based framework for understanding RA-SHG and provide a comprehensive overview of RA-SHG studies across a broad range of condensed matter systems. We begin with basic theoretical background for the multipole origins of SHG radiation, the construction of nonlinear susceptibility tensors, and a group-theoretical framework connecting tensor components to order parameters. We then review the applications of RA-SHG to polar materials, magnetic orders, and other hidden electronic orders. Finally, we outline challenges and future research directions for using SHG to reveal, image, and control hidden, intertwined, and nonequilibrium phases in quantum and functional materials.

# 1. Introduction

The rapid growth of quantum and functional materials research has revealed that many important phases of matter are often defined not only by their electronic band structures, but also by their symmetry properties. Therefore, identifying the symmetry of an ordered phase has stood as a pivotal task in condensed matter physics and materials science. Second harmonic generation (SHG), a nonlinear optical process in which two photons at frequency $\omega$ generate one photon at frequency $2\omega$, provides a key solution to this problem. SHG responses can be described by nonlinear susceptibility tensors of the material, whose allowed components are constrained by the crystallographic and magnetic symmetries. In particular, polarization-resolved rotational-anisotropy SHG (RA-SHG) directly probes these tensor elements by measuring the angular dependence of the nonlinear optical response with respect to crystallographic axes in different polarization channels. The resulting RA-SHG patterns encode the symmetry-allowed tensor components and can be compared directly with group-theoretical predictions. This capability makes RA-SHG particularly useful for tracking reduced crystallographic and magnetic point groups across phase transitions, identifying orientations of polar axis, distinguishing domain variants, and unveiling hidden order parameters that are otherwise challenging to detect.

Several previous reviews have established important foundations for nonlinear optical studies of ordered materials. Early reviews developed the microscopic and phenomenological description of SHG in magnetic and ferroic systems, including electric-dipole (ED), magnetic-dipole (MD), electric-quadrupole (EQ), bulk, and surface contributions [1,2]. Subsequent reviews introduced rotational-anisotropy nonlinear harmonic generation and highlighted its growing application to quantum materials [3–6]. These works provide essential foundations for the field, but they were either primarily focused on the general nonlinear-optical response of ferroic materials or on the RA-SHG technique and selected applications. A systematic framework connecting RA-SHG measurements to the identification and characterization of order parameters across the rapidly expanding landscape of quantum and functional materials remains comparatively less developed. In particular, there is a need to bring together fundamental concepts, including nonlinear susceptibility tensors, crystallographic and magnetic symmetry and Landau theory, along with representative applications of RA-SHG to quantum and functional materials within a unified symmetry-based perspective.

This review therefore focuses specifically on RA-SHG as a symmetry-resolved probe of order parameters and phase transitions in quantum and functional materials. Rather than providing a general review of nonlinear optics or SHG spectroscopy, we emphasize how experimentally measured SHG anisotropies can be translated into symmetry constraints and ultimately into microscopic information about ordered phases. In Section 1, we first introduce the theoretical basis of SHG radiation sources, including ED, MD, and EQ contributions, and describe how nonlinear susceptibility tensors are derived using Neumann's principle, magnetic symmetry, and Landau theory. We then review experimental approaches based on RA-SHG, SHG scanning microscopy, and wide-field SHG imaging. From Section 2 to Section 4 we discuss representative applications of RA-SHG to polar, magnetic, and other hidden electronic and ferroic orders. In Section 5, we conclude by discussing future challenges and opportunities for using RA-SHG to study quantum and functional materials.

## 1.1. Theoretical descriptions of SHG sources and susceptibility tensors

The SHG process can be viewed as involving two stages: *generation* and *radiation*. The generation stage arises from the nonlinear optical response of the material, described by the macroscopic nonlinear susceptibility tensor and induced nonlinear radiation sources. The radiation stage is governed by emission and detection geometry, where these sources are expressed through a multipole expansion and determine their emission into the far field. In this section, we summarize the essential theoretical framework for analyzing and simulating RA-SHG patterns for comparison with experiments. Section 1.1.1 introduces the multipole expansion of SHG sources and derives the polarization properties of far-field radiation. Section 1.1.2 discusses the SHG susceptibility tensors and their correspondence to different optical transition mechanisms. Section 1.1.3 presents a quantum mechanical formulation of the SHG susceptibilities.

### 1.1.1. Multipole expansion of SHG radiation sources

When an intense optical field interacts with a nonlinear crystal, SHG can arise from multiple microscopic processes. In general, the far-field SHG radiation can be described by a multipole expansion of the nonlinear charge and current distributions oscillating at frequency $2\omega$ generated by the incident optical field at frequency $\omega$. The multipole expansion organizes the radiation in terms of progressively more spatially structured source moments, whose leading contributions are the ED, MD, and EQ moments. Physically, ED radiation originates from an oscillating separation of positive and negative charge, MD radiation from circulating or loop-like currents, and EQ radiation from a spatially nonuniform charge distribution in which the net electric dipole vanishes while a finite quadrupolar spatial arrangement remains. One may visualize the EQ source as two neighboring local dipoles pointing in opposite directions, so that the total dipole cancels even though the charge distribution is still spatially structured. Although ED radiation is generally the leading contribution, it vanishes under the presence of inversion symmetry, in which case the nominally higher-order MD and EQ terms become important.

Mathematically, the total nonlinear radiation source $\mathbf{S}$ can be expressed as the sum of three leading contributions: the ED moment $\mathbf{P}$, the MD moment $\mathbf{M}$, and the EQ moment $\mathbf{Q}$

$$\mathbf{S}(2\omega) = \mu_0 \frac{\partial^2 \mathbf{P}(2\omega)}{\partial t^2} + \mu_0 \left( \nabla \times \frac{\partial \mathbf{M}(2\omega)}{\partial t} \right) - \mu_0 \left( \nabla \cdot \frac{\partial^2 \mathbf{Q}(2\omega)}{\partial t^2} \right) \tag{1}$$

where $t$ is time, $\mu_0$ is the vacuum permeability, and $\omega$ and $2\omega$ are the fundamental and SHG light frequencies. A concise derivation of Eqn. (1) follows from the macroscopic Maxwell equations in matter with a multipole expansion of the displacement field $\mathbf{D}$ and the auxiliary field $\mathbf{H}$ [7]:

$$D_\alpha = \epsilon_0 E_\alpha + P_\alpha - \sum_\beta \frac{\partial Q_{\alpha\beta}}{\partial x_\beta} + \cdots \tag{2}$$

$$H_\alpha = \frac{1}{\mu_0} B_\alpha - M_\alpha + \cdots \tag{3}$$

where $\epsilon_0$ is vacuum permittivity, $x_\beta$ is the real space coordinate, $E_\alpha$ and $B_\alpha$ are the electric and magnetic fields, respectively. We emphasize that the expression for $\mathbf{D}$ extends beyond the ED approximation by including the EQ term, $\sum_\beta \frac{\partial Q_{\alpha\beta}}{\partial x_\beta}$ (compactly written as $\nabla \cdot \mathbf{Q}$ in Eqn. (1)). Combining Maxwell equations with Eqns. (2) and (3), we obtain the following relation whose right-hand side defines the source term introduced in Eqn. (1):

$$\nabla^2 \mathbf{E} - \nabla(\nabla \cdot \mathbf{E}) - \frac{\partial^2 \mathbf{E}}{c^2 \partial t^2} = \mu_0 \frac{\partial^2 \mathbf{P}}{\partial t^2} + \mu_0 \left( \nabla \times \frac{\partial \mathbf{M}}{\partial t} \right) - \mu_0 (\nabla \cdot \frac{\partial^2 \mathbf{Q}}{\partial t^2}) \tag{4}$$

The second term on the left-hand side, $\nabla(\nabla \cdot \mathbf{E})$, is purely longitudinal and does not contribute to far-field radiation, and therefore can be neglected. Assuming a harmonic form for the far-field electric field (and similarly for $\mathbf{P}$, $\mathbf{Q}$ and $\mathbf{M}$ ):

$$\mathbf{E} = (\mathbf{E}_{\parallel} + \mathbf{E}_{\perp})e^{i(\mathbf{kr}-\omega t+\phi)} \tag{5}$$

where $\mathbf{E}_{\parallel}$ and $\mathbf{E}_{\perp}$ are the transverse and longitudinal components of the light electric fields, $\mathbf{k}$ is the light wave vector, $\mathbf{r}$ is the distance and $\phi$ is the phase of the electric field, the corresponding far-field radiation can be decomposed according to the ED, MD, and EQ source terms:

$$E_P \propto P_{\parallel} \tag{6}$$

$$E_Q \propto (\nabla \cdot \mathbf{Q})_{\parallel} \tag{7}$$

$$E_M \propto (\nabla \times \mathbf{M})_{\parallel} \tag{8}$$

Eqns. (6) to (8) are used for simulating far-field RA-SHG patterns in various polarization channels.

### 1.1.2 Nonlinear source terms and optical transitions

Each source term in Eqn. (1) can be expanded as a series of products of the incident electromagnetic fields, including the electric field $E_a$, the auxiliary magnetic field $H_a$, and the electric field gradient $\nabla_a E_b$, with coefficients given by the corresponding nonlinear optical susceptibility tensors [6]. Considering SHG, we have

$$P_i = \chi_{ijk}^{eee} E_j E_k + \chi_{ijk}^{eem} E_j B_k + \chi_{ijkl}^{eeq} E_j \nabla_k E_l \tag{9}$$

$$M_i = \chi_{ijk}^{mee} E_j E_k \tag{10}$$

$$Q_{ij} = \chi_{ijkl}^{qee} E_k E_l \tag{11}$$

For each susceptibility tensor $\chi$, the superscript indices label the nature of the optical transitions. The first letter denotes the radiation source and the corresponding emission process at frequency $2\omega$ ($e$ for ED transitions, $m$ for MD transitions, and $q$ for EQ transitions), while the second and third letters denote the excitation processes at frequency $\omega$. Among these terms, the ED contribution $P_i = \chi_{ijk}^{eee} E_j E_k$is typically dominant. However, it vanishes in centrosymmetric systems and the remaining terms, $\chi^{eem}$, $\chi^{eeq}$, $\chi^{mee}$, and $\chi^{qee}$, must be considered on equal footing, as they can produce comparable SHG intensities. Experimentally, the contributions $P_i = \chi_{ijkl}^{eeq} E_j \nabla_k E_l$and $Q_{ij} = \chi_{ijkl}^{qee} E_k E_l$ are generally difficult to distinguish and are often grouped together as EQ SHG. Similarly, $P_i = \chi_{ijk}^{eem} E_j H_k$and $M_i = \chi_{ijk}^{mee} E_j E_k$are commonly grouped as MD SHG.

The subscript indices $i$, $j$, and $k$ for the ED contribution (similarly for MD and EQ contributions) contain directional information. The first index $i$ gives the direction of the generated polarization at $2\omega$, while the second and third indices $j$ and $k$ give the directions of the two fundamental fields that drive the nonlinear response. For example, $\chi_{yxx}$ describes an $x$-polarized fundamental field that produces an SHG polarization along $y$, which may occur when the crystal symmetry allows nonlinear optical coupling between the $x$- and $y$-directions. In this way, the susceptibility tensor can be viewed as a directional map of all possible nonlinear optical conversion pathways available in the material. The magnitude of $\chi_{ijk}$ determines how strongly a particular pathway contributes. Its sign or, more generally, its complex phase determines the phase of the generated SHG field and therefore how it interferes with SHG generated through other pathways.

### 1.1.3 Quantum mechanical formulation for SHG susceptibilities

Each of the contributions discussed above corresponds to a distinct sequence of optical transitions. To illustrate this, consider a three-level system, with the ground state $|0\rangle$, intermediate state $|1\rangle$, and the final state $|2\rangle$. As indicated in Eqs. (9)-(11), the field factors in each term ($E_i$, $\nabla_i E_j$, and $B_i$) describe the excitation processes in which two fundamental photons at frequency $\omega$ are absorbed, promoting the system from $|0\rangle$ to $|2\rangle$ via virtual or real intermediate states. The multipole moments in each term ($P_i$, $Q_{ij}$, and $M_i$) correspond to the subsequent de-excitation from $|2\rangle$ back to $|0\rangle$, accompanied by the emission of a photon at frequency $2\omega$. The specific sequences of optical transitions associated with each contribution are schematically illustrated in Fig. 1 [8].

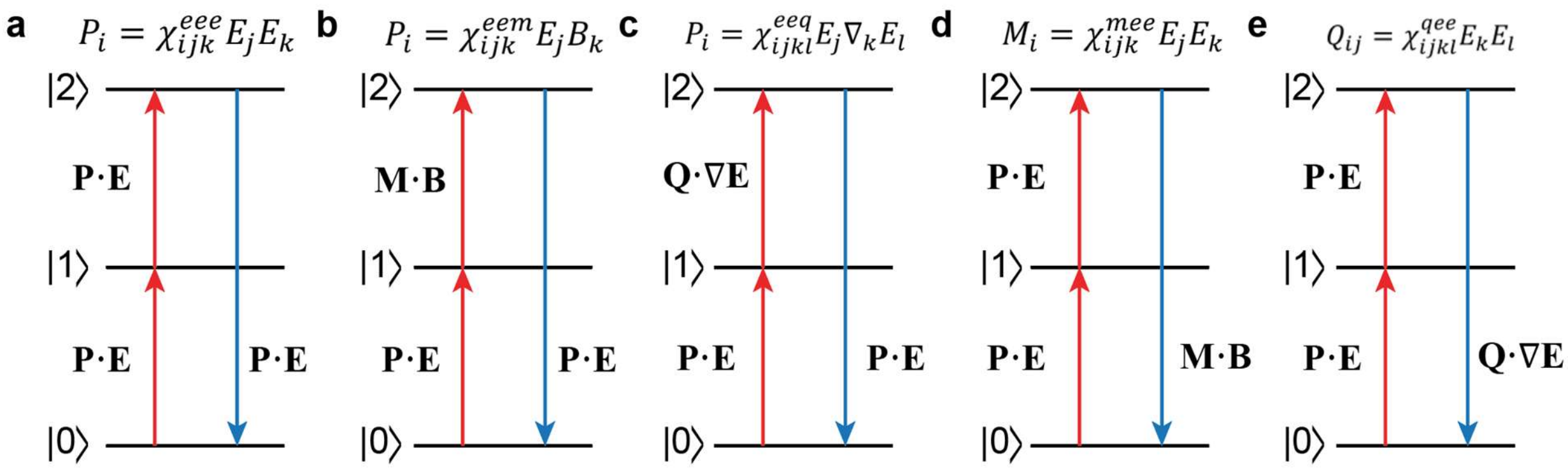

Fig. 1. Optical transitions and the corresponding radiation sources

The role of symmetry in determining selection rules can be directly understood within this framework. Consider the ED contribution $P_i = \chi_{ijk}^{eee} E_j E_k$ as an example. In a centrosymmetric system, electronic eigenstates have well-defined parity (even or odd) and ED transitions couple states of opposite parity. However, SHG via the ED channel requires three successive dipole matrix elements connecting the initial and final states. It is straightforward to verify that no combination of parity-even and parity-odd states can satisfy all three transitions simultaneously, leading to a vanishing ED SHG response in the presence of inversion symmetry.

The SHG susceptibility tensors corresponding to the optical processes illustrated in Fig. 1 can be derived using time-dependent perturbation theory. Taking the ED susceptibility $\chi^{eee}$ as an example, the light-matter interaction Hamiltonian is

$$V(t) = -\mathbf{P} \cdot \mathbf{E}(t) \tag{12}$$

where $\mathbf{P} = -e\mathbf{r}$ is the ED moment operator and $-e$ is the charge of the electron. The second-order induced dipole moment is given by

$$\langle \vec{P}^{(2)} \rangle = \langle \psi^{(0)} | \mathbf{P} | \psi^{(2)} \rangle + \langle \psi^{(1)} | \mathbf{P} | \psi^{(1)} \rangle + \langle \psi^{(2)} | \mathbf{P} | \psi^{(0)} \rangle, \tag{13}$$

where $|\psi^{(0)}\rangle$ is the unperturbed ground state, and $|\psi^{(1)}\rangle$, $|\psi^{(2)}\rangle$ are the first- and second-order corrections. For SHG, the induced polarization at frequency $2\omega$ can be written as

$$P_i^{(2)}(2\omega) = \epsilon_0 \sum_{jk} \chi_{ijk}^{(2)}(2\omega, \omega, \omega)\, E_j(\omega) E_k(\omega) \tag{14}$$

Assuming the system is initially in the ground state, the ED susceptibility takes the form

$$\chi_{ijk}^{eee}(2\omega) = \frac{N}{\epsilon_0 \hbar^2} \sum_{nm} \left[ \frac{P_{gn}^i P_{nm}^j P_{mg}^k}{(\omega_{ng} - 2\omega)(\omega_{mg} - \omega)} + (j \leftrightarrow k) \right] \tag{15}$$

where $N$ is the number density of electrons, $n$ and $m$ label excited states, and $\hbar\omega_{ab} = E_a - E_b$ is the energy difference between two levels. The term $(j \leftrightarrow k)$ denotes symmetrization with respect to the two input optical fields [6,8].

For other multipole processes, the dipole operator $\mathbf{P}$ is replaced by the MD operator $\mathbf{M}$ or EQ operator $\mathbf{Q}$. The selection rules are determined by the corresponding matrix elements. A transition is allowed only when the relevant operator has a nonzero matrix element between the involved states. The denominators describe resonance enhancement when the photon energies approach the energy differences between ground state and intermediate excited states.

This quantum mechanical formalism provides a rigorous foundation for computing SHG responses from first principles. However, it is not always directly applicable to the analysis of experimental data. In complex quantum materials, such as those with strong electronic correlations, significant spin-orbit coupling, and substantial orbital hybridization, the microscopic contributions to the susceptibility tensors become difficult to disentangle, limiting their practical utility. In such cases, a more effective approach is to combine this formalism with SHG spectroscopy [5] and use symmetry-based analysis to interpret experimental observations.

## 1.2. Experimental geometries for symmetry-resolved SHG

RA-SHG and SHG microscopy provide complementary views of symmetry information. RA-SHG constrains symmetry-allowed nonlinear susceptibility components through their angular dependence, whereas SHG microscopy visualizes the spatial organization of symmetry-related domains. Together, these approaches connect nonlinear optical selection rules with point-group symmetry and real-space domain textures.

### 1.2.1. Rotational anisotropy SHG geometries

RA-SHG measures the angular dependence of the SHG intensity to determine the symmetry-allowed tensor components (Fig. 2). In a typical RA-SHG measurement, the relative orientation between the light electric field polarization and the crystallographic axes is systematically varied, and the resulting SHG intensity is recorded as a function of the rotation angle $\varphi$. The symmetry of the measured anisotropic patterns reflects the symmetry-allowed tensor components and therefore constrains the possible point-group symmetries of the material.

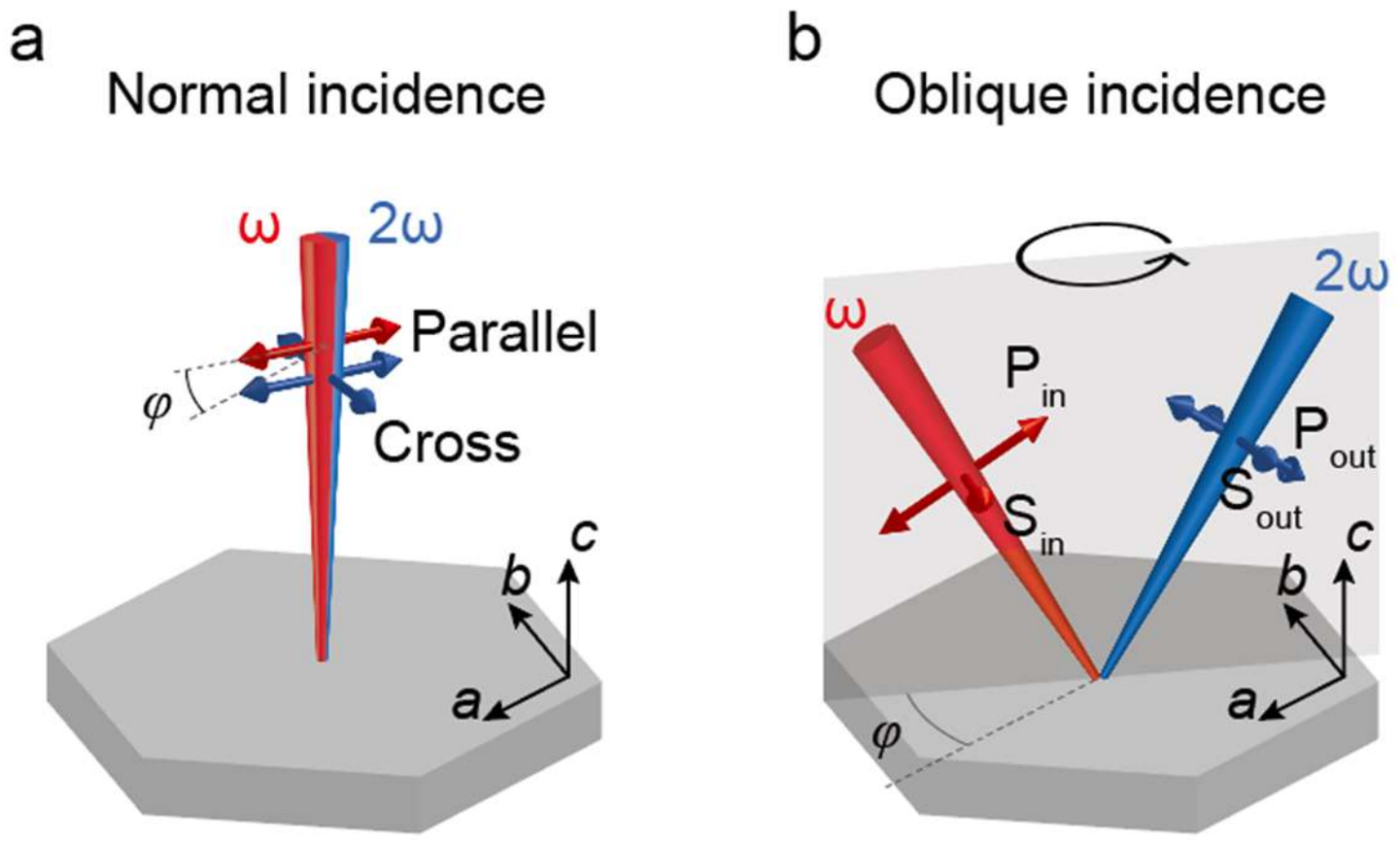

Fig. 2. Schematic illustration of RA-SHG measurement geometries. (a) Normal-incidence RA-SHG geometry, where the fundamental beam at $\omega$ is incident normal to the sample surface and the reflected SHG signal at $2\omega$ is collected along the same optical path. The incident and detected polarizations are rotated by an azimuthal angle $\varphi$ with respect to the crystallographic axes. Parallel and cross channels denote configurations in which the two polarizations are parallel and perpendicular, respectively. (b) Oblique-incidence RA-SHG geometry, where the incident and reflected beams propagate at finite angles from the surface normal. This geometry provides access to both in-plane and out-of-plane tensor elements through four polarization channels: SS, SP, PS, and PP. Here, S and P denote polarizations perpendicular and parallel to the scattering plane, respectively, and the first and second letters indicate the incident fundamental and reflected SHG polarizations.

In general, RA-SHG experiments are performed in two representative optical geometries: normal incidence and oblique incidence, as illustrated in Fig. 2. These configurations provide complementary sensitivity to different nonlinear susceptibility tensor components. In a normal-incidence geometry, the propagation direction of the incident beam is perpendicular to the sample surface, such that the electric field of the incident fundamental light lies within the sample plane, as illustrated in Fig. 2a. The accessible polarization selection channels in normal incidence RA-SHG are commonly denoted as parallel and cross channels. In the parallel channel, the polarization states of incident fundamental and detected SHG fields are parallel, whereas they are perpendicular in the cross channel. In this geometry, the fundamental field can directly drive nonlinear polarization components through tensor elements of the form $\chi_{xxx}$, $\chi_{xxy}$, $\chi_{xyx}$, $\chi_{xyy}$, $\chi_{yxx}$, $\chi_{yxy}$, $\chi_{yyx}$, and $\chi_{yyy}$, for the ED process. Since the fundamental electric fields are identical for SHG, the last two tensor indices are symmetric, $\chi_{ijk} = \chi_{ikj}$. With this intrinsic permutation symmetry, the independent in-plane tensor components accessible at normal incidence are reduced to $\chi_{xxx}$, $\chi_{xyy}$, $\chi_{xxy}$, $\chi_{yxx}$, $\chi_{yyy}$, and $\chi_{yxy}$. Thus, normal incidence RA-SHG is highly useful for probing in-plane anisotropy through the angular dependence of the accessible in-plane tensor elements.

On the other hand, tensor components involving an out-of-plane $z$-index, such as $\chi_{zxx}$, $\chi_{zyy}$, $\chi_{xxz}$, or $\chi_{zzz}$, are inaccessible in normal incidence geometry. Access to these components generally requires oblique incidence. In the oblique incidence geometry, the fundamental light and detected SHG light have a non-zero angle with respect to the sample surface normal, thereby forming a scattering plane, as shown in Fig. 2b. The polarization of light is then defined with respect to this scattering plane, denoting it as S- and P-polarized when its electric field is perpendicular to and lies within the scattering plane. While S-polarized light is typically associated with an in-plane electric-field component, P-polarized light generally contains both in-plane and out-of-plane electric-field components. This mixed in-plane and out-of-plane character of P-polarized light provides selective sensitivity to nonlinear susceptibility tensor elements containing a $z$ index. As a result, oblique incidence RA-SHG enables access to a broader set of nonlinear susceptibility tensor elements and can help distinguish candidate point-group symmetries when combined with crystallographic and complementary experimental information.

In oblique-incidence RA-SHG, the scattering plane introduces an additional laboratory-frame reference direction. Therefore, rotating only the incident and detected polarizations is not generally equivalent to rotating the sample azimuth unless the scattering plane is also rotated relative to the crystallographic axes. This distinction should be considered when interpreting the angular dependence of oblique-incidence RA-SHG patterns.

Reliable symmetry assignment requires careful control of systematic errors and optical artifacts that may distort the measured angular dependence. Polarization optics should therefore be calibrated for alignment, polarization-dependent transmission or retardance, and leakage between nominally orthogonal channels. Data averaging, suppression of stray and background light, and tracking the onset of a symmetry-breaking contribution across a phase transition provide additional confidence in the assignment.

A practical limitation of RA-SHG is that a measured angular dependence does not necessarily provide a unique point-group assignment because it represents only a projection of the full nonlinear susceptibility tensor in a particular experimental geometry. Different point groups may therefore produce identical or nearly identical patterns when their distinguishing tensor components are not accessed. This ambiguity can be reduced by simultaneously analyzing multiple polarization channels, varying the incidence angle [9], combining normal- and oblique-incidence measurements, measuring different crystallographic facets, and, where possible, using photon-energy-dependent [10], phase-sensitive, or domain-resolved SHG.

If several candidate symmetries remain consistent with the SHG measurements, complementary probes and microscopic theory provide additional constraints. Diffraction measurements can determine translational symmetry and atomic displacements [11], while resonant X-ray or neutron scattering can constrain magnetic or multipolar structures. When the observed SHG response cannot be reconciled with established structural or magnetic symmetry after such possibilities are considered, first-principles calculations or microscopic modeling can evaluate candidate electronic, magnetic, or multipolar states [12,13]. Thus, symmetry analysis determines which SHG responses are allowed, whereas complementary experiments and microscopic theory help establish their physical origin.

Throughout this review article, we use the notation parallel and cross to denote the selectable polarization channels for the normal incidence geometry. For the oblique-incidence geometry, we use SS, SP, PS, and PP to indicate the polarization channels, unless otherwise specified. Here, the first and second letters indicate the polarization states of the incident light and the detected light, respectively. For example, PS denotes P-polarized incident fundamental light and S-polarized detected SHG light.

### 1.2.2. SHG scanning microscopy and wide-field imaging

High-resolution optical imaging is essential for resolving the spatial distribution of physical properties such as ferroelectric, ferroelastic, magnetic, chiral, and other multipolar domains. In many ordered phases, domains are not merely morphological features, but symmetry-related variants of the same broken symmetry. When a material undergoes a phase transition from a high-symmetry parent phase to a low-symmetry ordered phase, the order parameter selects one of several symmetry-equivalent orientations. These degenerate states are transformed into one another by the symmetry operations that are present in the parent phase but lost in the ordered phase. Therefore, identifying the number of domains, their spatial distributions, and the symmetry operations relating them is central to determining the symmetry and microscopic nature of the order parameter.

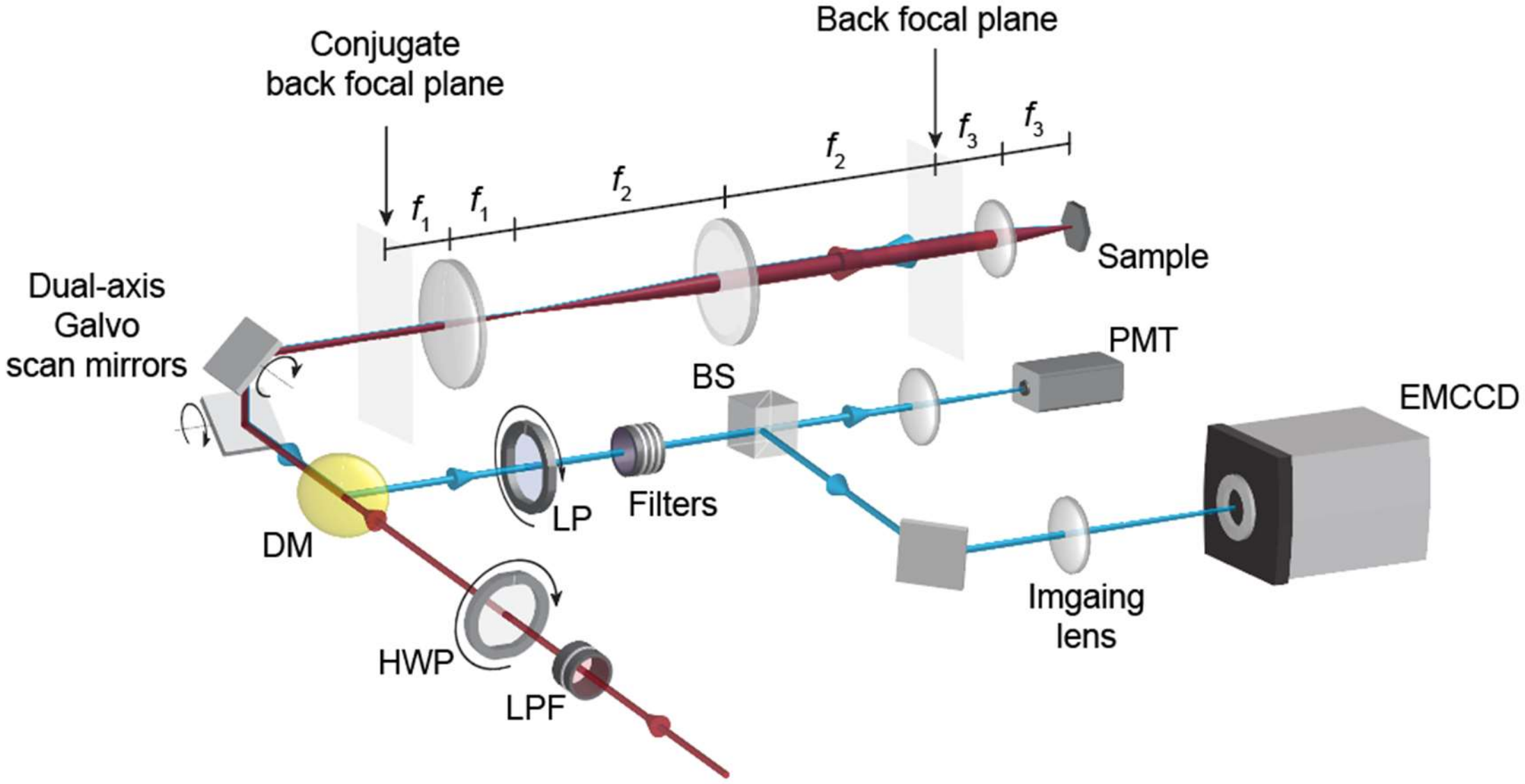

Fig. 3. Schematic of the optical platform of RA-SHG scanning microscopy and wide-field SHG imaging. The incident fundamental beam is conditioned by a low-pass filter (LPF), and its polarization is varied by half-wave plate (HWP). The galvo mirrors scan the focused beam across the sample. The reflected SHG signal is separated by a dichroic mirror (DM) and analyzed by a linear polarizer (LP). The remaining fundamental light is further removed by a set of filters. The signal is then directed to a photomultiplier tube (PMT) for scanning SHG microscopy or to an electron-multiplying charge-coupled device (EMCCD) for wide-field imaging. BS: beam splitter.

In scanning SHG microscopy, a tightly focused fundamental beam is raster-scanned across the sample while the SHG intensity is recorded at each position [14,15]. Scanning can be implemented either by translating the sample or by steering the beam using galvo mirrors and a 4f relay system, as illustrated in Fig. 3. Beam scanning is particularly advantageous in cryogenic or field-dependent measurements because the sample remains stationary, reducing mechanical drift and maintaining the optical geometry during imaging.

This 4f-based beam-scanning geometry offers several advantages over sample translation. Since the sample is not mechanically moved during image acquisition, this method reduces mechanical drift, vibration, positioning error, and beam-sample registration changes. This is particularly important for materials mounted in constrained experimental environments, such as being at cryogenic temperatures and under applied fields.

Wide-field SHG imaging provides a complementary route to domain visualization. Instead of raster-scanning a focused beam spot, a larger region of the sample is illuminated, and the emitted

SHG light is imaged onto a two-dimensional pixel-array detector (Fig. 3). The SHG signal emitted from the illuminated region is collected while being spectrally separated from the fundamental beam using filters and projected onto a camera through an imaging lens. Each detector pixel records the SHG intensity from a corresponding position on the sample, allowing rapid acquisition of two-dimensional SHG images without point-by-point scanning. This method is particularly useful for visualizing domain patterns over a large field of view, monitoring domain evolution in real time, and capturing spatially extended interference effects.

Moreover, wide-field SHG imaging enables direct visualization of domains related by inversion symmetry, such as polar and ferroelectric domains [14]. In inversion-related domains, the ED SHG susceptibility has the opposite sign, corresponding to a 180° phase shift of the emitted SHG field. Although the two domains may show similar SHG intensity away from the domain boundary, their SHG fields interfere destructively where they overlap optically near a domain wall. As a result, inversion-related ferroelectric or polar domains can produce dark contrast at the domain wall in wide-field SHG images. This destructive interference provides direct phase-sensitive evidence that the adjacent domains are related by inversion symmetry. Therefore, wide-field SHG imaging can reveal not only the spatial arrangement of domains, but also the relative phase of the nonlinear optical response, which is directly connected to the sign of the underlying polar order parameter.

For raster-scanning SHG microscopy, the lateral resolution is primarily determined by the optical wavelength $\lambda$ and numerical aperture NA, with a diffraction-limited scale of approximately $0.61 \times \lambda /$ NA. Wide-field SHG resolution is similarly limited by diffraction and the collection numerical aperture, with additional constraints from magnification and detector pixel size. In practice, low-NA focusing lenses produce micrometer-to-tens-of-micrometers spatial resolution, whereas high-NA objectives can approach the optical diffraction limit.

### 1.3. Symmetry-based derivation of SHG tensor elements

SHG directly connects crystal symmetry to experimentally measurable nonlinear susceptibility tensor elements because the allowed components are constrained by point-group symmetry. In this section, we first describe how the symmetry-allowed SHG tensor elements are derived by applying Neumann's principle to polar and axial tensors.

#### 1.3.1. Neumann's principle and polar SHG tensors

To derive the symmetry-allowed elements of a third-rank tensor, such as that for ED SHG, one begins with a general tensor $\chi_{ijk}$ and imposes an invariance constraint under all symmetry operations of the relevant crystallographic point group. For each symmetry operation represented by a matrix $R$, the tensor transforms as

$$\chi'_{i'j'k'} = \sum_{i,j,k} R_{i'i} R_{j'j} R_{k'k} \chi_{ijk} \tag{16}$$

According to Neumann's principle, any physical tensor describing a crystal property must remain unchanged under the symmetry operations of the crystal. Therefore, the transformed tensor must satisfy

$$\chi' = \chi \tag{17}$$

This condition is then imposed for each symmetry operation of the point group. This procedure produces a set of linear equations among the tensor components, from which the zero elements and symmetry-related nonzero elements are determined. In the following, examples of how to derive polar and axial SHG tensor elements are given by considering the $4mm$ point group.

The point group $4mm$ has a four-fold rotation $C_4$ around the crystallographic *c*-axis as a symmetry operation, which is parallel to the $z$-axis and two vertical mirror planes $m_{xz}$ and $m_{yz}$ that contain $x/z$- and $y/z$-axes, respectively. The corresponding symmetry matrices are given by

$$C_4 = \begin{pmatrix} \cos(\pi/2) & -\sin(\pi/2) & 0 \\ \sin(\pi/2) & \cos(\pi/2) & 0 \\ 0 & 0 & 1 \end{pmatrix}, m_{xz} = \begin{pmatrix} 1 & 0 & 0 \\ 0 & -1 & 0 \\ 0 & 0 & 1 \end{pmatrix}, m_{yz} = \begin{pmatrix} -1 & 0 & 0 \\ 0 & 1 & 0 \\ 0 & 0 & 1 \end{pmatrix} \quad (18)$$

A general third-rank tensor matrix, $\chi_{ijk}$, without any symmetry constraints is given as below.

$$\chi_{ijk} = \begin{pmatrix} \begin{pmatrix} \chi_{xxx} \\ \chi_{xxy} \\ \chi_{xxz} \end{pmatrix} & \begin{pmatrix} \chi_{xyx} \\ \chi_{xyy} \\ \chi_{xyz} \end{pmatrix} & \begin{pmatrix} \chi_{xzx} \\ \chi_{xzy} \\ \chi_{xzz} \end{pmatrix} \\ \begin{pmatrix} \chi_{yxx} \\ \chi_{yxy} \\ \chi_{yxz} \end{pmatrix} & \begin{pmatrix} \chi_{yyx} \\ \chi_{yyy} \\ \chi_{yyz} \end{pmatrix} & \begin{pmatrix} \chi_{yzx} \\ \chi_{yzy} \\ \chi_{yzz} \end{pmatrix} \\ \begin{pmatrix} \chi_{zxx} \\ \chi_{zxy} \\ \chi_{zxz} \end{pmatrix} & \begin{pmatrix} \chi_{zyx} \\ \chi_{zyy} \\ \chi_{zyz} \end{pmatrix} & \begin{pmatrix} \chi_{zzx} \\ \chi_{zzy} \\ \chi_{zzz} \end{pmatrix} \end{pmatrix} \quad (19)$$

By applying $\chi'_{i'j'k'} = \sum_{i,j,k} R_{i'i} R_{j'j} R_{k'k} \chi_{ijk}$ with $R = m_{xz}$ as described above, one obtains

$$\chi_{i'j'k'} = \begin{pmatrix} \begin{pmatrix} \chi_{xxx} \\ -\chi_{xxy} \\ \chi_{xxz} \end{pmatrix} & \begin{pmatrix} -\chi_{xyx} \\ \chi_{xyy} \\ -\chi_{xyz} \end{pmatrix} & \begin{pmatrix} \chi_{xzx} \\ -\chi_{xzy} \\ \chi_{xzz} \end{pmatrix} \\ \begin{pmatrix} -\chi_{yxx} \\ \chi_{yxy} \\ -\chi_{yxz} \end{pmatrix} & \begin{pmatrix} \chi_{yyx} \\ -\chi_{yyy} \\ \chi_{yyz} \end{pmatrix} & \begin{pmatrix} -\chi_{yzx} \\ \chi_{yzy} \\ -\chi_{yzz} \end{pmatrix} \\ \begin{pmatrix} \chi_{zxx} \\ -\chi_{zxy} \\ \chi_{zxz} \end{pmatrix} & \begin{pmatrix} -\chi_{zyx} \\ \chi_{zyy} \\ -\chi_{zyz} \end{pmatrix} & \begin{pmatrix} \chi_{zzx} \\ -\chi_{zzy} \\ \chi_{zzz} \end{pmatrix} \end{pmatrix} \quad (20)$$

Since $\chi' = \chi$ , $\chi_{xxy} = \chi_{xyx} = \chi_{xxy} = \chi_{xyz} = \chi_{xzy} = \chi_{yxx} = \chi_{yxz} = \chi_{yyy} = \chi_{yzx} = \chi_{yzz} = \chi_{zxy} = \chi_{zyx} = \chi_{zyz} = \chi_{zzy} = 0$, and thus

$$\chi_{ijk} = \begin{pmatrix} \begin{pmatrix} \chi_{xxx} \\ 0 \\ \chi_{xxz} \end{pmatrix} & \begin{pmatrix} 0 \\ \chi_{xyy} \\ 0 \end{pmatrix} & \begin{pmatrix} \chi_{xzx} \\ 0 \\ \chi_{xzz} \end{pmatrix} \\ \begin{pmatrix} 0 \\ \chi_{yxy} \\ 0 \end{pmatrix} & \begin{pmatrix} \chi_{yyx} \\ 0 \\ \chi_{yyz} \end{pmatrix} & \begin{pmatrix} 0 \\ \chi_{yzy} \\ 0 \end{pmatrix} \\ \begin{pmatrix} \chi_{zxx} \\ 0 \\ \chi_{zxz} \end{pmatrix} & \begin{pmatrix} 0 \\ \chi_{zyy} \\ 0 \end{pmatrix} & \begin{pmatrix} \chi_{zzx} \\ 0 \\ \chi_{zzz} \end{pmatrix} \end{pmatrix} \quad (21)$$

By following the same procedure for $R = m_{yz}$ and $R = C_4$, the final non-zero polar SHG tensor elements of the $4mm$ point group are obtained as

$$\chi^{4mm\,(polar)} = \begin{pmatrix} \begin{pmatrix} 0 \\ 0 \\ \chi_{xxz} \end{pmatrix} & \begin{pmatrix} 0 \\ 0 \\ 0 \end{pmatrix} & \begin{pmatrix} \chi_{xzx} \\ 0 \\ 0 \end{pmatrix} \\ \begin{pmatrix} 0 \\ 0 \\ 0 \end{pmatrix} & \begin{pmatrix} 0 \\ 0 \\ \chi_{xxz} \end{pmatrix} & \begin{pmatrix} 0 \\ \chi_{xzx} \\ 0 \end{pmatrix} \\ \begin{pmatrix} \chi_{zxx} \\ 0 \\ 0 \end{pmatrix} & \begin{pmatrix} 0 \\ \chi_{zxx} \\ 0 \end{pmatrix} & \begin{pmatrix} 0 \\ 0 \\ \chi_{zzz} \end{pmatrix} \end{pmatrix} \tag{22}$$

### 1.3.2. Derivation of axial SHG tensors

The axial SHG tensor must be considered when the SHG effects arise from MD contributions or other axial-vector-related optical processes. Unlike the ED contributions, axial tensors follow a different transformation rule because they change signs under operations that reverse handedness, such as mirror reflection and spatial inversion. Therefore, the axial tensor transformation contains an additional factor given by the determinant of the operation matrix

$$\chi_{i'j'k'} = \sum_{i,j,k} |R| R_{i'i} R_{j'j} R_{k'k} \chi_{ijk} \tag{23}$$

Here $|R| = -1$ for mirror and inversion operations, whereas $|R| = 1$ for proper rotations. This additional determinant factor distinguishes axial tensors from polar tensors and allows axial SHG contributions to survive under symmetry conditions where polar ED SHG may be forbidden.

The procedure for deriving the axial SHG tensor for the $4mm$ point group is described as follows.

By applying $\chi_{i'j'k'} = \sum_{i,j,k} |R| R_{i'i} R_{j'j} R_{k'k} \chi_{ijk}$ (with $R = m_{xz}$ and $|R| = -1$ as described above, one obtains

$$\chi_{i'j'k'} = \begin{pmatrix} \begin{pmatrix} -\chi_{xxx} \\ \chi_{xxy} \\ -\chi_{xxz} \end{pmatrix} & \begin{pmatrix} \chi_{xyx} \\ -\chi_{xyy} \\ \chi_{xyz} \end{pmatrix} & \begin{pmatrix} -\chi_{xzx} \\ \chi_{xzy} \\ -\chi_{xzz} \end{pmatrix} \\ \begin{pmatrix} \chi_{yxx} \\ -\chi_{yxy} \\ \chi_{yxz} \end{pmatrix} & \begin{pmatrix} -\chi_{yyx} \\ \chi_{yyy} \\ -\chi_{yyz} \end{pmatrix} & \begin{pmatrix} \chi_{yzx} \\ -\chi_{yzy} \\ \chi_{yzz} \end{pmatrix} \\ \begin{pmatrix} -\chi_{zxx} \\ \chi_{zxy} \\ -\chi_{zxz} \end{pmatrix} & \begin{pmatrix} \chi_{zyx} \\ -\chi_{zyy} \\ \chi_{zyz} \end{pmatrix} & \begin{pmatrix} -\chi_{zzx} \\ \chi_{zzy} \\ -\chi_{zzz} \end{pmatrix} \end{pmatrix} \tag{24}$$

Since $\chi' = \chi$ , $\chi_{xxx} = \chi_{xxz} = \chi_{xyy} = \chi_{xzx} = \chi_{xzz} = \chi_{yxy} = \chi_{yyx} = \chi_{yyz} = \chi_{yzy} = \chi_{zxx} = \chi_{zxz} = \chi_{zyy} = \chi_{zzx} = \chi_{zzz} = 0$, and thus

$$\chi_{ijk} = \begin{pmatrix} \begin{pmatrix} 0 \\ \chi_{xxy} \\ 0 \end{pmatrix} & \begin{pmatrix} \chi_{xyx} \\ 0 \\ \chi_{xyz} \end{pmatrix} & \begin{pmatrix} 0 \\ \chi_{xzy} \\ 0 \end{pmatrix} \\ \begin{pmatrix} \chi_{yxx} \\ 0 \\ \chi_{yxz} \end{pmatrix} & \begin{pmatrix} 0 \\ \chi_{yyy} \\ 0 \end{pmatrix} & \begin{pmatrix} \chi_{yzx} \\ 0 \\ \chi_{yzz} \end{pmatrix} \\ \begin{pmatrix} 0 \\ \chi_{zxy} \\ 0 \end{pmatrix} & \begin{pmatrix} \chi_{zyx} \\ 0 \\ \chi_{zyz} \end{pmatrix} & \begin{pmatrix} 0 \\ \chi_{zzy} \\ 0 \end{pmatrix} \end{pmatrix} \tag{25}$$

Applying the same procedure to $R = m_{yz}$ and $R = C_4$ results in the final form of the non-zero axial SHG tensor element of $4mm$ as given below.

$$\chi^{4m \quad (axial)} = \begin{pmatrix} \begin{pmatrix} 0 \\ 0 \\ 0 \end{pmatrix} & \begin{pmatrix} 0 \\ 0 \\ \chi_{xyz} \end{pmatrix} & \begin{pmatrix} 0 \\ \chi_{xzy} \\ 0 \end{pmatrix} \\ \begin{pmatrix} 0 \\ 0 \\ -\chi_{xyz} \end{pmatrix} & \begin{pmatrix} 0 \\ 0 \\ 0 \end{pmatrix} & \begin{pmatrix} -\chi_{xzy} \\ 0 \\ 0 \end{pmatrix} \\ \begin{pmatrix} 0 \\ \chi_{zxy} \\ 0 \end{pmatrix} & \begin{pmatrix} -\chi_{zxy} \\ 0 \\ 0 \end{pmatrix} & \begin{pmatrix} 0 \\ 0 \\ 0 \end{pmatrix} \end{pmatrix} \tag{26}$$

### 1.4. Landau- and group-theoretical interpretation of SHG

Landau theory provides a symmetry-based framework for understanding how optical responses emerge at phase transitions [16]. In this approach, the free energy of a material is expanded as a polynomial in the relevant order parameters and external fields. The fundamental requirement is that every term in the free-energy expansion must transform as the totally symmetric representation of the high-symmetry phase. In other words, each term must remain invariant under all symmetry operations of the parent crystallographic point group.

This constraint is especially powerful for nonlinear optics because optical susceptibility tensors are themselves symmetry-governed response functions. For example, ED SHG is described by the nonlinear polarization, $P_i(2\omega) = \chi_{ijk}E_j(\omega)E_k(\omega)$. Since the electric field is a polar vector, both $P_i$ and $E_{j/k}$transform according to irreducible representations (irreps) of the crystallographic point group whose basis functions are $x$, $y$ or $z$. Therefore, a SHG tensor component $\chi_{ijk}$ is allowed only when it shares the same irrep with the full coupling term involving $P_iE_jE_k$ of the symmetry group.

From the Landau-theory perspective, the coupling between an internal order parameter and optical fields can be described through symmetry-allowed terms in the free energy [6]. For example, one may write a hypothetical free-energy expansion as $F = F_0 + \frac{a}{2}P^2 - \boldsymbol{E} \cdot \boldsymbol{P} - b\psi P_x\left(E_x^2 + E_y^2 - E_z^2\right) + \cdots$ where $P$ is the induced electric polarization, $E_x, E_y$, and $E_z$ are electric-field components, $\psi$ is an internal order parameter of the material, and $a$ and $b$ are expansion coefficients. The nonlinear coupling term $-b\psi P_x\left(E_x^2 + E_y^2 - E_z^2\right)$ is included only when the product $\psi P_x\left(E_x^2 + E_y^2 - E_z^2\right)$ transforms as the totally symmetric representation of the parent point group. In group theory, this condition can be expressed as $\Gamma_\psi \otimes \Gamma_{P_x} \otimes \Gamma_{\left(E_x^2+E_y^2-E_z^2\right)} \supset A_1$ where $\Gamma_a$ denotes the irrep associated with the order parameter $a$, and $A_1$ denotes the identity representation.

Minimizing the free energy with respect to $P_x$ gives $P_x = \frac{1}{a}E_x + \frac{b}{a}\psi\left(E_x^2 + E_y^2 - E_z^2\right) + \cdots$. This expression directly connects the Landau expansion to optical susceptibilities. The first term corresponds to the linear dielectric response, $\chi_{xx} \propto \frac{1}{a}$, while the second term corresponds to a second-order nonlinear optical response, $\chi_{xxx} = \chi_{yyy} = -\chi_{xzz} \propto \frac{b}{a}\psi$. Thus, the SHG susceptibility is proportional to the order parameter $\psi$, implying that the SHG susceptibility can serve as a direct optical probe of the symmetry, sign, and magnitude of an underlying order parameter.

***Example: $B_2$ order parameter in the $4mm$-to-$mm2$ phase transition.*** As an example, we consider how an internal order parameter $\phi$ that transforms as the $B_2$ irrep of the $4mm$ point group can be detected by SHG response. The character table for the $4mm$ point group is given in Table 1.

Table 1. Symmetry character in $C_{4v}$ ($4mm$) point group

| $C_{4v}$ $(4mm)$ | E | $2C_4$ | $C_2$ | $2\sigma_v$ | $2\sigma_d$ | Basis functions |
|---|---|---|---|---|---|---|
| $A_1$ | 1 | 1 | 1 | 1 | 1 | $z, x^2 + y^2, \ldots$ |
| $A_2$ | 1 | 1 | 1 | -1 | -1 | $R_z, \ldots$ |
| $B_1$ | 1 | -1 | 1 | 1 | -1 | $x^2 - y^2, \ldots$ |
| $B_2$ | 1 | -1 | 1 | -1 | 1 | $xy, xyz, \ldots$ |
| $E$ | 2 | 0 | -2 | 0 | 0 | $\{R_x, R_y\}, \{x, y\},, \ldots$ |

From the character table, one may find that the relevant symmetry assignments for polar vector fields are

$$P_z, E_z \sim \boldsymbol{A_1}, \quad \{P_x, P_y\}, \{E_x, E_y\} \sim \boldsymbol{E}, \quad E_x^2 + E_y^2, E_z^2 \sim \boldsymbol{A_1}, \quad E_x^2 - E_y^2 \sim \boldsymbol{B_1}, \quad E_x E_y \sim \boldsymbol{B_2}$$

Since the internal order parameter transforms as $\phi \sim \boldsymbol{B_2}$, then the conjugate field to $\phi$ must also transform as $\boldsymbol{B_2}$ because the free energy must transform as $\boldsymbol{A_1}$. Thus, one needs to construct field-polarization combinations that also transform as $\boldsymbol{B_2}$.

From the direct product table for the $4mm$ point group given in Table 2,

Table 2. Direct product table for $C_{4v}$ ($4mm$) point group

| | $A_1$ | $A_2$ | $B_1$ | $B_2$ | $E$ |
|---|---|---|---|---|---|
| $A_1$ | $v$ | $A_2$ | $B_1$ | $B_2$ | $E$ |
| $A_2$ | $A_2$ | $A_1$ | $B_2$ | $B_2$ | $E$ |
| $B_1$ | $B_1$ | $B_2$ | $A_1$ | $A_2$ | $E$ |
| $B_2$ | $B_2$ | $B_1$ | $A_2$ | $A_1$ | $E$ |
| $E$ | $E$ | $E$ | $E$ | $E$ | $A_1 \oplus [A_2] \oplus B_1 \oplus B_2$ |

In the $4mm$ point group, one may find $P_z E_x E_y$ and $E_z(P_x E_y + P_y E_x)$ transform as $\boldsymbol{B_2}$. Therefore, the lowest-order SHG-related coupling terms can be written as $-b_1 \phi P_z E_x E_y$ and $-b_2 \phi E_z(P_x E_y + P_y E_x)$. Including the ordinary polarization energy and the linear coupling to the electric field, the symmetry-allowed free energy terms may be written as

$$F = F_0 + \frac{a_1}{2}(P_x^2 + P_y^2) + \frac{a_2}{2}P_z^2 - \boldsymbol{E} \cdot \boldsymbol{P} - b_1 \phi P_z E_x E_y - b_2 \phi E_z(P_x E_y + P_y E_x) + \cdots \quad (27)$$

Minimizing with respect to $P_x$, $P_y$, and $P_z$ leads to

$$P_x = \frac{1}{a_1}E_x + \frac{b_2}{a_1}\phi E_y E_z + \cdots, \; P_y = \frac{1}{a_1}E_y + \frac{b_2}{a_1}\phi E_x E_z + \cdots, P_z = \frac{1}{a_2}E_z + \frac{b_1}{a_2}\phi E_x E_y + \cdots \quad (28)$$

which gives $\chi_{xyz} = \chi_{xzy} \propto \phi$, $\chi_{yxz} = \chi_{yzx} \propto \phi$, and $\chi_{zxy} = \chi_{zyx} \propto \phi$. Thus, these SHG components provide direct symmetry-sensitive nonlinear optical signatures of the $B_2$ order parameter. This example illustrates how the Landau-invariant approach connects group-theoretical selection rules with experimentally measurable SHG tensor components.

## 2. SHG probes of polar orders in functional materials

SHG is particularly sensitive to inversion-symmetry breaking because its leading ED contribution is forbidden in centrosymmetric media and becomes active only when inversion symmetry is broken. In this section, we discuss how RA-SHG has been used to study polar materials. We summarize the symmetry principles that distinguish proper, improper, surface/interface-induced, and domain-wall polar orders, followed by reviewing representative SHG studies of polar order in heterostructures, atomically thin materials, oxide membranes, and domain walls.

### 2.1. Symmetry principles of polar and ferroelectric phase transitions

A central question in SHG studies of polar materials is how a measured nonlinear optical response relates to crystallographic symmetry and electric polarization. According to Neumann's principle, the physical property tensors of a crystal must be invariant under the symmetry operations of its point group [17]. Crystal structures in three dimensions are organized into seven crystal systems: cubic, hexagonal, trigonal, tetragonal, orthorhombic, monoclinic, and triclinic. Under crystallographic restriction, the crystallographic lattices are further classified into 32 point groups [18]. Among these, 21 point groups lack inversion symmetry and can support ED SHG. 20 of them are piezoelectric, while 10 point groups permit spontaneous polarization. Ferroelectricity requires the additional condition that this electric polarization be switchable by an external electric field. This hierarchy is useful because finite SHG response first establishes the loss of inversion symmetry, whereas RA-SHG further constrains the reduced point group and the symmetry-allowed polar axis.

#### 2.1.1. Pure ferroelectric phase transitions

A pure ferroelectric phase transition develops spontaneous polarization without producing distinct ferroelastic strain variants. In Aizu's notation [19,20], two ferroelectric states connected by the lost symmetry have different polarization states, $P_i \neq P_j$, but indistinguishable spontaneous strain states $\varepsilon_i = \varepsilon_j$. Equivalently, in Toledano's classification [21], the ferroelectric and paraelectric phases remain within the same crystalline system, and the polarization appears along a single symmetry-allowed direction of the parent phase. This restriction permits 14 parent paraelectric point groups and 23 point group-symmetry changes for pure ferroelectric phase transitions, with no such changes allowed in the cubic system. In the most common case, the primary order parameter is the polarization itself, producing $+P$ and $-P$ polar domains related by the lost inversion symmetry.

A representative case is $LiNbO_3$, where the transition occurs within the trigonal system ($\bar{3}m \rightarrow 3m$). The two orientation states correspond to $+P_z$ and $-P_z$ and are connected by inversion symmetry of the paraelectric phase. Because the strain state does not define additional variants, the domain variants are set by the sign of the polar order parameter rather than ferroelastic reorientation.

#### 2.1.2. Coupled ferroelectric-ferroelastic phase transitions

A ferroelectric-ferroelastic transition occurs when the symmetry lowering produces both spontaneous polarization and an orientation-dependent strain. The domain states are therefore distinguished not only by the sign or direction of the spontaneous polarization, but also by the orientation of strain. In Aizu's notation, both $P_i \neq P_j$ and $\varepsilon_i \neq \varepsilon_j$ can occur. When polarization is the primary order parameter, strain appears as a secondary order parameter through electrostrictive couplings.

The classic example is the cubic-to-tetragonal transition of $BaTiO_3$ ($m\bar{3}m \rightarrow 4mm$). The six orientation states, $+P_x, -P_x, +P_y, -P_y, +P_z$, and $-P_z$, include 180° ferroelectric pairs ($+P_z$ and $-P_z$) with the same tetragonal strain axis and 90° ferroelastic variants (e.g. $P_z \leftrightarrow P_x$ or $P_z \leftrightarrow P_y$) in which the tetragonal axis rotates. Thus, $BaTiO_3$ illustrates how a single polar instability can simultaneously generate switchable polarization and mechanically distinct domain variants.

#### 2.1.3. Improper and hybrid improper ferroelectricity

Improper ferroelectricity is defined by the identity of the primary order parameter rather than by the polar point group. In a proper ferroelectric, the polar vector itself is the primary order parameter. In an improper ferroelectric, one or more nonpolar structural distortions condense first, and polarization appears only as a secondary order parameter through symmetry-allowed coupling in the Landau free energy.

Two common routes are quadratic (inherent improper ferroelectricity) and trilinear coupling (hybrid improper ferroelectricity). In inherent improper ferroelectricity, a nonpolar primary order parameter, $\boldsymbol{\phi} = (\phi_1, \phi_2)$, produces a polar composite such as $\phi_1\phi_2$ or $\phi_1^2 - \phi_2^2$. By group theory, the polar vector irrep, $\Gamma_P$, must be contained in the symmetrized product $\Gamma_\phi \otimes \Gamma_\phi$. The quadratic composite then acts as a conjugate field to $P$ and fixes the sign of polarization after $\boldsymbol{\phi}$ condenses. This mechanism explains why the polar distortion can be secondary while still being locked to the primary structural domain state.

In hybrid improper ferroelectricity [22–25], the polarization is induced by two distinct nonpolar distortions, $\eta$ and $\psi$. Neither distortion alone carries a macroscopic polar moment, but their product transforms as a polar vector, allowing a trilinear invariant $\eta\psi P$. The symmetry condition is $\Gamma_P$ being contained in $\Gamma_\eta \otimes \Gamma_\psi$, with $\Gamma_\eta$ and $\Gamma_\psi$ generally distinct. A representative example is the layered Ruddlesden-Popper oxide $Ca_3Ti_2O_7$, where out-of-phase octahedral tilts and in-phase octahedral rotations together generate in-plane polarization [25]. The direct product of the corresponding tilt and rotation irreps contains the in-plane polar-vector irrep. Therefore, polarization appears only when both structural distortions are present.

#### 2.1.4. Boundary-induced polar order at surfaces and interfaces

A surface or interface becomes polar naturally, even when the bulk crystal is nonpolar because the boundary removes symmetry operations that exchange the two sides of the crystal. Inversion, horizontal mirrors, or twofold rotations that reverse the surface-normal are generally lost at a surface, and chemical inequivalence at a heterointerface further lowers the symmetry.

The surface or interface point group is obtained by reducing the bulk point group $G$ to the subgroup that leaves the boundary plane invariant. For a boundary with normal $n$, operations that send $n$ to $-n$ are removed unless the two sides of the boundary remain equivalent. The allowed polarization components are then those transforming as the totally symmetric representation of the reduced point group. Typically, the surface-normal component $P_n$ is allowed, whereas in-plane components require additional in-plane symmetry breaking. The commonly encountered polar point groups at surfaces and interfaces include $1$, $m$, $2$, $mm2$, $3$, $3m$, $4$, $4mm$, $6$, and $6mm$.

For example, an ideal (001) surface of a cubic centrosymmetric crystal with bulk point group $m\bar{3}m$ has the reduced surface point group of $4mm$. The bulk polar-vector ($T_{1u}$ under $m\bar{3}m$) decomposes into an out-of-plane component transforming as $A_1$ and in-plane components transforming as $E$ in $4mm$. Since $A_1$ is totally symmetric, $P_z$ is allowed at the surface, while a uniform in-plane polarization remains forbidden unless rotational or mirror symmetries are lost.

#### 2.1.5. Polar order confined to domain walls

A domain wall is not a homogeneous phase but a spatially confined boundary between two ordered states. Its symmetry is determined jointly by the symmetries of the adjacent domains, the operation that relates them, and the crystallographic orientation of the wall plane. Therefore, a domain wall can allow polar components that are forbidden in either the paraelectric parent phase or in the surrounding polar domains.

One route that leads to polar domain walls is geometric symmetry reduction. If the two adjacent domains are represented by order-parameter vectors $V_1$ and $V_2$, the intrinsic wall symmetry can be associated with the isotropy subgroup of the state $V_{DW} = V_1 + V_2$ [26]. For example, tetragonal $BaTiO_3$ with $4mm$ point group has $a_1/a_2$ domains with $V_1 = (P, 0,0)$ and $V_2 = (0, P, 0)$ connected by a 90° domain wall whose combined order parameter direction is [110] since $V_{DW} = V_1 + V_2 = (P, P, 0)$. This state has an orthorhombic polar symmetry $mm2$ and allows polarization at domain walls along the bisector direction. In this view, the domain wall polarity follows from the domain-pair geometry even before considering microscopic gradients.

A second route is gradient-induced improper polarity. When a nonpolar order parameter $\eta$ varies across the wall, Lifshitz-type invariants of the form $P\Phi(\eta, \nabla\eta)$ can become symmetry-allowed [27]. Here $\Phi(\eta, \nabla\eta) = \eta\frac{\partial\eta}{\partial s}$ where $s = x, y,$ or $z$, is a composite field built from $\eta$ and its spatial gradient. Since the gradient transforms as a polar vector, $\Phi$ may transform as the polar component $P$ even if $\eta$ itself is nonpolar. This mechanism is especially important in ferroelastic or antiferrodistortive materials, where octahedral rotations or tilts vary strongly at a wall. In nonpolar $CaTiO_3$ [28,29], for example, antiferrodistortive octahedral-tilt order parameters can stabilize an in-plane polar component confined to the wall. The resulting domain-wall polarity is improper, indicating that it is not a bulk instability, but a secondary response required by the symmetry of the inhomogeneous order parameter texture.

### 2.2. SHG studies of polar order beyond conventional bulk ferroelectrics

The use of SHG to study ferroelectricity dates back to the early development of nonlinear optical probes of noncentrosymmetric crystals [30]. Its particular strength is not only that ED SHG is activated by inversion-symmetry breaking, but also that RA-SHG provides access to the nonlinear susceptibility tensor governed by point-group symmetry. Thus, a measured polar plot can be compared with the tensor forms allowed for candidate point groups, providing a direct route from SHG intensity to symmetry reduction, polar-axis orientation, and domain selection.

Early studies used this principle to identify ferroelectric phase transitions, polar-axis reorientation, and domain switching in bulk materials [31–34]. In proper ferroelectrics such as $BaTiO_3$, the SHG tensor follows the macroscopic polar axis and distinguishes the tetragonal $4mm$, orthorhombic $mm2$, or rhombohedral $3m$ variants [35–37]. In improper and geometric ferroelectrics, tensor analysis is especially useful because the observed polar point group is secondary to nonpolar structural distortions. SHG then detects the polar irrep generated by quadratic or trilinear coupling to those distortions [23,38–40].

Recent work has extended this symmetry analysis beyond conventional bulk ferroelectrics to polar order emerging at surfaces [41], interfaces [42,43], domain walls [14,44–46], freestanding membranes [47–49], and two-dimensional van der Waals (vdW) materials [50–52]. In these systems, SHG is valuable because the polar volume may be too small for conventional diffraction or electrical measurements, while RA-SHG can still identify the local point group and the allowed tensor components. Since SHG studies of conventional bulk ferroelectrics have been extensively reviewed elsewhere [32,34], we here focus on cases where polarity is induced by local symmetry breaking, boundary conditions, dimensional confinement, or coupling to structural order parameters.

In the following Sections from 2.2.1 to 2.2.4, the representative SHG studies of polar and inversion-symmetry-breaking material systems are discussed. Table 3 summarizes representative applications of SHG to polar materials discussed as follows, highlighting their relevant symmetry information and principal findings.

Table 3. Summary of representative SHG studies of polar material systems

| Material/system | Relevant symmetry and main finding | Refs. |
|---|---|---|
| $SrRuO_3/SrTiO_3$ | Reduced surface/interface symmetry; RA-SHG separates surface and buried-interface polar reconstruction. | [41,53] |
| $LaAlO_3/SrTiO_3$ | Centrosymmetric bulk→noncentrosymmetric interface; SHG tracks interfacial electronic reconstruction. | [42,43,54] |
| $RuO_2$ thin film | Strain lowers centrosymmetric rutile symmetry to a polar state; RA-SHG identifies interface-induced polar metallicity. | [55] |
| $MoS_2$ | Odd layers: $\bar{6}m2$; even layers: $\bar{3}m$; SHG determines layer parity, orientation, stacking, and strain. | [56–60] |
| $CuInP_2S_6$, $In_2Se_3$ | Ferroelectric symmetry breaking to polar $m$ or $3m$; SHG tracks polar order and its thickness/temperature dependence. | [61,62] |
| Oxide membranes | Strain and curvature lower local symmetry; SHG images induced polarity in freestanding membranes. | [47–49] |
| Ferroelectric/polar domain walls | Domain walls possess lower local symmetry than surrounding domains; SHG resolves wall-specific polar components. | [14,45] |

### 2.2.1. Surface and interface polarity in oxide heterostructures

Oxide heterostructures provide a versatile platform for emergent interfacial phases because discontinuities at the boundary between two materials can drive reconstruction of degrees of freedom, such as lattice, charge and spin [63]. Such reconstruction can reduce the bulk point group to a surface or interface subgroup, as operations that exchange the two sides of the interface are generally removed. RA-SHG can then be used to determine the point group of the film, surface, substrate, or interface.

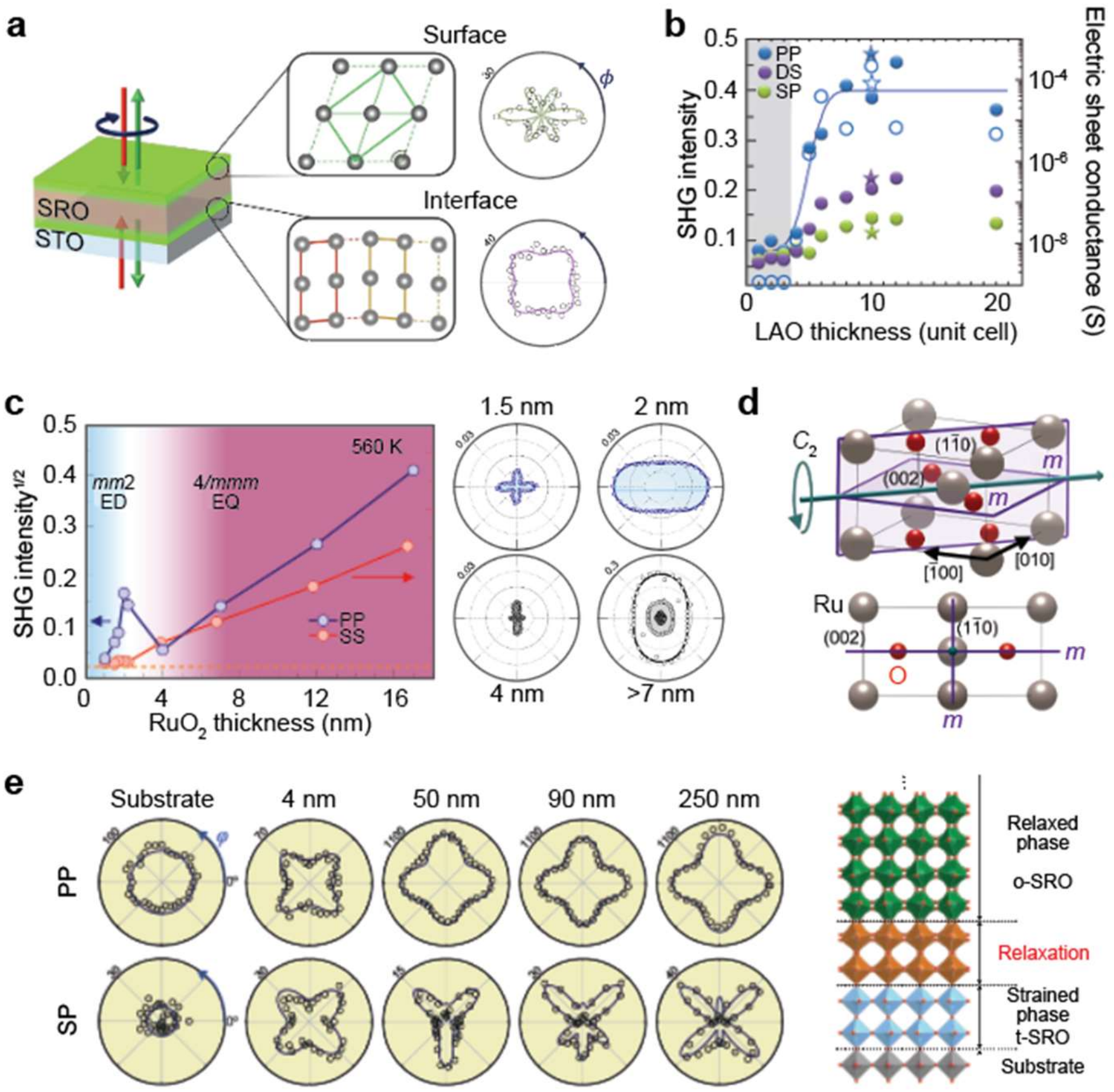

Fig. 4. Surface and interface symmetry characterization of oxide heterostructures using SHG. (a) Schematic of front-side and back-side reflection geometries used to distinguish surface and interface SHG responses in $SrRuO_3/SrTiO_3$ heterostructures. The corresponding RA-SHG patterns reveal different local crystalline symmetries at the film surface and interface. (b) Thickness-dependent SHG response of $LaAlO_3/SrTiO_3$, showing the onset of interface SHG near the critical $LaAlO_3$ thickness where interfacial conductivity emerges, as indicated by open circles. (c) Thickness-dependent SHG intensity and RA-SHG patterns of $RuO_2$ thin films, indicating symmetry lowering from $4/mmm$ phase to a polar $mm2$ phase in the ultrathin limit. (d) Reconstructed crystal structure of strained $RuO_2$, illustrating the interface-induced polar distortion and associated symmetry reduction. (e) RA-SHG patterns of $SrRuO_3$ films with different thicknesses, showing the evolution of surface/interface symmetry with strain relaxation. The schematic at right illustrates the emergence of a polar interface between strained and relaxed $SrRuO_3$ regions.

Figs. (a), (b), (c) and (d), and (e) are adapted from Refs. [41], [54], [55], and [53], respectively.

This principle has been used to distinguish surface and interface symmetries in epitaxial films. Roh *et al*. measured RA-SHG responses from $SrRuO_3$ thin films grown on an $SrTiO_3$ substrate using front-side and back-side reflection geometries to isolate the surface and interface responses, respectively (Fig. 4a) [41]. The surface signal shows a twofold RA pattern consistent with a monoclinic surface group $m$, whereas the back-side geometry shows a four-fold rotational pattern. The comparison demonstrates how RA-SHG separates two local symmetry environments that would be averaged together in thickness-dependent measurements.

Interface SHG has also been applied to the two-dimensional electron gas (2DEG) at oxide interface $LaAlO_3/SrTiO_3$ [42,43,54]. Bulk $SrTiO_3$ has the centrosymmetric cubic point group $m\bar{3}m$, but a (001) polar interface lowers the local symmetry to a noncentrosymmetric subgroup, described by $4mm$. In this group, tensor components such as $\chi_{zzz}$ and $\chi_{zxx} = \chi_{zyy}$ are allowed, whereas they are forbidden in the bulk. The SHG response follows the critical $LaAlO_3$ thickness for the onset of conductivity as shown in Fig. 4b, while polarization-resolved patterns show no additional symmetry lowering, indicating that the 2DEG mainly reflects electronic reconstruction within the same interfacial point group rather than a structural phase transition [43].

RA-SHG also revealed that polar symmetries are stabilized only in ultrathin epitaxial limits. In strained $RuO_2$ films on $TiO_2$, the bulk rutile symmetry $4/mmm$ is centrosymmetric and forbids ED SHG [55]. Below a few nanometers, epitaxial strain along [110] and an interface-induced polar field remove inversion and reduce the symmetry to the polar subgroup as $mm2$ (Figs. 4c and 4d). The appearance of PP channel components and the angular dependence of the RA pattern therefore identify not only inversion breaking, but the direction of the polar axis and the strain-selected orientation of the ultrathin altermagnetic polar phase.

Another example is the emergence of a polar phase in epitaxially grown $SrRuO_3$ thin films [53]. RA-SHG measurements on a series of $SrRuO_3$ thin films at various thicknesses directly revealed the emergence of a polar point group (Fig. 4e). As strain relaxes with increasing thickness, the symmetry-lowering from centrosymmetric $4/m$ to noncentrosymmetric $1$ emerges as the thickness changes from 4 nm to 50 nm, indicating the stabilization of a polar metallic state between the fully strained and the strain-relaxed phases. Revealed by the changed SHG tensor elements, RA-SHG directly links oxygen-octahedral-network relaxation to the emergence of a polar metallic state, rather than treating the thickness dependence as a simple change in optical absorption or film volume.

#### 2.2.2. Symmetry and ferroelectricity in atomically thin materials

SHG has become a standard symmetry probe for atomically thin transition-metal dichalcogenides such as $MoS_2$ [56,57]. Odd-layer 2H-$MoS_2$ lacks inversion symmetry and belongs to the noncentrosymmetric $\bar{6}m2$, whereas even-layer samples recover inversion symmetry and belong to $6/mmm$. The characteristic six-lobed RA-SHG pattern follows directly from the $\bar{6}m2$ tensor constraint (Fig. 5a), with the relation among in-plane tensor components $\chi_{yyy} = -\chi_{yxx} = -\chi_{xxy} = -\chi_{xyx}$ under an appropriate coordinate convention [58]. Figure 5b shows the SHG intensity dependence on layer number of $MoS_2$, demonstrating the restoration of the inversion symmetry in even-layer $MoS_2$ due to the stacking order. Building upon these foundational works, SHG has been extensively applied to investigate layer-dependent crystal orientation [56], stacking order [59], and strain-induced structural modifications (Figs. 5c and 5d) [60].

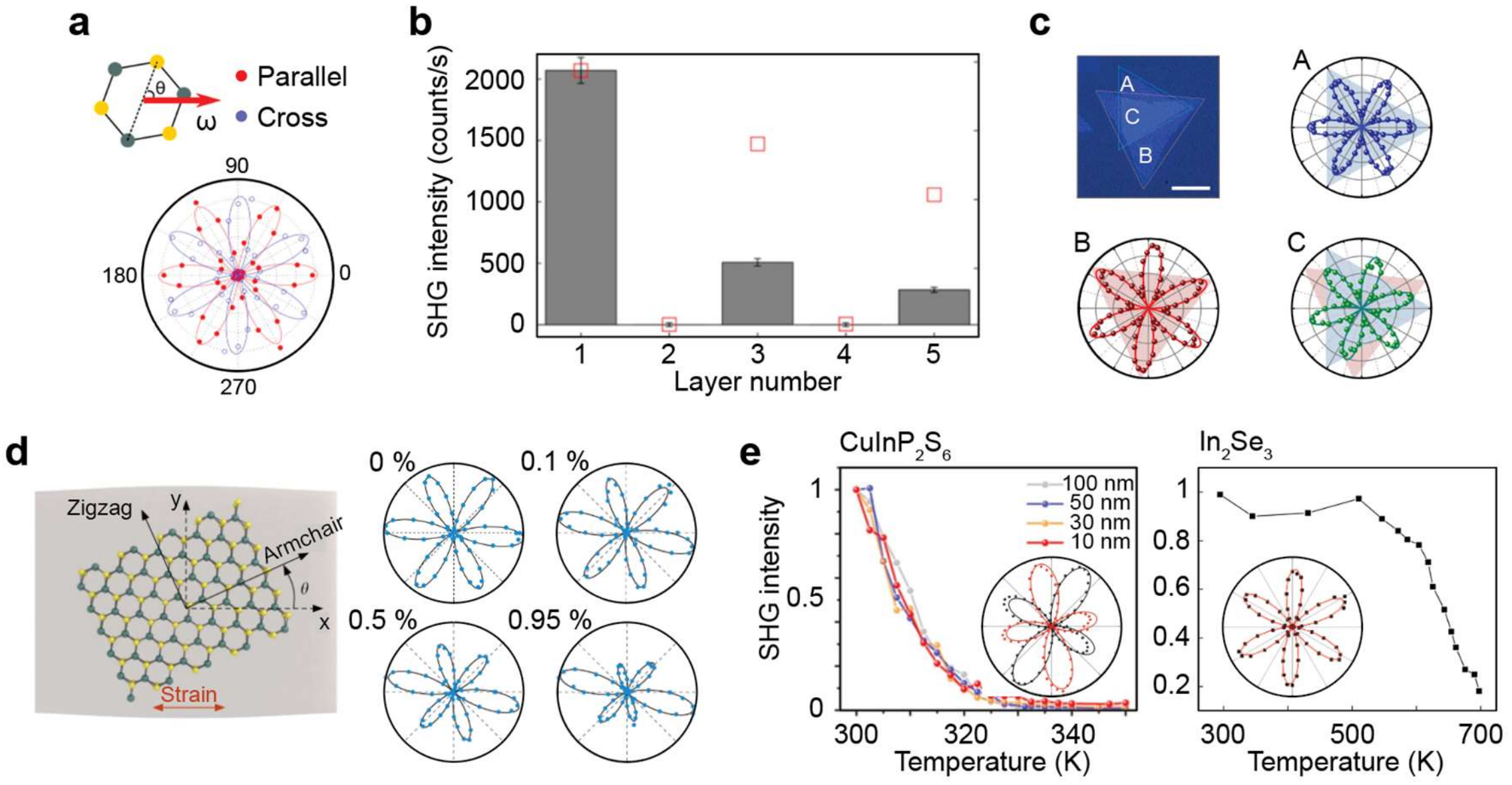

Fig. 5. Crystal symmetry characterization in vdW materials using RA-SHG. (a) Crystal structure of monolayer $MoS_2$ and representative RA-SHG patterns measured in parallel and cross polarization channels, consistent with $\bar{6}m2$ point group. (b) Layer-number dependence of SHG intensity, showing finite SHG responses from odd-layer $MoS_2$ and suppressed SHG responses from even-layer $MoS_2$. (c) RA-SHG patterns from $MoS_2$ flakes with different in-plane crystal orientations. (d) Schematic of uniaxially strained monolayer $MoS_2$, with strain applied along $x$-axis, and corresponding RA-SHG patterns under different strain magnitudes. (e) Temperature- and thickness-dependent SHG intensity from ferroelectric $CuInP_2S_6$ and $In_2Se_3$, showing the emergence of the ED SHG below their ferroelectric transition temperatures. Insets show representative RA-SHG patterns, consistent with polar monoclinic $m$ and polar trigonal $3m$ point groups for $CuInP_2S_6$ and $In_2Se_3$, respectively.

Figs. (a and b), (c), (d), and (e) are adapted from Refs. [58], [59], [60], and [61,62], respectively.

Two-dimensional ferroelectricity is challenging because depolarization fields and reduced dimensionality destabilize long-range polarization [64–66]. On the other hand, vdW ferroelectrics relax this constraint through weak interlayer bonding, and SHG has been used to follow symmetry changes in materials such as $CuInP_2S_6$ [61] and $In_2Se_3$ [62]. Temperature- and thickness-dependent SHG intensity identifies the ferroelectric transition (Fig. 5e), whereas RA-SHG determines whether the polar axis is primarily out-of-plane, in-plane, or coupled to the stacking order. Improper ferroelectricity in $NiI_2$ [67] further illustrates that SHG can detect a polar tensor generated by coupling to a nonpolar magnetic or structural order parameter, even when the final polar state is not a simple displacive ferroelectric.

### 2.2.3. Strain- and curvature-induced polarity in oxide membranes

Advances of epitaxial lift-off and transfer techniques of epitaxially grown thin film now allow freestanding oxide membranes to be fabricated from perovskite and related complex oxides [68]. Unlike vdW crystals, these membranes retain strong lattice connectivity and long-range crystalline order, but removal of the substrate changes the mechanical boundary conditions. In group-theoretical terms, curvature of membrane can reduce the parent point group by selecting a direction or bending axis, thereby making polar-vector components symmetry-allowed.

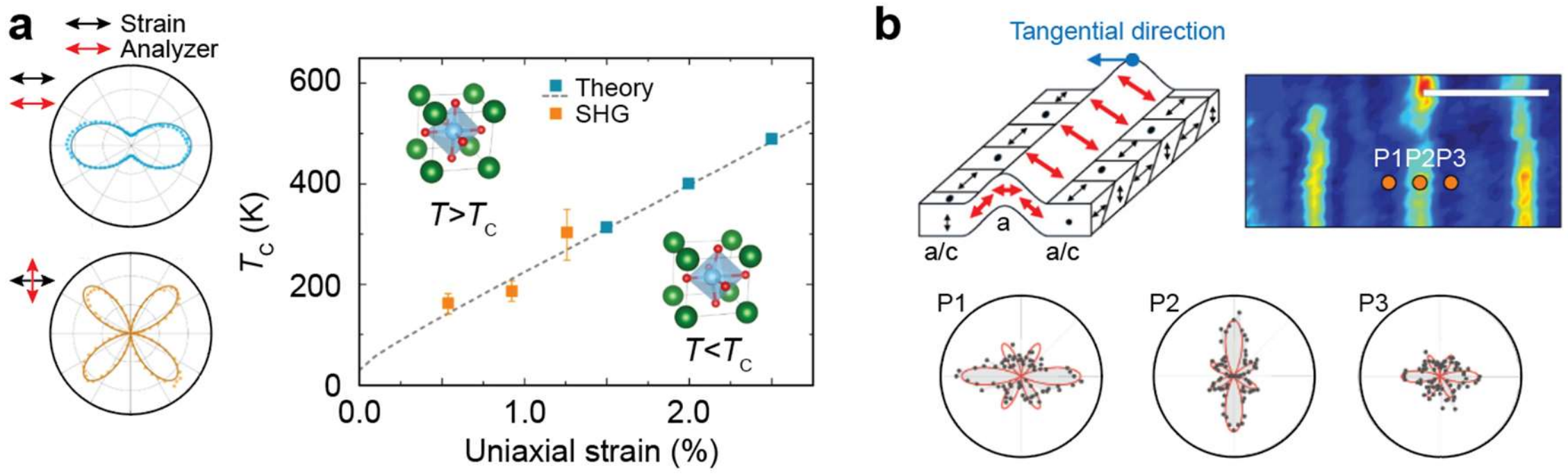

Fig. 6. Strain- and curvature-induced ferroelectricity in oxide membranes probed by SHG. (a) Polarization-resolved SHG patterns at left reveal strain-induced inversion-symmetry breaking and the emergence of a polar orthorhombic $mm2$ phase. The transition temperature increases with uniaxial tensile strain, demonstrating strain control of the ferroelectric instability. (b) Curvature-modified ferroelectric symmetry in rippled PbTiO3 membranes. The ripple geometry locally changes the strain and polarization configuration, producing locally distinct RA-SHG patterns at different positions along the ripple.

Figs. (a) and (b) are adapted from Refs. [48] and [49], respectively.

Xu *et al*. used SHG microscopy to demonstrate room-temperature ferroelectricity in tensilely strained $SrTiO_3$ membranes [48]. Under a tensile strain of approximately 2%, the crystal symmetry is lowered to the polar point group $mm2$ from bulk $m\bar{3}m$, with the polar axis along the strain direction. The RA-SHG patterns distinguish this $mm2$ state from a generic surface SHG contribution and agree with the observed 180° domain structure (Fig. 6a). Xu *et al*. also reported symmetry lowering in ultrathin $NaNbO_3$ membranes [47], where surface reconstruction and finite thickness stabilize a polar subgroup that is absent in the bulk antiferroelectric structure. In $PbTiO_3$ membranes [49], SHG microscopy results verify that the curved structure of the ripple modifies the ferroelectric property while the local tensor remains close to the tetragonal $4mm$ structure, illustrating how membranes can change ferroelectric polarization without necessarily changing the underlying polar point group (Fig. 6b).

### 2.2.4. Local polar symmetry at ferroelectric domain walls

Domain walls are internal interfaces whose symmetry can be lower than that of adjacent domains, allowing symmetry breaking and corresponding non-zero nonlinear susceptibility, which is forbidden in the domains. Such symmetry lowering explains why far-field SHG can detect polar domain walls even when the wall width is below the optical diffraction limit. Under normal incidence, the SHG response is predominantly governed by the in-plane components of the nonlinear susceptibility tensor, because the incident electric field is mostly in plane. When ferroelectric domains consist of oppositely polarized out-of-plane *c*-domains, the domain walls with non-zero polarization component can be specifically imaged via SHG microscopy.

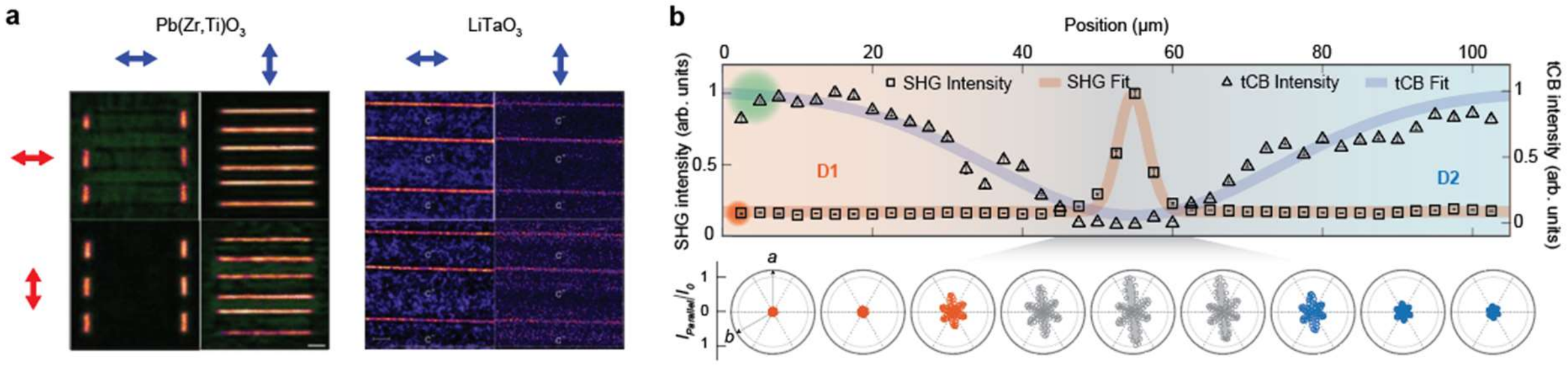

Fig. 7. Optical probing of polar domain walls by SHG. (a) Polarization-resolved SHG microscopy images of Pb(Ti,Zr)$O_3$ thin film and $LiTaO_3$ bulk crystal. Red and blue arrows indicate the polarization orientations of the incident fundamental and detected SHG light, respectively. For Pb(Ti,Zr)$O_3$, the SHG response at 180°-type domain walls strongly depends on both the wall orientation and polarization configuration, consistent with Néel-type domain walls. In contrast, domain walls in $LiTaO_3$ exhibit SHG responses for multiple polarization configurations, indicating mixed Néel-Bloch-type wall character. (b) Spatial profile of SHG and transmission circular birefringence (tCB) intensity across a domain wall in $Ni_3TeO_6$. The enhanced SHG intensity indicates local in-plane polar vector at the domain wall, whereas the suppressed tCB signal reflects reduced chirality. The RA-SHG patterns shown below the line profile reveal broken threefold rotation symmetry and emergent in-plane polarization at the wall.

Figs. (a) and (b) are adapted from Refs. [45] and [14], respectively.

Cherifi-Hertel *et al*. used SHG microscopy to study the polar domain walls in Pb(Ti,Zr)$O_3$ thin films and $LiTaO_3$ bulk crystals (Fig. 7a). Polarization-resolved SHG distinguished the wall-specific polar symmetry and showed that Pb(Ti,Zr)$O_3$ supports predominantly Néel-type domain walls, while $LiTaO_3$ exhibited mixed Bloch-Néel nature [45]. These results demonstrate that pre-existing structural order parameters, such as oxygen octahedral distortions, can vary spatially across a domain wall and generate symmetry-allowed polar components. In this sense, the polar domain wall is not simply associated with a remnant of the bulk ferroelectric polarization, but can be viewed as an improper polar order emerging from the reduced symmetry and spatial variation of the underlying structural order parameters at the wall.

A recent multimodal optical study of $Ni_3TeO_6$ [14] further illustrates this symmetry-based capability of RA-SHG. Bulk $Ni_3TeO_6$ belongs to the polar-chiral point group $3$, where polarity, chirality, and ferro-rotational order are interlocked. At the wall, RA-SHG changes from threefold rotation-symmetric pattern to asymmetric lobes as shown in Fig. 7b, evidencing the broken threefold rotation symmetry and emergent in-plane polarization. Combined with suppressed chirality from transmission circular birefringence, this reveals a mixed Néel-Bloch wall in which both parallel and perpendicular in-plane polar components are symmetry-allowed. This example demonstrates that SHG enables the determination of how the local wall point group reorganizes coupled polar, chiral, and ferro-rotational orders.

## 3. SHG probes of magnetic symmetry and spin order

SHG has historically focused on studying ferroelectric materials, owing to its sensitivity to inversion symmetry breaking (Section 2) [34]. Specifically, the polar order breaks the inversion symmetry of the electronic state wave functions, opening ED transition pathways allowed by selection rules. Recently, with advances in detection sensitivity, higher order SHG processes, such as EQ and MD, have been observed and investigated [69–73]. In addition, subtle symmetry breakings arising not from crystal structure but from magnetic order can now be probed. Unlike the ferroelectric and ferro-rotational orders (Section 4), magnetic orders break time-reversal (TR) symmetry, giving rise to distinct signatures in SHG radiation [74]. This TR-sensitive tensor response allows SHG to identify subtle magnetic orders and disentangling complex magnetic structures from other SHG sources.

Recently, magnetic SHG has proven to be highly effective for characterizing low-dimensional magnetic materials, such as vdW magnets, where traditional magnetic probes are challenging to implement. Going beyond intensity measurements, polarimetric techniques, namely RA-SHG, enable detection not only of long-range magnetic order, but also of detailed spin textures and magnetic fluctuations [75,76].

In this section, we introduce the general features of RA-SHG in magnetic materials, together with group theory analysis of SHG susceptibility tensors. In Section 3.1, we outline the group theory framework for analyzing RA-SHG data, including the treatment of TR symmetry breaking and magnetic point groups. In Section 3.2, we review the development from early studies of SHG in magnetic structures to state-of-the-art work on detecting magnetic order parameters, spin textures, and spin fluctuations in both three-dimensional (3D) ionic and two-dimensional (2D) vdW materials, highlighting the strength of SHG in studying magnetic systems. In Section 3.3, we present two examples: the 2D vdW magnet CrSBr and the 3D Weyl semimetal $Co_3Sn_2S_2$, to demonstrate how SHG can be used to study magnetic materials.

## 3.1. Magnetic point groups and time-reversal symmetry in SHG

Based on different transformations under TR operation, the SHG susceptibility tensors can be classified into two groups [77]. The first group consists of tensor elements that are invariant under TR operation (TR even, or *i*-type) and typically arise from crystal structures or charge orders that preserve TR symmetry. The second group consists of tensor elements that change sign under TR operation (TR odd, or *c*-type), which are commonly associated with magnetic orders. Depending on the nature of spin interactions, magnetism can contribute to both *i*- and *c*-type SHG: linear spin terms break TR and contribute *c*-type SHG, whereas bilinear spin-spin interactions preserve TR and contribute to *i*-type SHG. This classification has been clearly demonstrated in bilayer CrSBr [78].

We next outline the procedure for determining the nonzero tensor elements in the *i*- and *c*-type nonlinear susceptibility tensors. The first step is to identify the magnetic point group of the material. Depending on the role of TR operation plays, magnetic point groups can be classified into three categories [79,80].

### 3.1.1. Grey, black-and-white, and colorless magnetic point groups

The first category is the *grey point groups*, in which TR operation itself is a symmetry operation. There are 32 such groups, equal in number to the crystallographic point groups. They are denoted by a crystallographic point group followed by $1'$ in Hermann–Mauguin notation (e.g., $mmm1'$, $mm21'$, $21'$). Their symmetry operations include those of the associated crystallographic point group, plus their combinations with TR operation. For example, the grey group $21'$ contains both $C_2$ (two-fold rotation, from the crystallographic point group 2) and $C_2$T (two-fold rotation plus TR). Collinear antiferromagnets (AFMs) often fall into this category, such as checkerboard AFM in cuprate parent compound $Sr_2CuO_2Cl_2$ (Fig. 8a) [10] and layered AFM in CrSBr (Fig. 8b) [81].

These structures preserve TR symmetry at the point group level because TR operation followed by a lattice translation restores the magnetic configuration. The second category is the *black-and-white (B&W) magnetic point groups*, comprising 58 groups (Fig. 8c). In these, TR itself is not a symmetry operation, but combinations of TR with point operations are. For example, in the magnetic point group $2'$, $C_2T$ is a symmetry operation, whereas neither $C_2$ nor TR alone is. These groups have lower symmetry than their corresponding grey groups (e.g., $m'm'2$ is a subgroup of $mm21'$ and has a lower symmetry). Even-layer $CrI_3$ is one of the examples with magnetic point group $2/m'$ that falls into this category (Fig. 8c) [82].

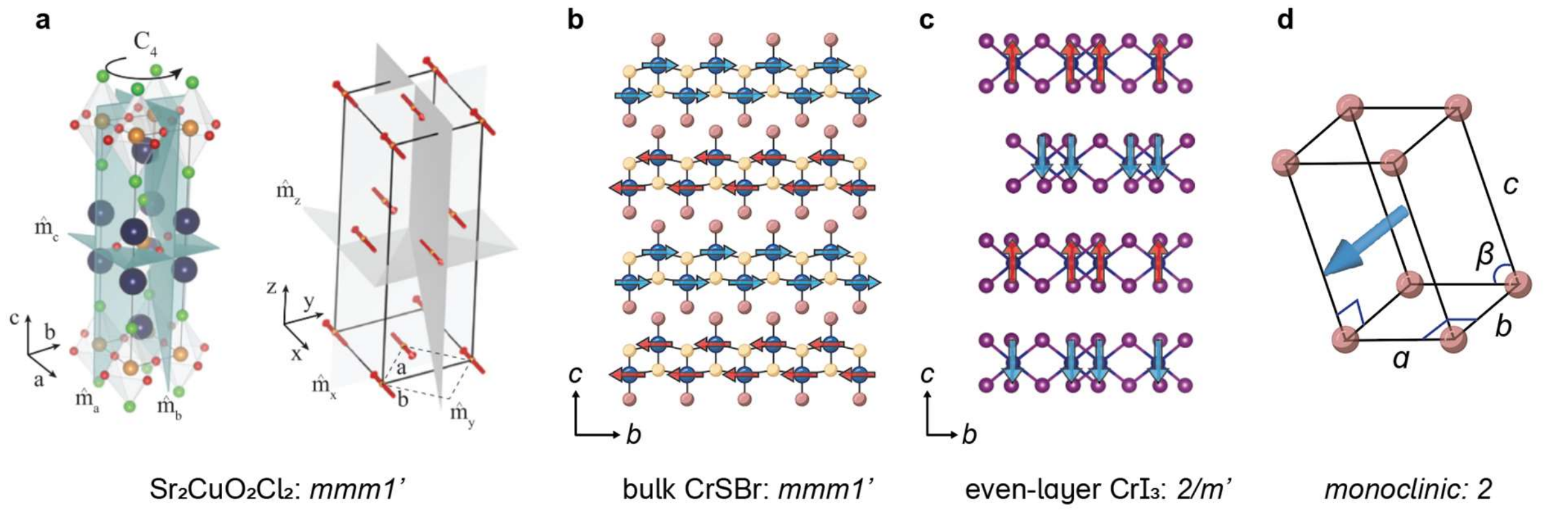

Fig. 8. Different categories of magnetic point groups and their examples. (a and b) Examples of grey point group $mmm1'$ $Sr_2CuO_2Cl_2$ (a) and bulk CrSBr (b). (c) An example of B&W point group $2/m'$, even layer $CrI_3$. (d) An example of a colorless monoclinic point group $2$ with ferromagnetism. Arrows indicate spins. Fig. (a) is adapted from [10].

The third category consists of the 32 crystallographic point groups themselves, referred to as *colorless* magnetic point groups. In these, neither TR nor its combination with spatial operations is a symmetry operation. Although they share the same notation as crystallographic point groups, they describe magnetic systems with rather low symmetry. For example, a monoclinic crystal with crystallographic point group $2$ and ferromagnetic (FM) order along the $C_2$ axis has a magnetic point group $2$ (Fig. 8d).

### 3.1.2. Tensor construction for magnetic SHG

We now determine the nonzero elements of the SHG susceptibility tensors for these magnetic point groups, as the basis for RA-SHG simulation and analysis. For colorless groups, the same procedure described in Section 1.3 can be applied following Neumann's principle. Taking the colorless group $mmm$ as an example,

$$\chi^{SHG}_{magnetic}(mmm) = \chi^{SHG}_{crystal}(mmm). \quad (29)$$

Here, the equality indicates identical positions of nonzero tensor elements.

For grey groups and B&W groups, the effect of TR operation must be treated explicitly. In grey groups, TR is a symmetry operation, meaning the system is TR-even and only *i*-type tensor elements are allowed. Since the system is invariant under TR, the $1'$ in the magnetic point group can be ignored when determining the tensor form. For example, the magnetic point groups $mmm$ and $mmm1'$ share the same pattern of nonzero tensor elements:

$$\chi_{magnetic}^{SHG}(mmm1') = \chi_{magnetic}^{SHG}(mmm). \tag{30}$$

The situation becomes more involved for B&W groups, as TR and point operations must be considered together. As noted earlier, a B&W group (*e.g.*, $2/m'$) is a subgroup of the corresponding grey group ($2/m1'$) and therefore has a lower symmetry. Consequently, it allows more nonzero tensor elements than the grey group. Among these, the elements inherited from the grey group are *i*-type (TR-even), by definition of grey groups. For example, for $2/m'$:

$$\chi_{magnetic}^{SHG,i-type}(2/m') = \chi_{magnetic}^{SHG}(2/m)\,. \tag{31}$$

The additional nonzero elements are *c*-type (TR-odd) and change sign under TR. To determine their form, we combine this property with Neumann's principle. Using $2/m'$ as an example, the symmetry constraints become:

$$O(C_2)\chi_{magnetic}^{SHG,c-type} = \chi_{magnetic}^{SHG,c-type} \tag{32}$$

$$(-1) \times O(m)\chi_{magnetic}^{SHG,c-type} = \chi_{magnetic}^{SHG,c-type} \tag{33}$$

where $O(x)$ denotes the action of symmetry operation $x$ on the tensor, i.e., Eqns. (16) and (23). The factor (-1) in Eqn. (33) reflects the TR-odd nature of the *c*-type susceptibility.

The full SHG tensor is then given by the sum of the two contributions:

$$\chi_{magnetic}^{SHG} = \chi_{magnetic}^{SHG,c-type} + \chi_{magnetic}^{SHG,i-type}. \tag{34}$$

Only the combination of the *i*- and *c*-type tensors fully captures the symmetry of the material system. Once the tensor form is obtained, radiation theory [7] can be used to simulate the RA-SHG patterns in different polarization channels [4].

To gain intuition, the distinction between *i*- and *c*-type tensors can be understood from how SHG responds to TR operation. The *i*-type (TR-even) contributions mostly originate from structural or charge-related effects that are insensitive to the direction of magnetic moments, remaining unchanged under TR. In contrast, the *c*-type (TR-odd) contributions are directly linked to magnetic order parameters and therefore reverse sign when all spins are flipped. As a result, *c*-type SHG carries explicit information about magnetic symmetry breaking, while *i*-type SHG usually serves as nonmagnetic background. Their interference is often essential in experiments, as it enables sensitivity to magnetic domains and subtle symmetry changes that would otherwise be inaccessible from intensity measurements alone.

A natural question is how these tensor elements relate to physical quantities such as crystal structure, magnetic order parameters, magnetic fluctuations and spin textures. In general, nonlinear susceptibility tensor elements reflect physical quantities with matching symmetry. For example, in a second-order phase transition, newly allowed tensor elements in the low-symmetry phase are proportional to the corresponding order parameter. More broadly, these tensor elements can be associated with different interaction terms in the underlying system [78].

### 3.2. Magnetic SHG from bulk, surface, and vdW materials

The study of magnetism by SHG dates back to the late 20$^{th}$ century. It was first observed in surfaces of 3d and 4d metals [83] and alloys [84] that magnetic long-range order can give rise to additional contributions to SHG radiation. With the implementation of RA-SHG and magnetic point group analysis, these studies expanded to various magnetic oxides [77,85] and magnetic thin film and interfaces [86,87], where SHG sources and their associated symmetries were identified through polarization-resolved measurements.

More recently, SHG has been applied beyond purely magnetic systems to probe magnetism embedded in or coupled to other exotic orders. These include Mott insulators and high-$T_c$ superconductors, where the strong electron-electron interactions play a crucial role [10,69–72,88], and magnetic materials with nontrivial electronic topology [76,89].

Besides, SHG has also become one of the mainstream techniques for investigating magnetism in vdW magnets since the discovery of vdW magnetism about a decade ago. This is because, first, the small size and sample volume of these 2D materials make traditional magnetic characterization techniques, such as magneto-transport or diffraction, difficult to implement. In contrast, optical probes are limited only by the optical diffraction, typically a few microns, making them well suited for studying 2D materials. Second, SHG is particularly powerful for probing AFM materials, in contrast to linear magneto-optical techniques such as Faraday rotation or magneto-optical Kerr effect (MOKE), which typically require nonzero net magnetization.

In this section, we survey the literature on SHG studies of magnetism over the past two decades, including both 3D ionic and 2D vdW materials. Table 4 summarizes representative magnetic material systems discussed from Sections 3.2.1 to 3.3.2, highlighting the relevant symmetry constraints and principal findings.

Table 4. Summary of representative SHG studies of magnetic states

| Material/system | Relevant symmetry and main finding | Refs. |
|---|---|---|
| $Cr_2O_3$; $HoMnO_3$ | Magnetic ordering changes the allowed SHG tensors; SHG tracks AFM order, domains, and spin reorientation. | [77,90] |
| $CrI_3$ | Even-layer: $2/m1' \rightarrow 2/m'$; magnetic ED SHG probes layered AFM order and magnetic domains. | [82] |
| $MnPS_3$; $CrPS_4$; $NiPS_3$ | Magnetic ordering lowers mirror/rotational symmetry; RA-SHG determines magnetic symmetry and Néel-vector orientation. | [91–94] |
| CrSBr | Surface and bulk have distinct magnetic symmetries; ED/EQ interference separates surface and bulk magnetic transitions. SHG captures phase transitions in few-layer CrSBr. | [81,95] |
| $MnBi_2Te_4$; $MnPSe_3$ | Surface-sensitive SHG reveals magnetic order and AFM domains distinct from the centrosymmetric bulk response. | [96,97] |
| Correlated oxides | Mirror-, inversion-, and/or time-reversal-related symmetry breaking reveals hidden magnetic/electronic orders. | [70–72,88,98,99] |
| $Co_3Sn_2S_2$ | Magnetic ordering breaks mirror symmetry in stages; EQ RA-SHG resolves distinct out-of-plane and in-plane magnetic components. | [89] |

### 3.2.1. Tracking magnetic phase transitions by SHG

Despite the absence of net magnetism in many AFMs, the long-range AFM order can break inversion symmetry, thereby opening additional ED SHG channels. Across such magnetic phase transitions, SHG typically shows a pronounced increase in signal and may exhibit symmetry changes in RA-SHG patterns. One example is $Cr_2O_3$, one of the first studied magnetic oxides by SHG: its paramagnetic state (crystallographic point group $\overline{3}m$) preserves inversion symmetry, whereas the AFM phase (magnetic point group $\overline{3}'m'$) breaks it, leading to an onset of SHG intensity at the Néel temperature $T_N$ (Fig. 9a) [77].

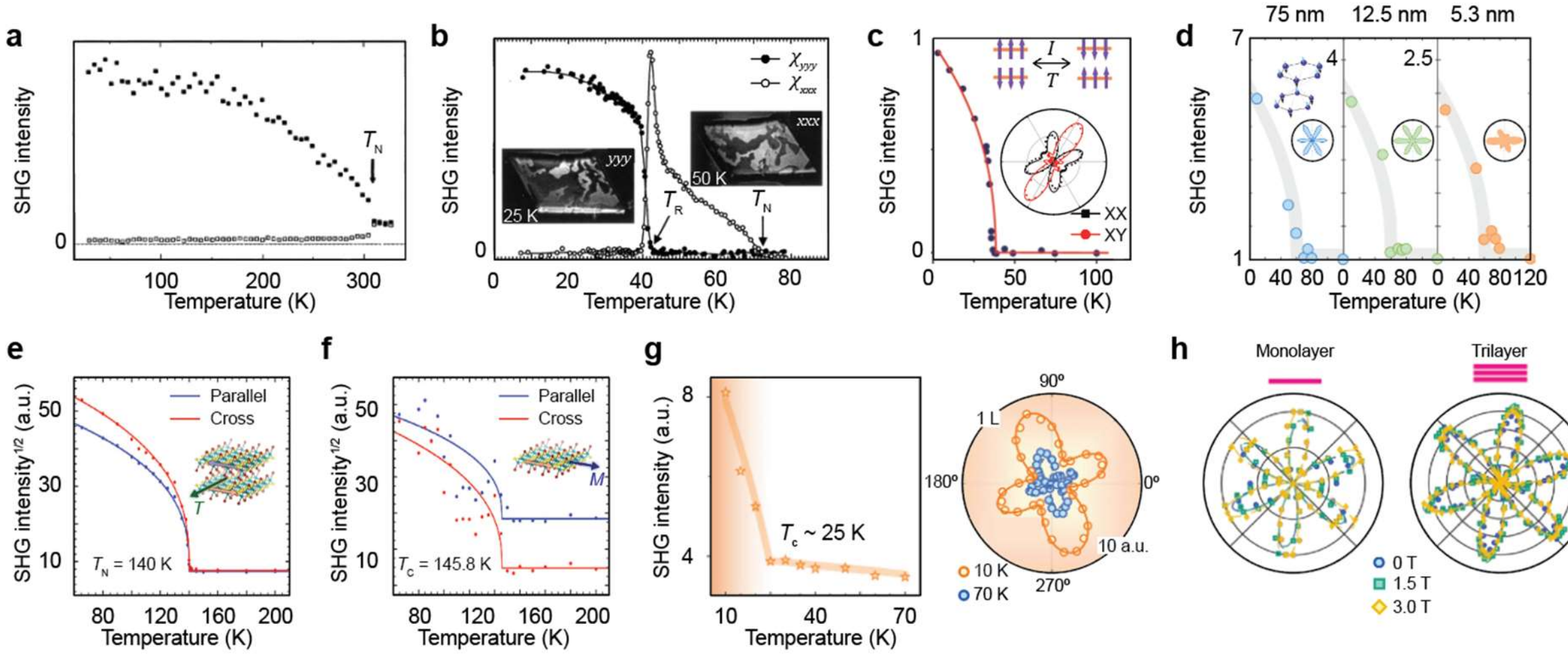

Fig. 9. SHG detection of magnetic order and spin-symmetry breaking in bulk and two-dimensional magnets. (a) Temperature dependence of the SHG intensity of $Cr_2O_3$ at 2.00 eV in one domain. Full and open squares refer to right and left circularly polarized light of the incoming laser beam, respectively. (b) Temperature dependence of SHG intensities from $\chi_{yyy}$ and $\chi_{xxx}$ of $HoMnO_3$. (c) Temperature-dependent SHG intensity of $CrI_3$ bilayer. Insets show a schematic of antiferromagnetic orders of bilayer $CrI_3$ and RA-SHG patterns for parallel (XX) and cross (XY) polarization setup. (d) Temperature dependent SHG intensity of MnPS3 samples with different thicknesses. Insets show an illustration of antiferromagnetic order of $MnPS_3$ and corresponding RA-SHG patterns at 10 K. (e and f) Square roots of the SHG intensity of the bilayer (e) and monolayer (f) CrSBr as a function of temperature. Inset shows the magnetic orders of bilayer and monolayer CrSBr. (g) Temperature-dependent SHG intensity of a monolayer $CrPS_4$ and RA-SHG patterns measured at 70 K (blue circle) and 10 K (orange circle). (h) RA-SHG patterns of monolayer (left) and trilayer (right) $MnBi_2Te_4$ samples under various applied magnetic fields.

Figs. (a), (b), (c), (d), (e and f), (g), and (h) are adapted from Refs. [77], [90], [82], [91], [95], [93], and [96], respectively.

SHG can also reveal magnetic phase transitions that involve more complex spin textures. For example, in the antiferromagnet $HoMnO_3$, both the crystal structure and magnetic order break inversion symmetry. Its point group symmetry reduces from $6mm$ at room temperature to $6'm'm$ below Néel temperature $T_N = 72$ K and further becomes $6'mm'$ at $T_R = 41$ K due to spin reorientation. These changes are reflected in the evolution of nonlinear susceptibility tensor elements (Fig. 9b) [90].

The even-layer form of one of the first discovered vdW magnets, $CrI_3$, serves as a 2D analogue of $Cr_2O_3$, where the crystal structure preserves inversion symmetry while magnetism breaks it. Even-layer $CrI_3$ has a monoclinic stacking with a crystallographic point group $2/m$. Below the Néel temperature $T_N \sim 40$ K, the layered AFM order with an out-of-plane easy axis leads to a magnetic point group of $2/m'$, breaking the inversion symmetry (Fig. 9c inset). Consequently, an SHG signal emerges below $T_N$, with RA-SHG patterns consistent with $2/m'$ (Fig. 9c) [82].

Another example is layered AFM $MnPS_3$. Neutron scattering shows that the spins in each layer are slightly tilted away from the crystallographic *c*-axis [100]. This tilting persists in few-layer samples and changes the point group symmetry from $2/m$ (inversion symmetric) in the paramagnetic phase to $2'/m$ (inversion broken) in the AFM phase, leading to an onset of SHG at $T_N$ across different thicknesses (Fig. 9d). Moreover, for thicknesses below 10 layers, the RA-SHG patterns exhibit additional mirror symmetry breaking inconsistent with $2'/m$, possibly due to strain effects (Fig. 9d inset) [91,92]. Similar behavior has been observed in other vdW magnets, including even-layer CrSBr, where the point group changes from $mmm$ to $mmm'$ (Fig. 9e) [95], and $CrPS_4$, where it changes from $2$ to $2'$ (Fig. 9g)[93].

Even when magnetic order does not break inversion symmetry, it can still contribute to higher-order SHG processes and lead to an enhancement of SHG signals across magnetic phase transitions, as observed in monolayer CrSBr (Fig. 9f) [95].

More recently, resonant SHG on atomically thin $NiPS_3$ shows how MD SHG can track the symmetry reduction associated with two-dimensional spin ordering [94]. In monolayer $NiPS_3$, the high-temperature SHG pattern with sixfold rotation symmetry evolves at low temperature into an anisotropic pattern with twofold rotation symmetry, indicating the selection of one magnetic domain orientation from symmetry-equivalent clock states. Gao *et al.* identified an intermediate BKT-like regime near $T_{BKT} \approx 141$ K from a lower-temperature pinned six-state clock AFM phase below $T_N \approx 120$ K, where discrete rotational symmetry is broken.

### 3.2.2. Surface magnetic orders probed by SHG

Interfaces between a material and its environment naturally break inversion symmetry, and the magnetic order at surfaces can therefore exhibit symmetries distinct from those in the bulk. However, surface and bulk SHG signals are often mixed. Two main approaches can be used to disentangle their contributions.

First, systematic thickness-dependent SHG measurements can help distinguish surface from bulk signals. If the SHG intensity is independent of thickness, it is likely dominated by surface contributions. For example, in $MnPS_3$ [101] and $MnBi_2Te_4$ [96] (Fig. 9h), SHG shows little dependence on layer number, indicating a surface origin.

Second, surface and bulk contributions can be separated based on their distinct symmetry properties. This approach will be illustrated in detail using bulk CrSBr in Section 3.3.1 [81].

### 3.2.3. Exotic stacking and hidden magnetic orders

Complex magnetic textures and hidden orders coupled to magnetism in correlated materials are often difficult to probe, especially beyond simple collinear FM or AFM states. SHG provides a powerful means to directly identify the symmetries of such exotic orders.

In underdoped cuprate high-$T_c$ superconductor $YBa_2Cu_3O_y$, an SHG signal emerges at temperature scales within the pseudogap regime. RA-SHG measurements indicate that this inversion-breaking phase has a magnetic point group $2'/m$ or $m1'$ (Fig. 10a) [71]. In another cuprate $Sr_2CuO_2Cl_2$, symmetry breaking near the Néel temperature was observed but found to be incompatible with a collinear AFM structure (Fig. 10b) [72]. Proposed explanations include FM components from spin canting, orbital currents and higher order magnetic multipoles [10]. At the same time, in the overdoped cuprate $(Bi,Pb)_2Sr_2CaCu_2O_{8+\delta}$, RA-SHG reveals a similar mirror-symmetry-breaking phase transition into $mm'm'$ symmetry near the crossover between Fermi-liquid-like and strange metal behavior (Fig. 10c) [98], which is absent in the underdoped regime. Together, these results suggest that the mirror and TR symmetry breaking may be intrinsic to the strange metal phase. In the Mott insulating iridate $Sr_2IrO_4$, a symmetry breaking into either $2'/m$ or $m1'$ is observed above the Néel temperature (Fig. 10d), and has been attributed to magneto-electric loop current order [70], surface magnetic order [88] or a nematic phase transition [99]. In Section 3.3.2, we will further show in detail how SHG captures multiple phase transitions and complex magnetic orders in the Weyl semimetal $Co_3Sn_2S_2$ [89].

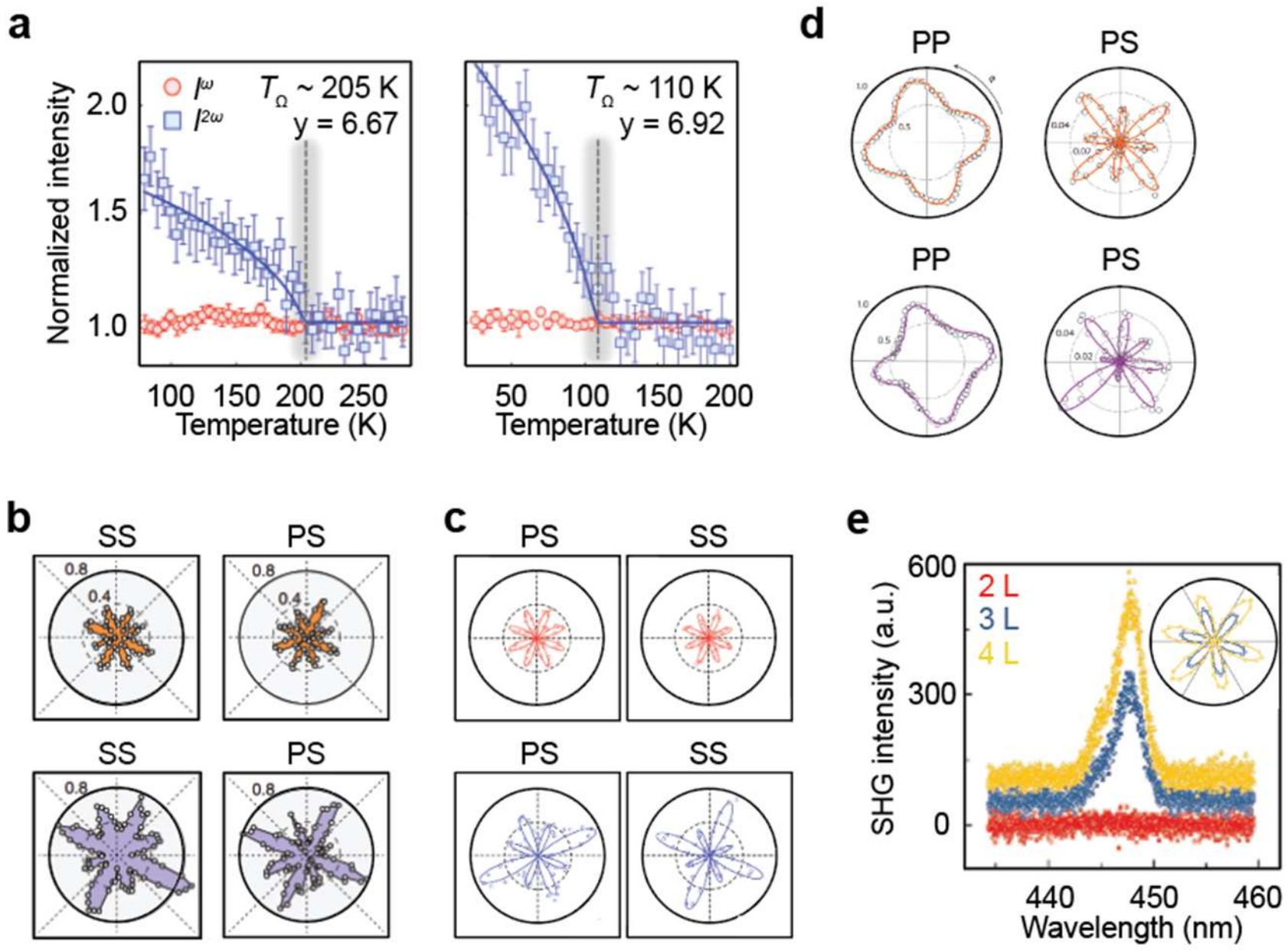

Fig. 10. RA-SHG detection of exotic magnetic orders in quantum materials. (a) Temperature dependence of the linear ($I^{\omega}$) and SHG ($I^{2\omega}$) intensity of $YBa_2Cu_3Oy$ with y = 6.67 (left) and y = 6.92 (right). (b) RA-SHG patterns in the SS and PS polarization channels of $Sr_2CuO_2Cl_2$ measured above (top) and below (bottom) a transition temperature. (c) RA-SHG patterns for Pb-Bi2212 collected under the PS and SS polarization channels measured at above (top) and below (bottom) a transition temperature. (d) RA-SHG data for $Sr_2IrO_4$ collected under PP and PS polarization channels at 295 K (top) and 170 K (bottom). (e) SHG spectra of the 2L (red), 3L (blue), and 4L (yellow) $VI_3$. Inset shows the RA-SHG patterns of the 3L and 4L $VI_3$ samples.

Figs. (a), (b), (c), (d), and (e) are adapted from Refs. [71], [72], [98], [70], and [102], respectively.

In vdW magnets, the stacking configuration of magnetic layers strongly influences exchange interactions and can determine the magnetic ground state. For example, bulk $CrI_3$ with rhombohedral stacking exhibits FM order, whereas few-layer $CrI_3$ adopts monoclinic stacking and favors an AFM ground state [103]. RA-SHG can be compared with expected magnetic point group symmetries, and deviations may indicate hidden stacking or magnetic orders. For instance, the hard vdW ferromagnet $VI_3$ shows six-fold RA-SHG patterns in trilayer and four-layer samples, but almost no SHG signal in bilayers. This behavior suggests unconventional chiral stacking and possibly nontrivial magnetic textures in thicker samples (Fig. 10e) [102]. Beyond identifying ground states, SHG tensor analysis can also reveal detailed spin textures and track their evolution under external fields, as demonstrated in bilayer CrSBr [78].

#### 3.2.4. Interference SHG imaging of magnetic domains

Degenerate magnetic domains in 2D magnets are experimentally challenging to distinguish, especially for AFM domains, which lack net magnetization and are therefore challenging to detect using magnetic circular dichroism or MOKE. These domains are related by time-reversal symmetry, and since most of the SHG detection schemes, such as PMT and CCD, measure only intensity without phase information, RA-SHG from both domains typically exhibits identical patterns.

This limitation can be overcome by leveraging interference between magnetic SHG and a reference SHG field. The reference can originate from two sources. First, it can arise from intrinsic, nonmagnetic SHG contributions of the material, such as those from the crystal structure. For example, in $Cr_2O_3$, interference between ED SHG from inversion-breaking magnetic order and higher-order SHG from the lattice enables imaging of AFM domains (Fig. 11a) [104]. Beyond AFM, SHG has also been used to study ferrotoroidal domains, another magnetic ferroic order characterized by a toroidal arrangement of magnetic moments, which breaks both inversion and time-reversal symmetry [105]. High-resolution SHG imaging of such domains has been demonstrated in $LiCoPO_4$ [106,107] and $NdB_4$ [108] (Figs. 11b and 11c).

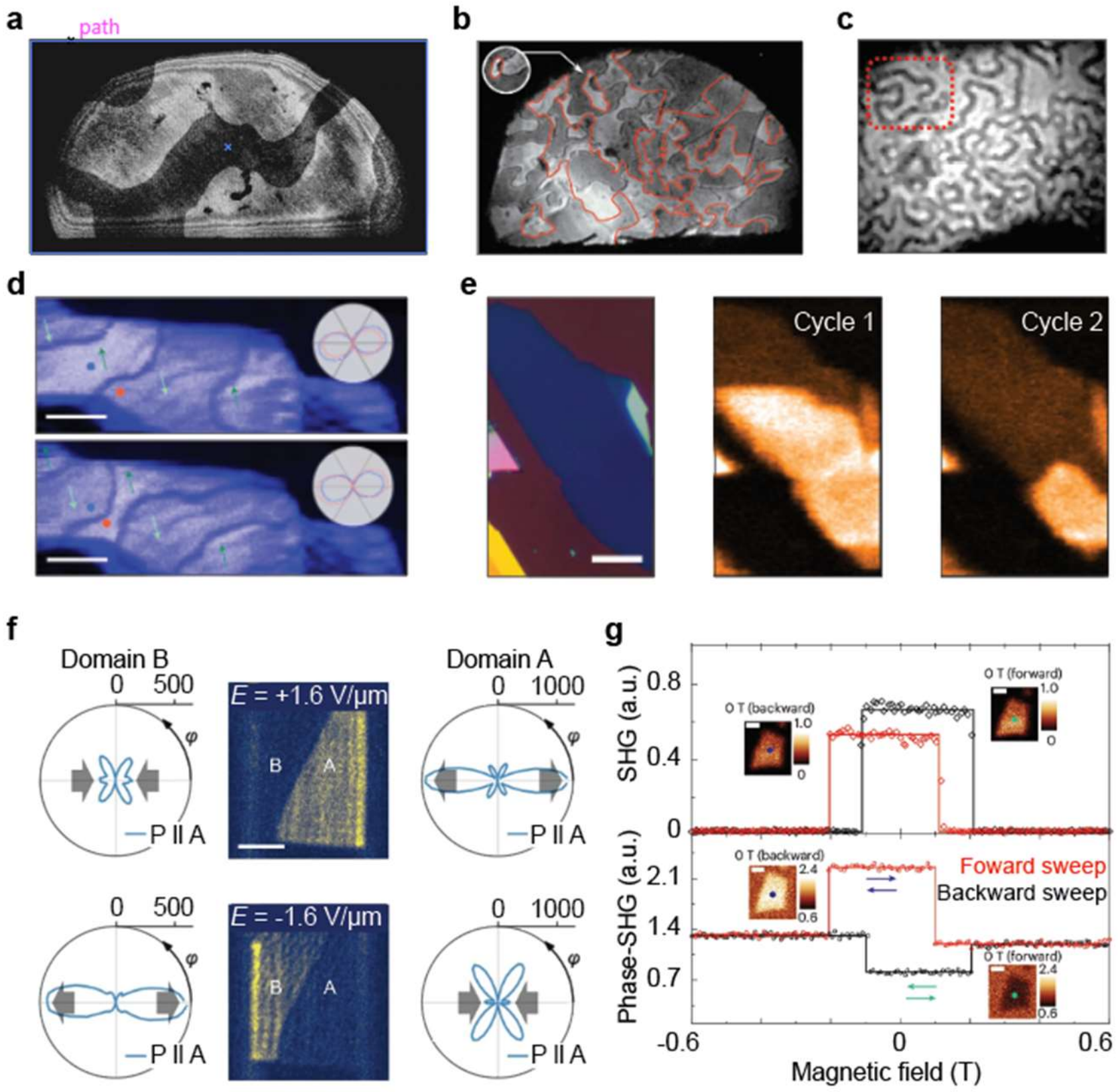

Fig. 11. SHG visualization and control of ferroic domains in magnetic materials. (a) SHG images of $Cr_2O_3$ measured at room temperature with right circularly polarized incident light. (b) Visualized coexisting AFM and ferrotoroidal (FT) domains of (100)-oriented $LiCoPO_4$ sample measured at 10 K. Intensity contrast is a consequence of the interference between AFM and FT SHG contributions. Red lines indicate FT domain walls. (c) SHG image of FT $NdB_4$. The curved dark lines correspond to the FT domain walls. (d) Spatial SHG mapping of $MnPSe_3$ measured at 5 K after the first (top) and the second (bottom) cooling cycles. Dark and bright green arrows indicate the directions of the Néel vector in AFM domains. Insets show RA-SHG patterns measured at the sample locations marked by blue and red dots. (e) Optical and SHG images of the 9L MnPS3 sample. SHG images (middle and right) are measured after the first and the second cooling cycles. (f) SHG microscopy images (middle) of $MnPS_3$ under applied electric fields of $E$ = +1.6 V/μm and $E$ = -1.6 V/μm. RA-SHG patterns measured for domains B and A in the parallel polarization channel are presented on the left and right sides of the microscopy images. (g) The magnetic-field hysteresis loop of SHG intensity (top) and phase-resolved SHG from 2L CrSBr sample. Arrows illustrate the magnetic structures in two AFM orders in 2L CrSBr.

Figs. (a), (b), (c), (d), (e), (f), and (g) are adapted from Refs. [104], [106], [108], [97], [92], [109], and [110], respectively.

In vdW 2D magnets, SHG remains effective for domain imaging despite the small sample size. Real-space SHG images of a thick $MnPSe_3$ sample (Fig. 11d) [97] and a nine-layer $MnPS_3$ sample (Fig. 11e) [92] show clear contrast between degenerate AFM domains. This contrast arises from interference between *c*-type ED SHG from AFM order and the *i*-type EQ SHG from the structure. The $\pi$ phase difference between ED SHG contributions from the two domains leads to distinct RA-SHG patterns and intensity contrast. Besides, the domain contrast can be tuned electrically: in a gated 90 nm $MnPS_3$ device, the SHG signal consists of *c*-type ED, *i*-type MD, and the electric-field induced contributions, with the latter enabling real-time control of domain contrast via gate voltage (Fig. 11f) [109].

Alternatively, the reference SHG field can be provided externally. For example, phase-resolved SHG measurements on bilayer CrSBr use a separate Y-cut quartz crystal as a reference, placed between the sample and the objective (Fig. 11g) [110].

## 3.3. Case studies of RA-SHG analysis of magnetic materials

In this section, we present two case studies with detailed analyses of RA-SHG data from magnetic materials, illustrating how RA-SHG can be used to map phase transitions, spin fluctuations and the evolution of spin textures. The first example focuses on bulk CrSBr, a van der Waals AFM. We show how SHG can detect both surface and bulk magnetic order parameters, as well as their fluctuations, and how these contributions can be disentangled based on symmetry (Section 3.3.1) [81]. The second example examines the Weyl semimetal $Co_3Sn_2S_2$, where RA-SHG captures two magnetic phase transitions and reveals the evolution of spin textures with temperature. We further show how the corresponding order parameters are encoded in the RA-SHG response (Section 3.3.2) [89].

### 3.3.1. Extraordinary bulk and surface magnetic phase transitions in bulk CrSBr

vdW materials have been recognized to host distinct surface properties compared to 3D ionic crystals, due to their much weaker interlayer coupling relative to intralayer interactions. In this example, we demonstrate RA-SHG measurements of bulk CrSBr that reveal distinct surface and bulk magnetic phase transitions. Notably, the surface exhibits an enhanced transition temperature compared to the bulk, with the bulk transition identified as an *extraordinary* phase transition [111,112].

CrSBr is a vdW antiferromagnet with a crystallographic point group $mmm$ in the paramagnetic state. Below $T_N = 132$ K, it develops layered antiferromagnetism along the *b*-axis, with a magnetic point group $mmm1'$. In the paramagnetic state, bulk CrSBr preserves inversion symmetry and thus, the leading order ED SHG radiation is forbidden. Besides, the two-fold rotational symmetry ($C_{2z}$) forbids SHG under normal incidence geometry. Oblique incidence RA-SHG overcomes this constraint and yields predominantly EQ SHG from a 0.5 mm thick CrSBr single crystal, as confirmed by comparison with simulations. The surface (point group $mm2$) contributes negligibly in this phase.

Below $T_N$, AFM order develops in both bulk and surface. The bulk has a magnetic point group of $mmm1'$ and according to Eqns. (29) and (30), the RA-SHG patterns should preserve the same symmetry as in the paramagnetic phase. However, a clear mirror symmetry breaking is observed experimentally, which arises from interference between the bulk and surface SHG. The magnetic surface has magnetic point group $m'm2'$, where inversion symmetry is broken, allowing *c*-type ED SHG. A summary of the relevant point groups, symmetry-allowed SHG channels, and crystal/magnetic structures is shown in Fig. 12a.

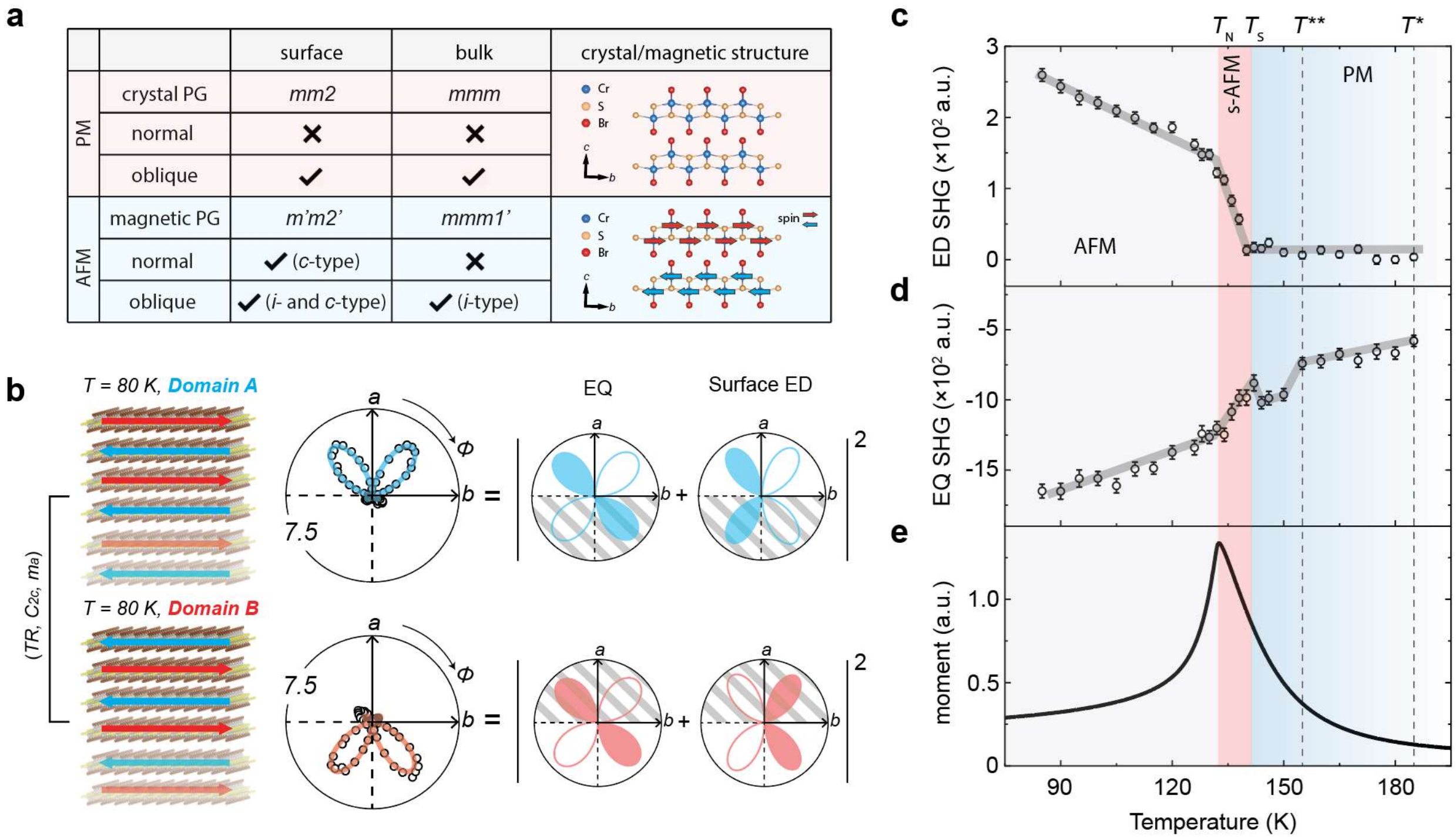

Fig. 12. Surface and extraordinary phase transitions in bulk CrSBr revealed by RA-SHG. (a) Summary of point group symmetries, SHG contributions and crystal/magnetic structures of paramagnetic (PM) and antiferromagnetic (AFM) bulk CrSBr. (b) Left: schematic of the layered crystal structure overlaid with the spin texture in domain states A and B, related by the time-reversal operation (TR), twofold rotation along the c-axis ($C_{2c}$) and mirror operation perpendicular to the *a*-axis ($m_a$). Right: RA-SHG patterns in the SS channel, resulting from the interference between the bulk EQ and the surface ED contributions. (c) ED and (d) EQ SHG tensors as a function of temperature. Gray curves serve as guides to the eyes. (e) Magnetization as a function of temperature measured under 1000 Oe magnetic field along the *b*-axis.

Figures are adapted from Ref. [81].

Under oblique incidence geometry, the interference between EQ and surface magnetic ED SHG leads to the observed mirror symmetry breaking. Because the surface ED SHG from the two degenerate AFM domains carries opposite phases, the resulting RA-SHG patterns are inverted between domains (Fig. 12b).

The interference model for the overall SHG radiation source is then constructed:

$$\vec{S}_{total} = \mu_0 \left( \frac{\partial^2 \vec{P}_{m'm2'}}{\partial t^2} - \nabla \cdot \frac{\partial^2 \overleftrightarrow{Q}_{mmm}}{\partial t^2} \right) \tag{35}$$

By fitting the temperature-dependent RA-SHG data with the simulated functional form, the evolution of both surface and bulk magnetic orders can be extracted. As shown in Fig. 12c, the surface ED SHG tensor onsets at $T_S = 140$ K, which is higher than the bulk Néel temperature $T_N = 132$ K. A kink is also observed at $T_N$, reflecting the influence of the bulk extraordinary phase transition on the surface order. The same $T_S$ is also captured in the EQ SHG channel as a peak (Fig. 12d), which is attributed to surface spin fluctuations $\langle \vec{S}_i \cdot \vec{S}_j \rangle$. This contribution is observed because it shares the same point group symmetry ($mmm$) as the bulk structure. A similar kink appears in the EQ SHG at $T_N$. The onset of bulk magnetic order at $T_N$ is independently confirmed by magnetization measurements on the same CrSBr sample, which shows a peak at $T_N$ and no feature at $T_s$ (Fig. 12e).

Taken together, the results in Figs. 12c-e reveal that surface magnetic order in CrSBr emerges at a higher temperature than the bulk, demonstrating distinct surface and bulk extraordinary phase transitions. This example highlights the effectiveness of RA-SHG in disentangling surface and bulk magnetic order through their distinct symmetry signatures.

### 3.3.2. RA-SHG detection of dual magnetic orders in $Co_3Sn_2S_2$

The second example focuses on the Weyl semimetal $Co_3Sn_2S_2$. The magnetic Co atoms form a kagome lattice, and the crystal structure has a crystallographic point group $\bar{3}m$. The system undergoes two phase transitions at $T_{c,1} = 175$ K and $T_{c,2} = 120$ K, with the associated spin textures still under debate [113–116]. In this section, RA-SHG is used to provide insight into the magnetic structure of $Co_3Sn_2S_2$ [89].

RA-SHG measurements at room temperature confirm the $\bar{3}m$ crystallographic point group. By comparing experiment data with simulations, EQ SHG has been identified as the dominant radiation source. Upon cooling, two notable changes in the RA-SHG response are observed. First, the overall SHG intensity increases. Second, the RA-SHG patterns rotate, indicating breaking of vertical mirror symmetry. Both the intensity and rotation angle exhibits clear onsets at the two transition temperatures $T_{c,1}$ and $T_{c,2}$ (Figs. 13a and 13b, left column).

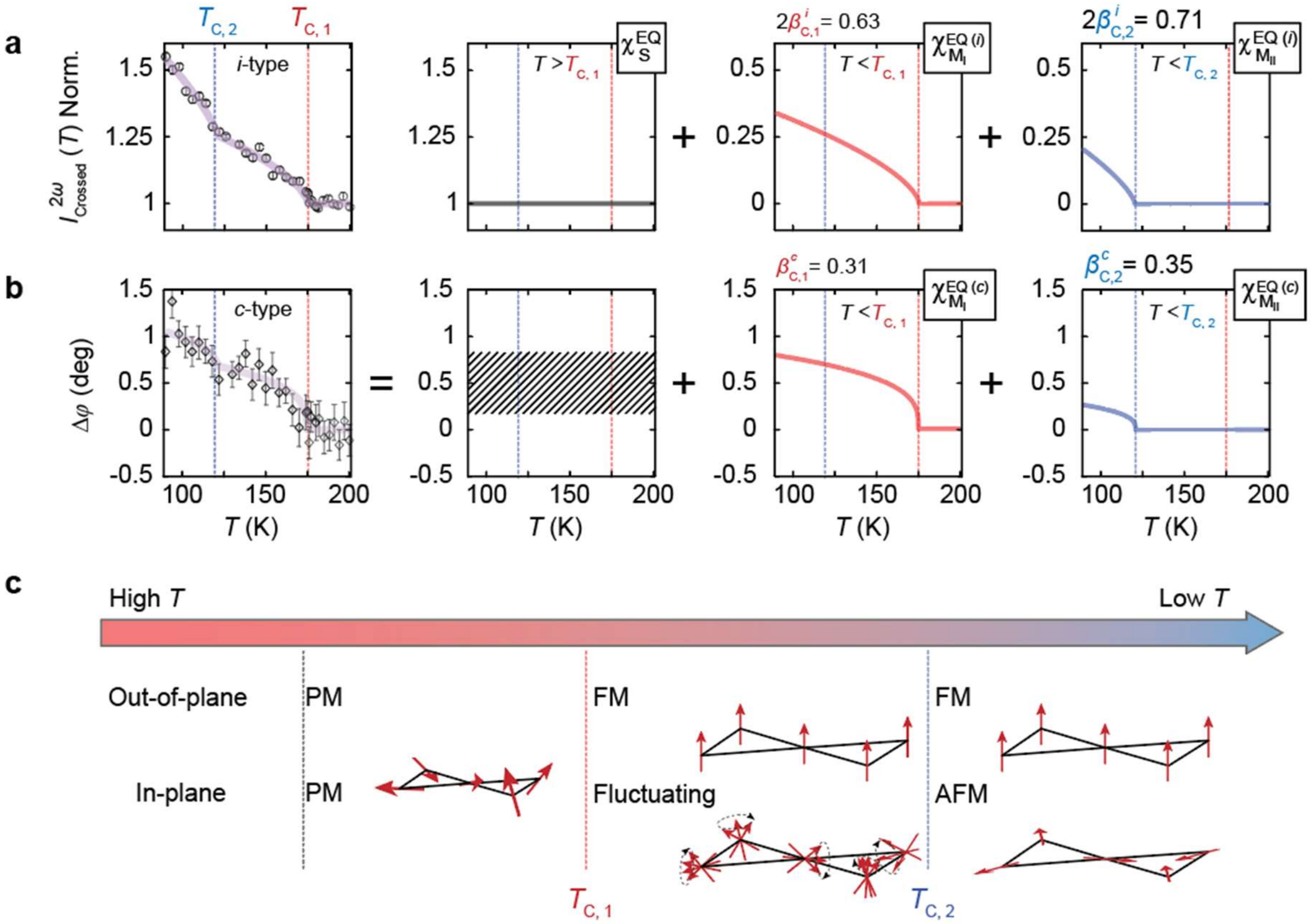

Fig. 13. Dual magnetic orders in $Co_3Sn_2S_2$ revealed by RA-SHG. Temperature dependence of the SHG intensity (a) and angular rotation $\Delta\varphi$ (b) from the TR-invariant *i*-type and TR-broken *c*-type contributions, respectively. $T_{C,1}$ and $T_{C,2}$ are marked by the dashed red line and blue line, respectively. (c) Diagram of magnetic phases across $T_{C,1}$and $T_{C,2}$ with spin components along the out-of-plane and in-plane directions.

Figures are adapted from Ref. [89].

The observed changes in RA-SHG are consistent with a symmetry reduction from $\bar{3}m$ to $\bar{3}m'$. Taking the crossed polarization channel as an example, Neumann's principle gives the functional form of the EQ SHG from the crystal structure in the $\bar{3}m$ paramagnetic phase as

$$I_{\text{Crossed}}^{\bar{3}m(s)}(\varphi) \propto \left|\chi_{yyzy}^{EQ(s)}\,\text{Sin}(3\varphi)\right|^2, \tag{36}$$

In the magnetic state under $\bar{3}m'$, the *i*-type tensor retains the same form as the structural contribution (Eqns. (30) and (31)):

$$\chi_{\bar{3}m}^{EQ(i)} = \chi_{\bar{3}m}^{EQ(s)} \tag{37}$$

In contrast, the *c*-type tensor $\chi_{\bar{3}m'}^{EQ(c)}$ has a different form. Following the procedure outlined in Section 3.2, the total crossed-channel RA-SHG can be written as the sum of structural, *i*-type and *c*-type contributions:

$$I_{\text{Crossed}}^{2\omega}(\varphi) \propto \left|\chi_{yyzy}^{\text{EQ}\,(s)}\sin(3\phi) + \chi_{yyzy}^{\text{EQ}\,(i)}\sin(3\phi) - \chi_{xxzx}^{\text{EQ}\,(c)}\cos(3\phi)\right|^2. \tag{38}$$

Equation (38) implies a rotation of the RA-SHG pattern from the paramagnetic to the magnetic states, with the rotation angle

$$\Delta\varphi = \frac{1}{3}\tan^{-1}\left(\frac{\chi_{xxzx}^{\text{EQ}\,(c)}}{\chi_{yyzy}^{\text{EQ}\,(s)} + \chi_{yyzy}^{\text{EQ}\,(i)}}\right). \tag{39}$$

The evolution of RA-SHG reflects two magnetic order parameters as follows:

**1) Above $T_{C,1}$**: only the structural term $\chi^{EQ(s)}$ contributes, yielding Eqn. (36) and preserving mirror symmetry.

**2) For $T_{C,2}$ < $T$ < $T_{C,1}$:** a magnetic order parameter $M_\text{I}$ emerges, introducing both *i*-type and *c*-type SHG contributions. The resulting RA-SHG follows Eqn. (38), exhibiting a finite rotation (Eqn. (39)) and mirror symmetry breaking. Since the rotation angle is small, $\chi_{xxzx}^{\text{EQ}\,(c,\text{M}_\text{I})} \ll \chi_{yyzy}^{\text{EQ}\,(s)} + \chi_{yyzy}^{\text{EQ}\,(i,\text{M}_\text{I})}$, and to leading order:

$$\Delta I_{\text{Crossed}}^{2\omega}(T) \propto \text{M}_\text{I} \propto \left|T - T_{c,1}\right|^{2\beta_{c,1}} \tag{40}$$

$$\Delta\varphi\,(T) \propto \text{M}_\text{I} \propto \left|T - T_{c,1}\right|^{2\beta_{c,1}}. \tag{41}$$

**3) Below $T_{C,2}$:** a second order parameter $M_{\text{II}}$ merges, contributing additional *i*-type and *c*-type terms. Keeping only leading-order contributions:

$$\Delta I_{\text{Crossed}}^{2\omega}(T) \propto A\left|T - T_{c,1}\right|^{2\beta_{c,1}} + B\left|T - T_{c,2}\right|^{2\beta_{c,2}}, \tag{42}$$

$$\Delta\varphi\ (T) \propto C\left|T - T_{c,1}\right|^{2\beta_{c,1}} + D\left|T - T_{c,2}\right|^{2\beta_{c,2}}, \tag{43}$$

where *A*, *B*, *C* and *D* are linear combinations of SHG tensor elements. A summary of the presence of SHG tensor elements and order parameters for different temperature ranges is given in Table 5.

Table 5. SHG tensor elements and order parameters of $Co_3Sn_2S_2$ in different temperature ranges

| **Temperature** | **T > 175 K** | **175 K > T > 120 K** | **120 K > T** |
|---|---|---|---|
| **Tensors** | $\chi^{\mathrm{EQ}(s)}$ | $\chi^{\mathrm{EQ}(s)}$, $\chi_{\mathrm{M_I}}^{\mathrm{EQ}\,(i)}$, $\chi_{\mathrm{M_I}}^{\mathrm{EQ}\,(c)}$ | $\chi^{\mathrm{EQ}(s)}$, $\chi_{\mathrm{M_I}}^{\mathrm{EQ}\,(i)}$, $\chi_{\mathrm{M_I}}^{\mathrm{EQ}\,(c)}$, $\chi_{\mathrm{M_{II}}}^{\mathrm{EQ}\,(i)}$, $\chi_{\mathrm{M_{II}}}^{\mathrm{EQ}\,(c)}$ |
| **Order parameters** | N/A | $M_{\mathrm{I}} \neq 0,\ M_{\mathrm{II}} = 0$ | $M_{\mathrm{I}} \neq 0,\ M_{\mathrm{II}} \neq 0$ |

Fitting the data in Figs. 13a and 13b using Eqns. (40)-(43) yields critical exponents $\beta_{C,1} = 0.31 \pm 0.02$ for $M_{\mathrm{I}}$ and $\beta_{C,2} = 0.35 \pm 0.02$ for $M_{\mathrm{II}}$. Based on these exponents, the first transition at $T_{c,1} = 175$ K is consistent with a 3D Ising universality class ($\beta_{3D,Ising} \approx 0.327$), indicating an easy-axis ferromagnetic order. The second transition at $T_{c,2} = 120$ K is consistent with a 3D XY universality class ($\beta_{3D,XY} = 0.348$), suggesting an in-plane 120° antiferromagnetic order. The resulting evolution of spin textures is illustrated in Fig. 13c.

## 4. SHG probes of hidden and multipolar orders

The tensor-symmetry framework for RA-SHG is particularly useful for hidden orders and multipolar orders, whose order parameters are not simple polar vectors or magnetic dipoles. Ferro-rotational order, for example, is described by an axial, time-even vector that preserves inversion symmetry and lacks a simple form of conjugate field like electric field to polarization or magnetic field to magnetic moment. It can nevertheless be detected by EQ SHG because the higher-order optical coupling can construct a composite field with the same symmetry as the ferro-rotational order parameter. Conversely, parity-odd multipolar nematic order in spin-orbit-coupled metals breaks inversion symmetry through an electronic pseudo-tensor order parameter. In that case, electric-dipole SHG is allowed in the ordered bulk and can distinguish the primary electronic hidden order from a secondary structural distortion.

In this section, we discuss these two representative examples to illustrate how SHG can reveal hidden order through symmetry analysis. The first case is ferro-rotational order, where EQ SHG detects an inversion-even axial order that is nearly invisible to conventional probes [73]. The second case is parity-odd electronic nematic order in $Cd_2Re_2O_7$, where RA-SHG separates a hidden multipolar nematic order parameter from the accompanying inversion-breaking lattice response [117]. Together, these examples establish SHG as a general experimental approach for identifying, imaging, and eventually controlling subtle ferroic and multipolar orders in quantum materials.

### 4.1. Ferro-rotational order revealed by higher-order SHG

Ferro-rotational order, also termed ferroaxial order, is a ferroic state whose order parameter is an axial vector generated by a collective rotational distortion of the crystal lattice. It is fundamentally different from conventional ferroelectric and ferromagnetic orders because it preserves both spatial-inversion and time-reversal symmetries. This symmetry property makes the order parameter difficult to detect by ordinary linear optical, electric, or magnetic probes, but it also gives ferro-rotational materials unusual stability because they do not carry a macroscopic polarization or magnetization. Recent progress in higher-order SHG response, circular-dichroic SHG, electrogyration, and ultrafast phonon control has transformed ferro-rotational order from a largely symmetry-based concept into an experimentally accessible ferroic state. This section introduces the symmetry basis of ferro-rotational order, explains why its conjugate field is nontrivial, and reviews representative optical approaches for observing, imaging, and controlling ferro-rotational domains.

### 4.1.1. Ferro-rotational order and relevant symmetry reduction

Ferro-rotational order can be viewed as a macroscopic vortex-like structural rotation described by an axial vector **A** as illustrated in Fig. 14. Microscopically, this axial moment may be represented as an electric toroidal moment **A** that is proportional to $\sum r_i \times P_i$, where $P_i$ is a local polar displacement or ED located at position $r_i$ with respect to the center of the structural unit. In this arrangement, the local dipoles form a head-to-tail loop, and the structure can possess a finite axial moment without producing a net polar vector. Because **A** is even under both spatial inversion and time reversal, ferro-rotational order does not by itself imply ferroelectricity, ferromagnetism, or chirality. Its essential symmetry signature is instead the loss of mirror planes that contain the rotation axis, while a mirror plane perpendicular to the axis may remain allowed (Fig. 14). This distinction is important because a ferroaxial crystal can be nonchiral even though it contains two domain states with opposite signs of **A**.

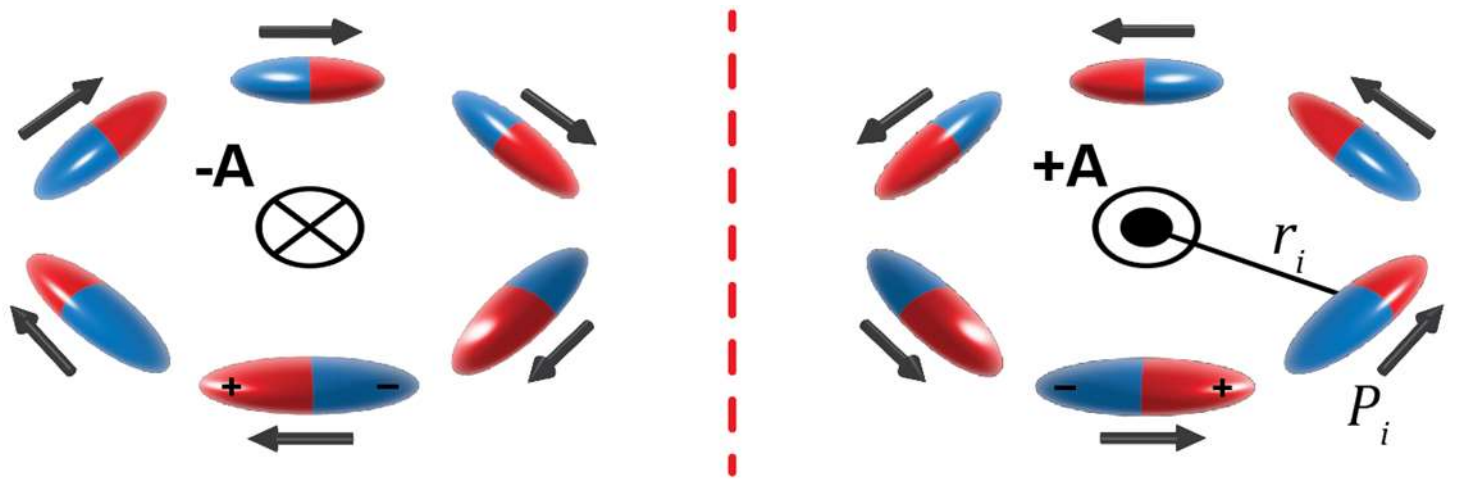

Fig. 14. Schematic illustration of two degenerate ferro-rotational domain states. Local electric dipoles $P_i$ located at position $r_i$ with respect to the center form a vortex-like head-to-tail arrangement. The clockwise and counterclockwise arrangements generate opposite ferroaxial moments **-A** and **+A**. The red dashed line denotes a vertical (diagonal) mirror plane that relates the two degenerate domain states.

Hlinka et al. provided a systematic symmetry classification of ferroaxial transitions within the Aizu-species framework [20,118]. Among the 212 nonmagnetic structural phase-transition species, 124 are ferroaxial. Many of them are also ferroelectric or ferroelastic, but a small subset are purely ferroaxial, meaning that the transition creates an axial order parameter without producing macroscopic polarization or spontaneous strain. The pure ferroaxial transitions are particularly important for isolating the intrinsic physics of ferro-rotational order because their domain states are distinguished primarily by the sign of **A**. Representative pure ferroaxial symmetry reductions are summarized in Table 6. In these cases, the transition lowers the parent symmetry by removing vertical or diagonal mirrors containing the principal axis, thereby generating two degenerate domain states related by the lost mirror operation.

Table 6. A list of purely ferro-rotational transitions with corresponding irreducible representations and relevant macroscopic symmetry breaking.

| Pure ferro-rotational transition pathway | Corresponding irreducible representation | Macroscopic symmetry breaking |
|---|---|---|
| $\bar{4}2m \rightarrow \bar{4}$ <br> $(D_{2d} \rightarrow S_4)$ | $A_2$ | Loss of two diagonal mirror planes |
| $4/mmm \rightarrow 4/m$ <br> $(D_{4h} \rightarrow C_{4h})$ | $A_{2g}$ | Loss of all vertical/diagonal mirror planes; inversion retained |
| $\bar{3}m \rightarrow \bar{3}$ <br> $(D_{3d} \rightarrow S_6)$ | $A_{2g}$ | Loss of three diagonal mirror planes; inversion retained |
| $\bar{6}2m \rightarrow \bar{6}$ <br> $(D_{3h} \rightarrow C_{3h})$ | $A_2'$ | Loss of three vertical mirror planes |
| $6/mmm \rightarrow 6/m$ <br> $(D_{6h} \rightarrow C_{6h})$ | $A_{2g}$ | Loss of vertical mirror planes; inversion retained |

#### 4.1.2. Conjugated fields for ferro-rotational order

A central difficulty in ferro-rotational physics is the absence of a simple uniform conjugate field. In Landau theory, a field can linearly couple to an order parameter only if it transforms in the same way under all symmetry operations of the high-symmetry phase. For ferroelectrics, this role is played by the electric field because both the polarization *P* and the electric field *E* are polar, time-even vectors. For ferromagnets, the magnetic field couples directly to the magnetization *M*. In contrast, the ferro-rotational order parameter **A** is an axial, time-even vector. A uniform electric field is polar, and a uniform magnetic field is time-odd. Therefore neither can serve as a direct conjugate field to **A**.

This distinction is clarified by comparing ferro-rotational order with magnetic ferrotoroidal order. A magnetic toroidal moment is commonly represented by a circulating arrangement of magnetic moments, **T** proportional to $\sum r_i \times M_i$, and it is odd under both spatial inversion and time reversal [119]. As a result, its symmetry and conjugate fields differ from those of the electric toroidal moment associated with ferro-rotational order. The ferro-rotational moment **A** proportional to $r \times P$ is instead inversion-even and TR-even. Its direct conjugate field must therefore be an axial, TR-even quantity, such as a curl-like or cross-product combination of two polar fields. This requirement explains why ferro-rotational domains are difficult to switch using conventional static electric or magnetic fields.

Recent theoretical work has made this idea more concrete by showing that electric toroidal dipoles generate antisymmetric, rotational components in response tensors (Fig. 15) [120]. In a simple

rank-2 response, the presence of **A** allows an input along one transverse direction to induce a conjugate response along the orthogonal direction, with the sign of the response reversing when **A** is reversed. This rotational-response viewpoint naturally connects ferro-rotational order to transverse electromagnetic, thermopolar, magnetostrictive, and nonlinear optical effects. Since a static macroscopic field with the symmetry of $\nabla \times E$ is not readily accessible in ordinary experiments, high-order optical processes and dynamically generated fields provide practical routes for probing and controlling **A**.

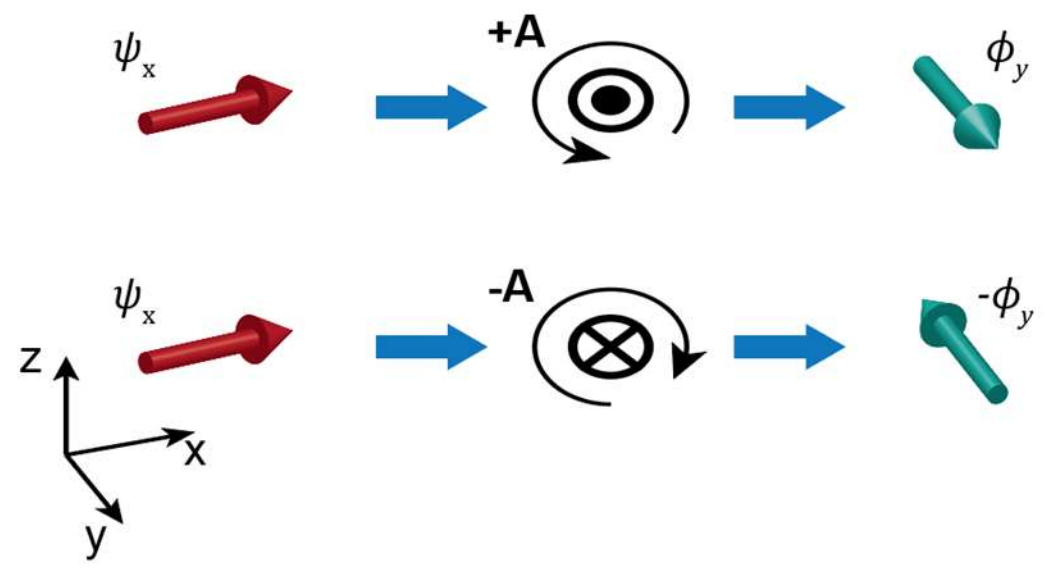

Fig. 15. Rotational response induced by an electric toroidal dipole. In a ferro-rotational state, the axial order parameter +**A** (-**A**) allows an applied field $\psi_x$ to induce a conjugate response $\phi_y$ $(-\phi_y)$ along the transverse direction. This illustrates that the axial order parameter can be detected through off-diagonal or rotational components of response tensors.

The same symmetry logic also motivates recently proposed transverse responses governed by electric toroidal invariance [121]. Unlike the familiar transverse responses associated with broken inversion or time-reversal symmetry, ferro-rotational systems can preserve both spatial inversion and time-reversal while still allowing Hall-like higher-order transverse susceptibilities. In this view, the sign of the transverse response is controlled by the sign of **A**, making response tensors themselves a possible probe of ferro-rotational domains.

### 4.1.3. Optical detection and imaging of ferro-rotational domains

The modern concept of ferroaxiality grew from earlier symmetry classifications of ferroic species and electrogyration [20], but it gained renewed importance through studies of multiferroics in which a structural axial degree of freedom mediates magnetoelectric coupling. In $Cu_3Nb_2O_8$ [122] and $CaMn_7O_{12}$ [123], for example, the electric polarization cannot be fully explained by the conventional cycloidal-spin mechanism. Instead, the coupling between magnetic chirality and a ferroaxial structural component provides a route for generating polarization perpendicular to the spin-rotation plane. These studies established the physical importance of axial structural order, but they did not by themselves provide a direct real-space image of ferro-rotational domains.

A decisive breakthrough was achieved in $RbFe(MoO_4)_2$ using high-sensitivity rotational-anisotropy SHG. Because the high-temperature phase is centrosymmetric, electric-dipole SHG is forbidden in the bulk and the observed signal must arise from higher-order multipoles, most importantly the electric-quadrupole contribution. Jin *et al*. [73] showed that the room-temperature RA-SHG pattern is consistent with the centrosymmetric $\bar{3}m$ point group, while cooling through $T_c$ produces a rotation of the RA-SHG pattern and removes the mirror symmetries expected for the low-temperature $\bar{3}$ ferroaxial phase. This is discussed in detail in the Section 4.1.4.

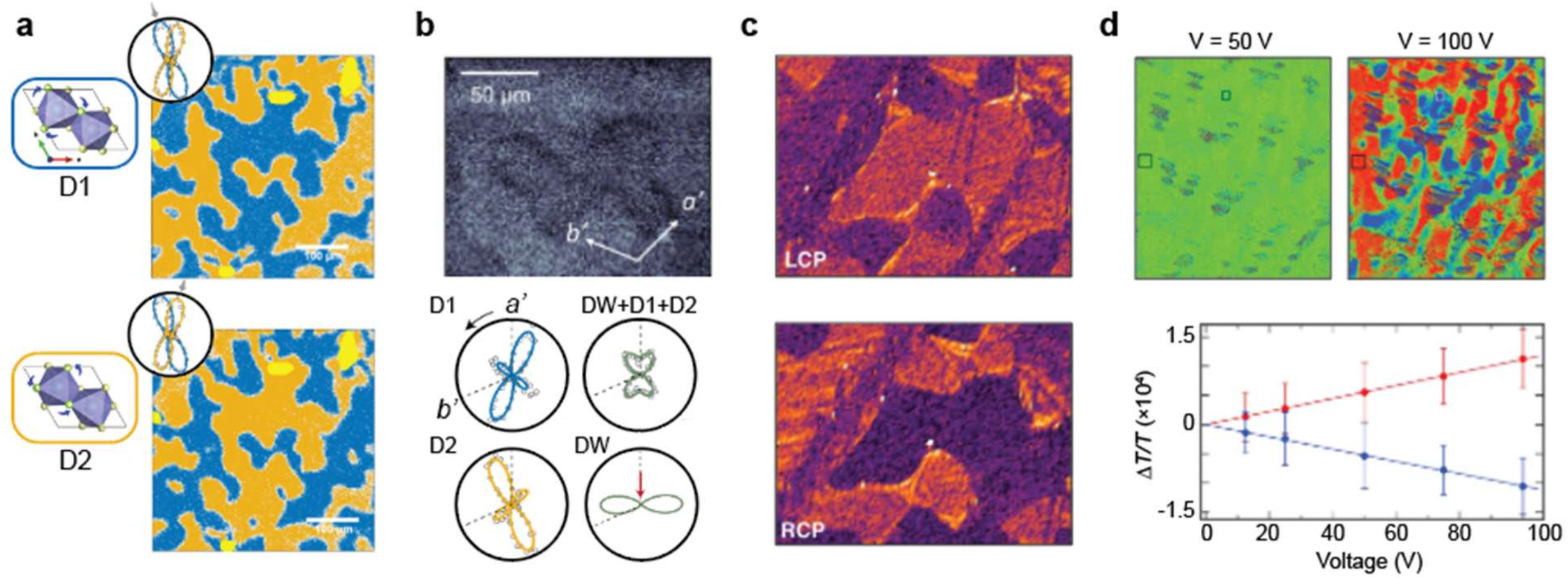

Fig. 16. Optical scanning probes and symmetry characterization of ferro-rotational domains in $NiTiO_3$. (a) EQ SHG microscopy resolves two ferro-rotational domain states, labeled D1 and D2. Local RA-SHG patterns show that the two domains are related by the mirror symmetry. (b) Domain-wall (DW) regions exhibit the reduced SHG intensity and distinct RA-SHG pattern attributed to restored mirror plane at the domain wall. (c) Circularly polarized SHG microscopy provides SHG intensity contrast arising from the opposite ferroaxial moments. (d) Electrogyration measurements show that the optical transmittance is coupled to the ferroaxial domain state and can be modulated by an applied electric field.

Figs. (a and b), (c) and (d) are adapted from Ref. [15], [124], and [125], respectively.

Subsequent work extended the detection of ferro-rotational domains from symmetry identification to real-space imaging. In $NiTiO_3$, Guo *et al*. used EQ SHG microscopy to image ferro-rotational domains and to resolve domain walls (Fig. 16a) [15]. Local RA-SHG measurements showed that the two domains are related by the mirror operations lost below the ferroaxial transition, while the domain walls exhibit suppressed SHG intensity and restoration of mirror symmetry (Fig. 16b). As illustrated in Fig. 16c, Yokota *et al*. further showed that circular intensity difference in SHG can distinguish the sign of the ferroaxial moment in $NiTiO_3$ [124]. Because the two ferroaxial domains reverse the sign of a relevant nonlinear susceptibility component, right- and left-circularly polarized fundamental light produce different SHG intensities in opposite domains.

Electrogyration provides an independent optical route to ferroaxial-domain imaging. Hayashida *et al*. [125] visualized ferroaxial domains in $NiTiO_3$ using the linear electrogyration effect, in which optical rotation is induced in proportion to an applied electric field. Follow-up study [126] showed that ferroaxial domain patterns form through the phase transition, depend on thermal history such as cooling rate, and can be correlated with scanning x-ray diffraction. These results established that ferroaxial domains are true ferroic domains rather than merely local optical contrast.

The coupling between ferroaxiality and chirality can also be exposed by electric-field-induced magnetochiral dichroism (Fig. 16d) [127]. In $NiTiO_3$, an applied electric field breaks the remaining mirror symmetry and converts the nonchiral ferroaxial state into an electrically controllable chiral optical medium. The resulting magnetochiral dichroism appears in the short-wavelength infrared region around $Ni^{2+}$ d-d magnetic-dipole transitions, providing a useful example of how ferroaxial order can act as a parent symmetry channel for field-induced chirality-related phenomena.

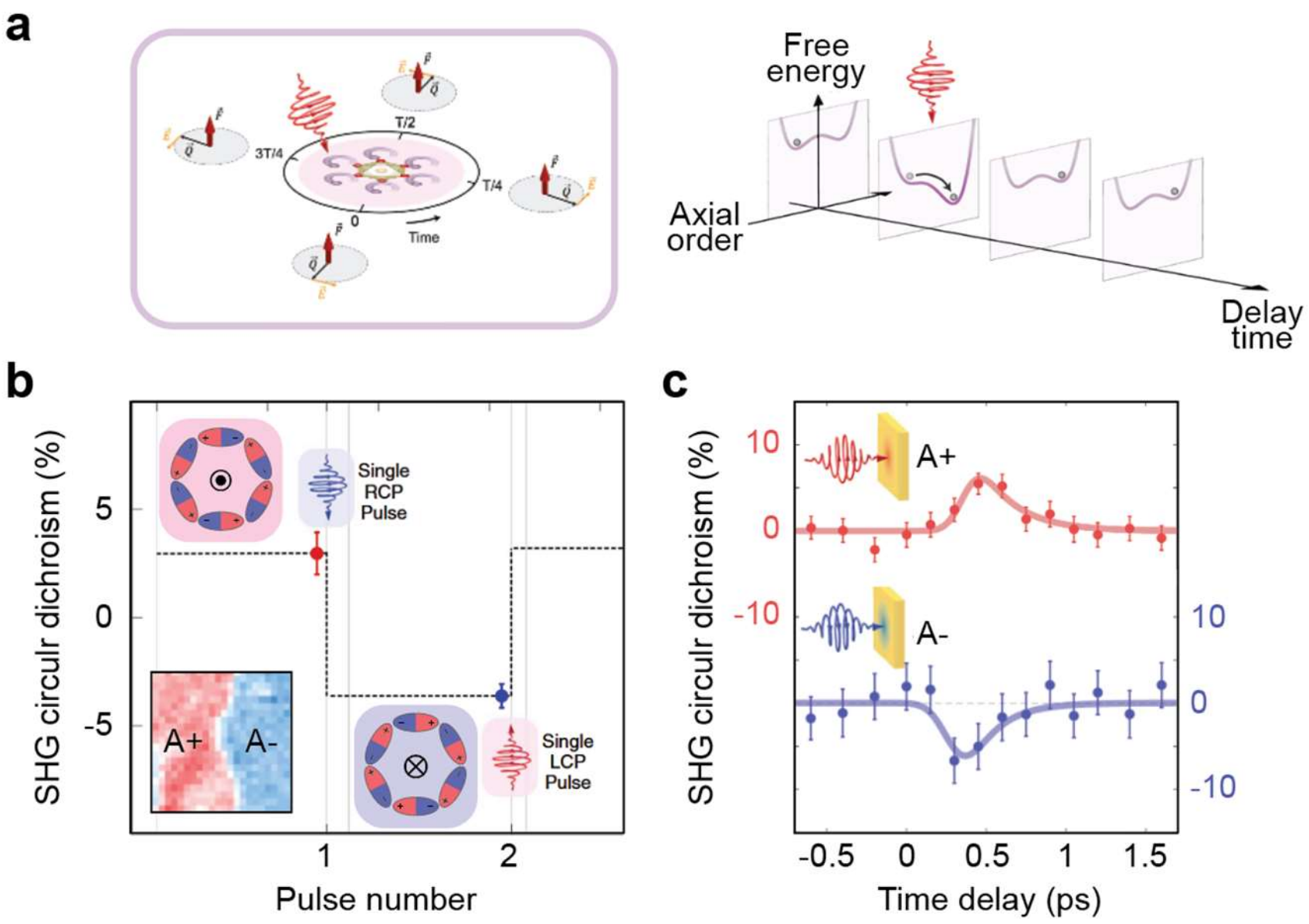

Fig. 17. Ultrafast optical control of ferro-rotational order by THz-driven phonon. (a) Schematic illustration of the dynamical conjugate field *Q*×*E* generated by a resonantly driven infrared-active displacement *Q* and the THz field *E*. The right panel shows a schematic illustration of time-dependent free energy landscape of the axial order after the phonon excitation biasing the double-well potential and switching the ferroaxial order parameter. (b) Demonstration of single-pulse switching of the ferroaxial order parameter as probed by the sign of SHG circular dichroism. (c) Time-resolved SHG circular dichroism above $T_c$ reveals the transient emergence of ferro-rotational states.

Figures are adapted from Ref. [128], respectively.

The remaining challenge is deterministic control. Zeng *et al*. [128] recently demonstrated nonvolatile, rewritable switching of ferroaxial order in $RbFe(MoO_4)_2$ by engineering a dynamic conjugate field with circularly driven terahertz phonons (Fig. 17a). In this mechanism, a resonantly driven infrared-active phonon displacement *Q* and the terahertz electric field *E* form an axial field proportional to $Q \times E$. Linearly polarized terahertz excitation does not generate such a field because *Q* and *E* are parallel, whereas circularly polarized excitation produces a fixed axial field whose sign is selected by the helicity of the pulse. This field couples to the ferroaxial soft mode, lifts the degeneracy between the two wells of the ferroaxial double-well potential, and enables single-shot switching between **A**+ and **A**- states, as verified by the reversal of SHG circular dichroism (Figs. 17b and 17c). This result provides a compelling route toward ultrafast and nonvolatile control of ferro-rotational order.

### 4.1.4. Ferro-rotational order in $RbFe(MoO_4)_2$ revealed by EQ SHG

$RbFe(MoO_4)_2$ is an archetypal triangular-lattice type-II multiferroic. At low temperature, its 120° antiferromagnetic order becomes incommensurate along the *c* direction and produces a ferroelectric polarization along the three-fold axis [129]. This polarization is perpendicular to the spin-rotation plane, which is difficult to explain by the conventional cycloidal or inverse-Dzyaloshinskii-Moriya picture. The later ferroaxial interpretation resolved this issue by recognizing that a macroscopic structural rotation can couple to magnetic chirality and thereby generate a polar vector along the axial rotation direction [122,123].

The relevant structural precursor is the transition near 190 K. Early x-ray diffraction and Raman studies identified a structural phase transition near 190 K and associated it mainly with rotations of tetrahedra, most likely lowering the high-temperature $P\bar{3}m1$ structure to a closely related low-temperature $P\bar{3}c1$ structure while their determination of space group was rather indecisive due to limited resolutions [129,130].

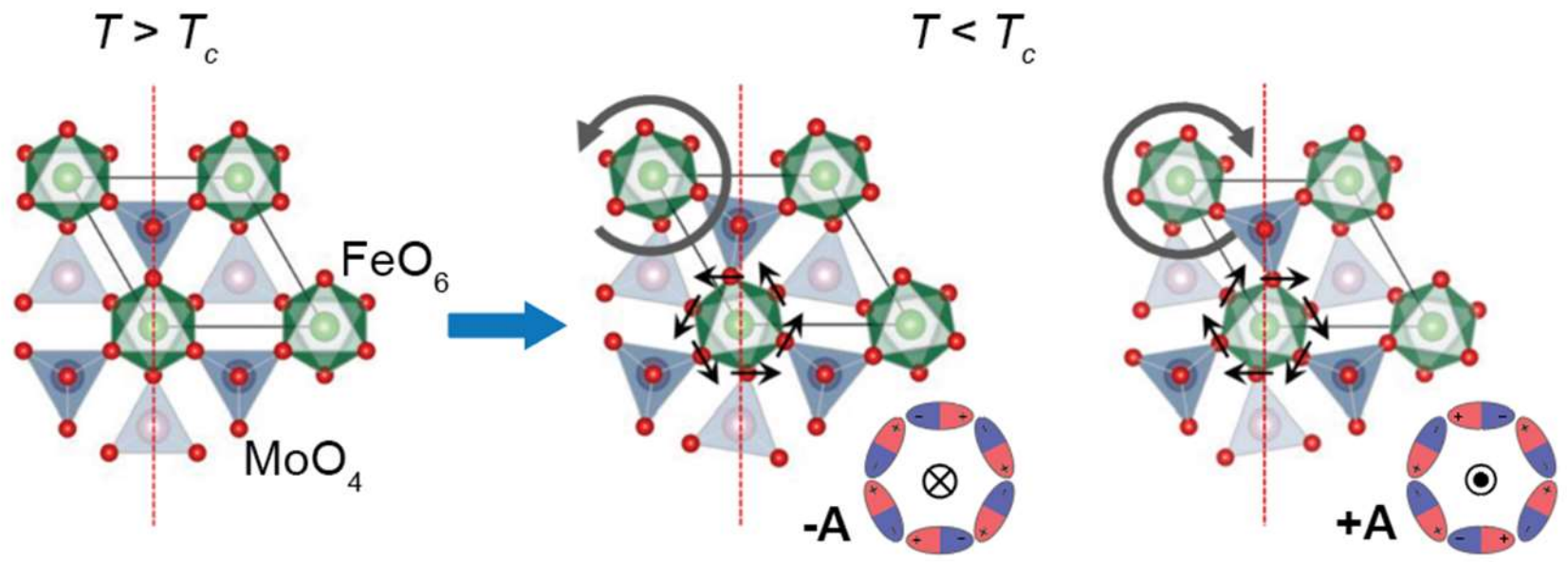

Fig. 18. Structural distortion corresponding to ferro-rotational orders in $RbFe(MoO_4)_2$. Above $T_c$, the crystal structure preserves mirror plane indicated by red-dotted line. Below $T_c$, counter-rotations of $FeO_6$ octahedra and $MoO_4$ tetrahedra generate rotational local polar displacements, breaking the mirror symmetry. Their rotational arrangements define two degenerate **-A** and **+A** domain states.

Figures are adapted from Ref. [73], respectively.

The symmetry lowering pathway is the essential ingredient to understand ferro-rotational order. Microscopically, the counter-rotation of $FeO_6$ octahedra and $MoO_4$ tetrahedra produces circulating local polar displacements around the cation sites (Fig. 18). The local dipoles cancel in pairs, so the crystal remains centrosymmetric and does not become ferroelectric at the structural transition. Nevertheless, their circulating arrangement defines an axial vector **A** parallel to the *c* axis. Because **A** is even under both time reversal and spatial inversion, it belongs to the same parity class as the ferro-rotational order parameter. In the high-temperature $\bar{3}m$ point group, **A** along *c* axis transforms as $A_{2g}$ (Table 7). The order parameter is therefore not a conventional electric polarization, but a time-even, inversion-even axial vector.

Table 7. Symmetry character in $D_{3d}$ ($\bar{3}m$) phase

| $D_{3d}$ ($\bar{3}m$) | E | $2C_3$ | $3C_2'$ | i | $2S_6$ | $3\sigma_d$ | Basis functions |
|---|---|---|---|---|---|---|---|
| $A_{1g}$ | 1 | 1 | 1 | 1 | 1 | 1 | $z^2, yz(3x^2-y^2), \ldots$ |
| $A_{2g}$ | 1 | 1 | -1 | 1 | 1 | -1 | $R_z, \boldsymbol{xz(x^2-3y^2)}, \ldots$ |
| $E_g$ | 2 | -1 | 0 | 2 | -1 | 0 | $\{R_x, R_y\}, \{x^2-y^2, xy\}, \ldots$ |
| $A_{1u}$ | 1 | 1 | 1 | -1 | -1 | -1 | $x(x^2-3y^2), \ldots$ |
| $A_{2u}$ | 1 | 1 | -1 | -1 | -1 | 1 | $z, y(3x^2-y^2), \ldots$ |
| $\boldsymbol{E_u}$ | 2 | -1 | 0 | -2 | 1 | 0 | $\{x, y\}, \{z(x^2 - y^2), xyz\}, \ldots$ |

This explains why ordinary linear optical probes are not well suited for detecting the ferro-rotational order parameter. The ferro-rotational vector **A** is inversion-even and therefore cannot be addressed as a simple polar response. ED SHG is also forbidden in the centrosymmetric phases. EQ SHG provides a natural route because its nonlinear polarization contains one additional spatial derivative, $P_i(2\omega) = \chi_{ijkl} E_j \partial_k E_l$. The product of the two fundamental electric fields, the optical wave vector, and the induced SHG polarization can form composite fields with the same $A_{2g}$ symmetry as given in basis functions of $xz(x^2 - 3y^2)$.

Jin *et al*. first established that the room-temperature SHG signal is dominated by the bulk EQ contribution [73]. In oblique-incidence rotational-anisotropy SHG, all four polarization channels showed the three-fold rotation and three mirror planes required by $\bar{3}m$ (Fig. 19a). They also compared possible ED, surface, electric-field-induced, and MD contributions and concluded that the measured signal is bulk EQ SHG. This step is important because the later temperature-dependent analysis relies on tracking symmetry-allowed EQ tensor elements rather than an unwanted surface or magnetic contribution.

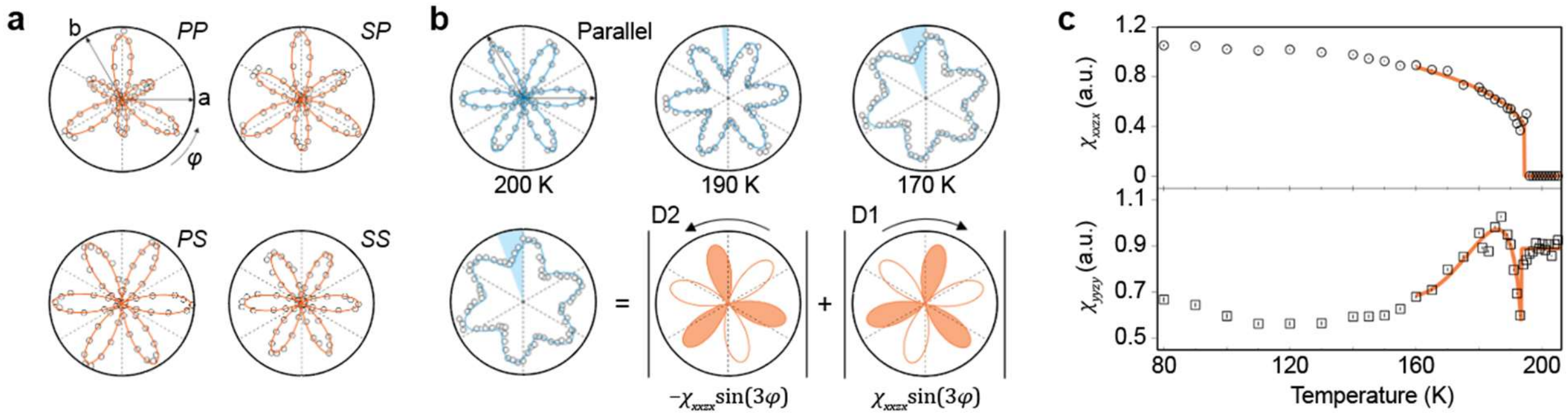

Fig. 19. The evidence for ferro-rotational order in $RbFe(MoO_4)_2$ revealed by EQ SHG. (a) Oblique-incidence RA-SHG patterns measured in four different polarization channels show three-fold rotational and mirror symmetries consistent with centrosymmetric $\bar{3}m$ point group. (b) Temperature-dependent RA-SHG patterns exhibit a rotation of the lobes with an angular-independent background signal above and below $T_c$. The emergent RA-SHG pattern is a weighted average of contributions from two domain states D1 and D2, which rotate clockwise and counterclockwise, respectively. (c) The extracted temperature dependence of EQ SHG susceptibility tensor elements $\chi_{xxzx}$ and $\chi_{yyzy}$, which are the emerging component below $T_c$ and the pre-existing component, respectively.

Figures are adapted from Ref. [73], respectively.

The decisive measurement was then performed at normal incidence as illustrated in Fig. 19b. In the high-temperature phase, the parallel channel selects a single EQ tensor element, $\chi_{yyzy}$, with angular forms $I_{parallel} \propto |\chi_{yyzy}\cos(3\varphi)|^2$. These patterns retain the mirror planes of $\bar{3}m$. Below $T_c \approx$ 195 K, the RA-SHG lobes rotate away from those mirror planes and develop an angular-independent background. The rotation requires an emergent tensor component, $\chi_{xxzx}$, which is symmetry-forbidden above $T_c$ but allowed in the $\bar{3}$ phase. The low-temperature point group is best described as $\bar{3}$ rather than $\bar{3}m$ or $32$ as suggested by Klimin *et al*. [130] because rotational-anisotropy SHG directly detects the loss of the three diagonal mirror planes while finding no large electric-dipole SHG enhancement across the transition [73]. In point-group language, the transition is therefore $\bar{3}m$ to $\bar{3}$, where the three-fold axis, inversion, and time-reversal symmetry remain, whereas the diagonal mirror symmetries are lost.

The low-temperature response can be written as $I_{D1} \propto \left|\chi_{xxzx}\sin(3\varphi) + \chi_{yyzy}\cos(3\varphi)\right|^2$ for one domain and $I_{D2} \propto \left|-\chi_{xxzx}\sin(3\varphi) + \chi_{yyzy}\cos(3\varphi)\right|^2$ for the mirror-related domain. Thus, the sign of $\chi_{xxzx}$ distinguishes the two ferro-rotational domains with opposite axial vectors. The measured pattern below $T_c$ is a weighted sum of these two domain contributions, consistent with a laser spot larger than individual domains. This domain-sensitive SHG analysis turns the otherwise subtle octahedral rotation into a directly measurable ferro-rotational order parameter.

The temperature dependence of the EQ SHG susceptibility further establishes its direct connection to the ferro-rotational order parameter (Fig. 19c). Upon cooling through $T_c$=195 K, the temperature dependence of $\chi_{xxzx}$ appears and grows with decreasing temperature. Since $\chi_{xxzx}$ transforms according to the same $A_{2g}$ irrep as the ferro-rotational order parameter, its onset provides an optical measure of the ferro-rotational phase transition. The abrupt emergence of $\chi_{xxzx}$, together with the rotation of the RA-SHG pattern below $T_c$, indicates that the transition has weakly first-order character. By contrast, $\chi_{yyzy}$ remains finite across the transition and mainly serves as the symmetry-allowed EQ SHG background of the high-temperature phase, while its anomaly near $T_c$ reflects the coupling between the pre-existing quadrupolar optical response and the newly developed ferro-rotational distortion.

This example also clarifies what is meant by a conjugate field for a hidden ferroic order. A true conjugate field must transform identically to the order parameter. For the ferro-rotational vector **A** in $\bar{3}m$, this means $A_{2g}$ symmetry. In the normal-incidence EQ SHG geometry, the combination of incoming optical fields $E_x$, $E_y$(transform as the $E_u$ irrep), the wave-vector component $k_z$ entering the EQ process (transforms as the $A_{2u}$ irrep), and the outgoing nonlinear polarization $P_x$, $P_y$ (transform as the $E_u$ irrep) supplies precisely such an $A_{2g}$ composite. Therefore, EQ SHG does more than visualize a structural distortion. It constructs an optical coupling channel that is symmetry-equivalent to the ferro-rotational order parameter.

The broader significance of $RbFe(MoO_4)_2$ is that it bridges ferroaxial multiferroicity and nonlinear optical detection of hidden order. Ferroaxial coupling had already explained why $Cu_3Nb_2O_8$ and $CaMn_7O_{12}$ develop polarization perpendicular to the spin-rotation plane. A magnetic chirality couples to a pre-existing structural axial vector **A**, producing $P \parallel \mathbf{A}$ [122,123]. In $RbFe(MoO_4)_2$, EQ SHG directly identifies the underlying ferro-rotational order and its two domains, providing a general strategy for detecting centrosymmetric axial orders whose conjugate fields are not available in ordinary static-field experiments.

## 4.2. Electronic nematic order revealed by nonlinear optical response

Electronic nematic order is a symmetry-broken electronic state in which the electronic fluid lowers the rotational symmetry of a crystal while preserving translational symmetry. Unlike a charge- or spin-density wave, it does not require a new ordering wave vector or an enlarged unit cell. Instead, crystallographically equivalent directions become electronically inequivalent, which can appear as resistivity anisotropy, orbital splitting, anisotropic spin fluctuations, Fermi-surface distortion, or a subtle symmetry-lowering lattice response.

Understanding electronic nematic order is important because it clarifies how electronic correlations can reduce crystal symmetry without forming a conventional density wave. Nematicity often couples strongly to orbital, spin, lattice, and superconducting degrees of freedom. One of the central issues is therefore to distinguish the primary order parameter that drives the transition from secondary responses that are induced through symmetry-allowed coupling. This distinction is essential for determining whether nematicity originates from electronic correlations [131,132], orbital order [133], spin fluctuations [134,135], lattice instability [136–138], or spin-orbit-coupled multipolar order [117,139].

SHG is a powerful probe for this purpose because its nonlinear susceptibility tensor is highly sensitive to point-group symmetry. RA-SHG can identify lost symmetry operations, newly allowed tensor components, inversion-symmetry breaking, and domain structures. It therefore provides a direct route for separating hidden primary nematic order from secondary structural distortions.

#### 4.2.1. Electronic nematicity and relevant symmetry reduction

The essential symmetry criterion for electronic nematic order is the spontaneous loss of rotational symmetry while translational symmetry remains intact. The defining signature is the loss of rotational equivalence between crystallographic directions that were symmetry-related in the high-temperature phase. In widely studied Fe-based and Cu-based superconductors [135,140], for example, a parity-even nematic transition commonly lowers four-fold rotational symmetry to two-fold one, $C_4 \rightarrow C_2$, while preserving spatial inversion (Fig. 20). Such in-plane nematic states are often described by $d_{x^2-y^2}$ or $d_{xy}$-type order parameters. The symmetry lowering primarily produces anisotropic electronic responses such as in-plane resistivity anisotropy, while a subtle lattice distortion of the same symmetry may appear as a secondary consequence [140,141].

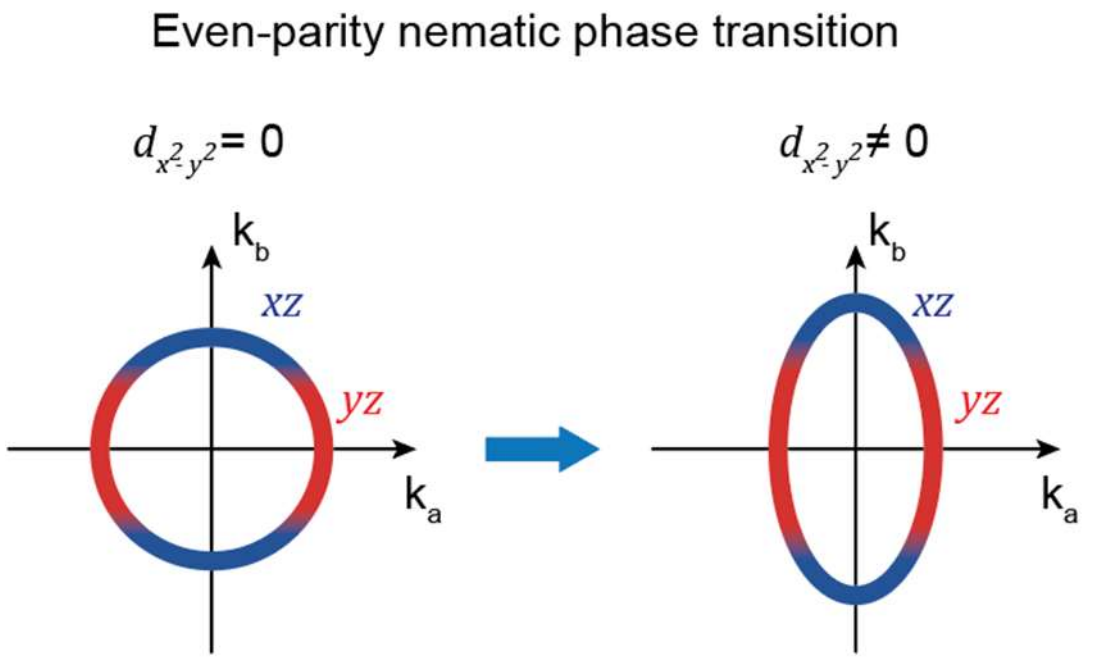

Fig. 20. Even-parity electronic nematic phase transition. The electronic structure represented by Fermi surface changes from a rotational symmetric state with $d_{x^2-y^2} = 0$ to an anisotropic nematic state with $d_{x^2-y^2} \neq 0$.

Hecker *et al*. classified even-parity electronic nematic phase transitions by considering all crystallographic point groups [142]. In this classification, a nematic state can be represented by five components of a symmetric traceless rank-2 tensor, $\boldsymbol{d}$ = ($d_{(2z^2-x^2-y^2)/\sqrt{3}}$, $d_{x^2-y^2}$, $d_{2yz}$, $d_{2xz}$, $d_{2x}$ ). This form emphasizes that electronic nematicity is not limited to in-plane $d_{x^2-y^2}$ or $d_{2xy}$ states. It can also involve out-of-plane states such as $d_{xz}$ or $d_{yz}$. Their analysis identified 59 nematic phase-transition species, including Potts- and clock-type electronic nematic states associated with different rotational-symmetry reductions. In spin-orbit-coupled metals, a distinct parity-odd form of nematic order can arise from spin-orbit coupling and Fermi-surface instability, as proposed by Fu [139]. The parity-odd nematic order parameter is described by a symmetric traceless pseudotensor $\boldsymbol{Q}$ = $(Q_{xx} - Q_{yy}\ ,\ 2Q_{zz} - Q_{xx} - Q_{yy}\ ,\ Q_{xy}\ ,\ Q_{yz}\ ,\ Q_{zx})$. In contrast to conventional parity-even nematicity, this order breaks both rotational symmetry and spatial inversion while preserving translational symmetry. Physically, the preferred electronic direction is tied not only to momentum-space anisotropy but also to the spin texture through spin-momentum locking.

### 4.2.2. SHG identification of the primary nematic order parameter in $Cd_2Re_2O_7$

$Cd_2Re_2O_7$ is a rare metallic pyrochlore in which a subtle structural transition near 200 K is accompanied by pronounced electronic anomalies. In the high-temperature phase, the crystal is cubic and centrosymmetric, with space group $Fd\bar{3}m$ and point group $m\bar{3}m$. Upon cooling through $T_c \approx 200$ K, diffraction and optical studies identify a transition into a noncentrosymmetric tetragonal structure, commonly described by space group $I\bar{4}m2$ [143–145]. Although this transition was traditionally attributed to the softening of an $E_u$ phonon mode (Fig. 21a), the associated lattice distortion is extremely small compared with the large changes in resistivity, magnetic susceptibility, and density of states [146,147]. This imbalance motivated the proposal that the transition is driven not by a simple lattice instability, but by a hidden electronic order that couples secondarily to the lattice.

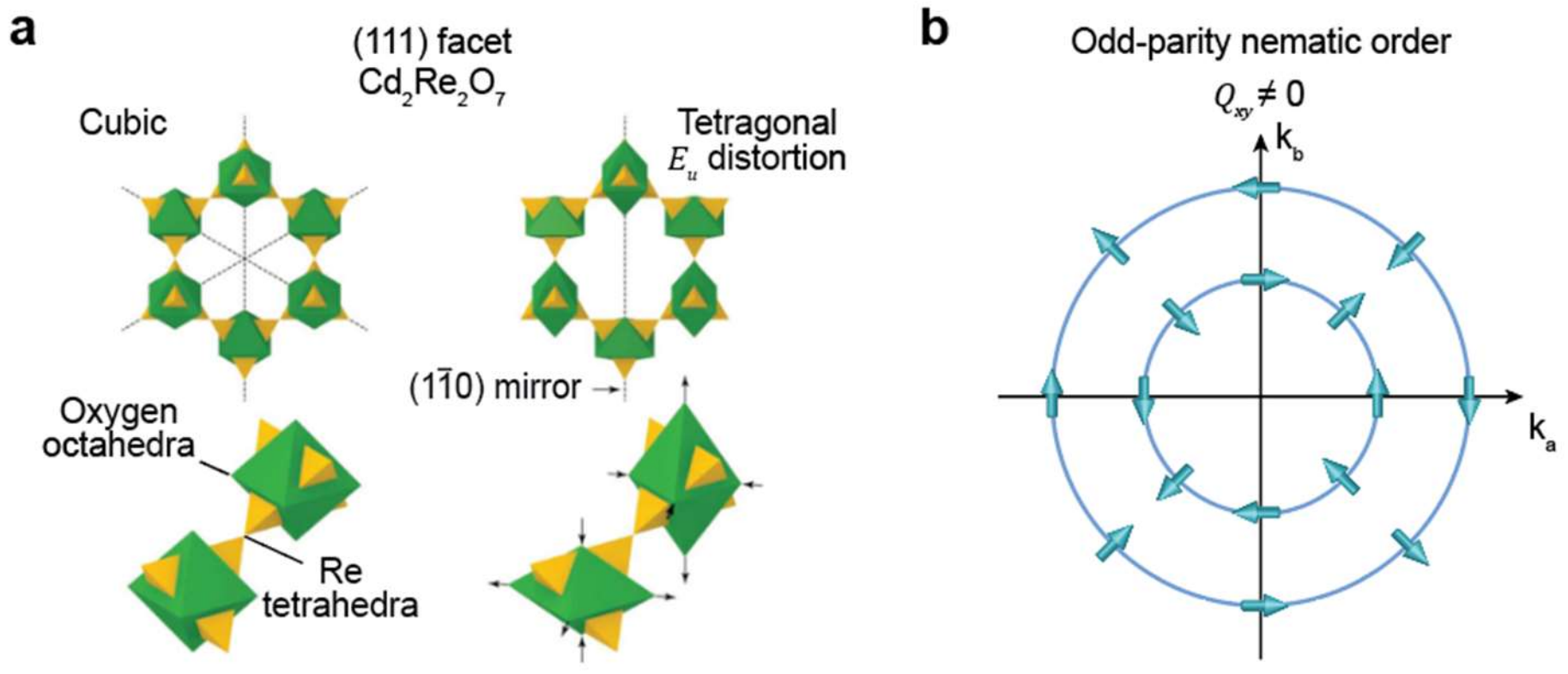

Fig. 21. Structural and electronic order parameters in pyrochlore $Cd_2Re_2O_7$. (a) Inversion-breaking structural phase transition from cubic to tetragonal crystal structures with the softening of an $E_u$ phonon mode. (b) The parity-odd nematic state illustrated with a spin-textured Fermi surface, which belongs to the $T_{2u}$ irrep with non-zero $Q_{xy}$.

Fig. (a) is adapted from Ref. [117], respectively.

Fu proposed that $Cd_2Re_2O_7$ may host an electronically driven parity-odd multipolar nematic order [139]. In this framework, an $E_u$ component distinguishes one cubic axis from the other two axes, whereas a $T_{2u}$ component, such as $Q_{xy}$, produces an anisotropy oriented along diagonal or shear-like cubic directions (Fig. 21b). Conventional electronic nematicity is often detected through transport anisotropy, sometimes with uniaxial strain or magnetic field used to align nematic domains. However, since $Q_{ij}$ is parity-odd, ordinary strain and magnetic fields do not couple linearly to it. ED SHG provides a more direct route because the bulk second-order susceptibility $\chi_{ijk}$ is forbidden in a centrosymmetric crystal but becomes allowed when inversion symmetry is broken. The key strength of RA-SHG is that the angular dependence resolves the tensor structure of $\chi_{ijk}$ and can therefore determine which irrep of the high-temperature point group is active.

Harter et al. first performed RA-SHG measurements on the (111) facet of single-crystalline $Cd_2Re_2O_7$ [117]. Above $T_c$, the SHG signal is dominated by the surface contribution because the surface naturally breaks the inversion symmetry even when the bulk remains centrosymmetric. Well below $T_c$, SHG microscopy revealed six distinct structural domains (Fig. 22a). This observation is consistent with a cubic-to-tetragonal structural phase transition associated with an $E_u$ distortion. The tetragonal axis can select any one of three equivalent cubic directions, and each inversion-breaking distortion can occur in two parity-related forms.

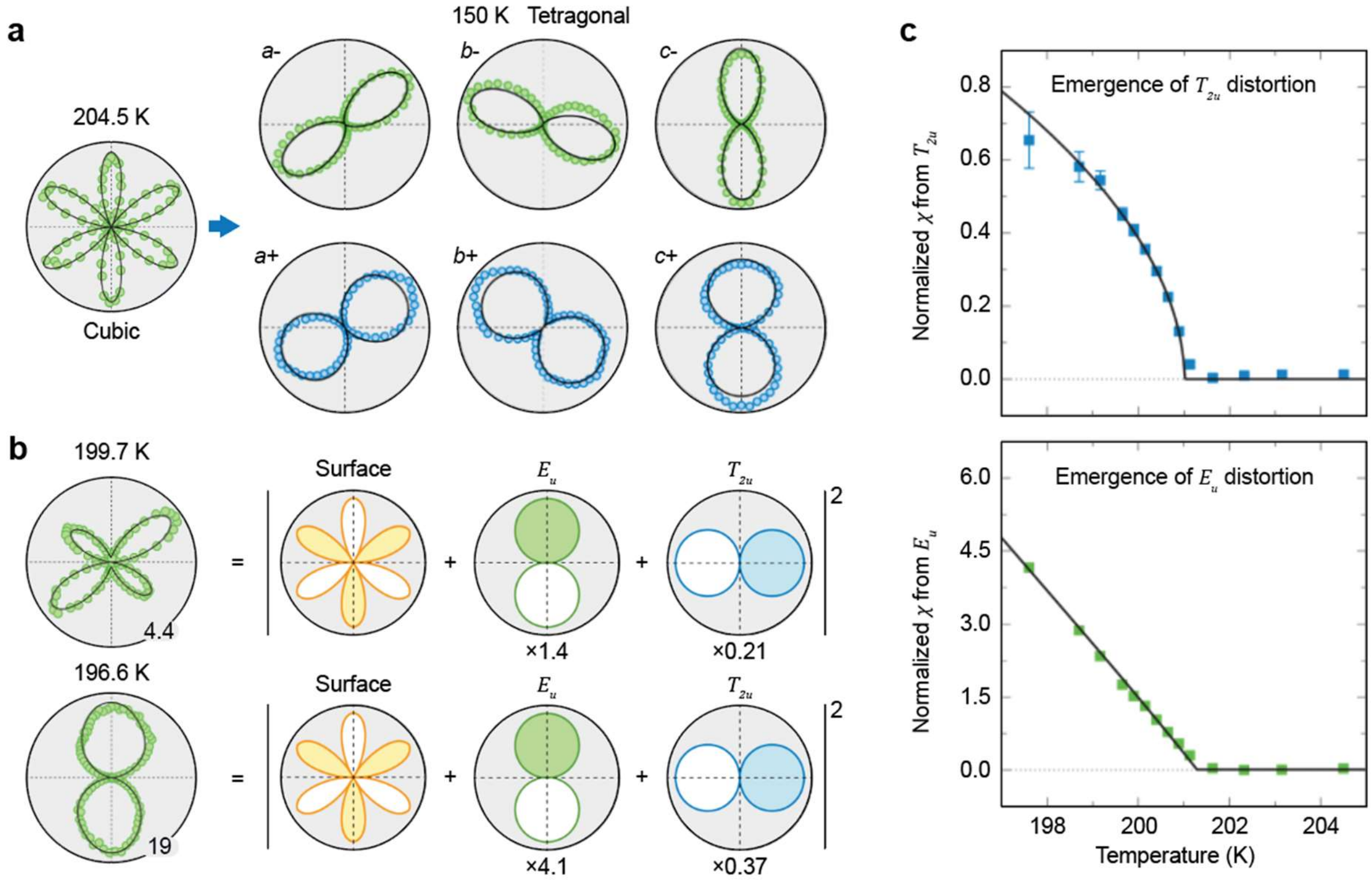

Fig. 22. RA-SHG identification of hidden parity-odd nematic state in $Cd_2Re_2O_7$. (a) RA-SHG patterns measured above and far below the transition temperature. Through the cubic-to-tetragonal phase transition, six different tetragonal domains are identified with distinct SHG patterns. (b) RA-SHG patterns measured just below the transition temperature demonstrate that the structural $E_u$ distortion cannot explain the observed loss of mirror symmetry by itself, revealing an emergence of the additional $T_{2u}$ multipolar nematic contribution. (c) Temperature-dependent measurements of SHG susceptibilities from $T_{2u}$ and $E_u$ contributions exhibit distinct critical behaviors, developing as the $T_{2u}$ primary and $E_u$ secondary orders.

Figures are adapted from Ref. [117], respectively.

The decisive observation was made in the vicinity of the transition temperature (Fig. 22b). RA-SHG patterns just below 200 K could not be reproduced using only with the temperature-independent surface contribution and the $E_u$ tensor contribution. In particular, the data could not be fit by the $\bar{4}2m$ point group expected for the $I\bar{4}m2$ structural distortion because the RA-SHG patterns show the loss of a mirror symmetry that should be retained in that point group. Therefore, an additional SHG contribution that both breaks this mirror symmetry and couples to the $E_u$ structural order is necessitated to explain the observation. Symmetry analysis shows that this requirement is satisfied by a $T_{2u}$ order parameter. The resulting interpretation is a symmetry reduction from the high-temperature $m\bar{3}m$ point group to the lower-symmetry $\bar{4}$ point group, with the $T_{2u}$ multipolar nematic order acting as the hidden primary order parameter. In the spin-orbit-coupled picture, this $T_{2u}$ order corresponds to a spin-textured Fermi-surface distortion.

The temperature dependence of the $E_u$ and $T_{2u}$ susceptibility components provides the central evidence for identifying the $T_{2u}$ component as primary (Fig. 22c). Near $T_c \approx 201$ K, the $T_{2u}$ susceptibility follows an order-parameter-like onset with a critical exponent close to $\beta \approx 1/2$. By contrast, the $E_u$ susceptibility grows approximately linearly with reduced temperature over a broad range, corresponding to $\beta \approx 1$. Because ED SHG susceptibility is linearly proportional to an inversion-odd order parameter near an inversion-breaking transition, the $T_{2u}$ behavior is consistent with a primary order parameter, whereas the $E_u$ behavior is characteristic of a secondary order induced by coupling to the primary order.

A symmetry-based Landau analysis further clarifies the coupling among the order parameters. The linear induction of the $E_u$ structural response by the $T_{2u}$ primary order requires an additional even-parity order parameter with $T_{1g}$ symmetry, so that a trilinear invariant can couple $T_{2u}$, $T_{1g}$, and $E_u$. Later theoretical work refined this tensor interpretation. Norman subsequently revisited the problem from the structural side and showed that group-subgroup paths from $m\bar{3}m$ to $\bar{4}$ can accommodate three related distortions: an odd $T_{2u}$ component associated with $I\bar{4}2d$, an even $T_{1g}$ component associated with $I4_1/a$, and an odd $E_u$ component associated with $I\bar{4}m2$ [11]. This structural scenario has the same symmetry content as the electronic interpretation, even though the microscopic origin remains a subject of discussion.

This example is conceptually important because it demonstrates how SHG can detect a hidden nematic order whose defining symmetry is parity-odd. In the ferro-rotational case discussed above, the axial order is parity-even and must be accessed through EQ SHG. $Cd_2Re_2O_7$ represents the complementary situation. The relevant order is parity-odd, and so ED SHG is directly allowed in the ordered bulk. The optical susceptibility itself therefore becomes a symmetry-selective probe of the order parameter. The experiment does more than establish that inversion symmetry is broken. It separates the symmetry of the primary hidden order from that of the secondary lattice response.

## 5. Outlook and challenges

The role of SHG has expanded from detecting noncentrosymmetric crystals to resolving tensor signatures of broken symmetry in quantum and functional materials [4,6,8,34]. As reviewed above, SHG is no longer limited to identifying whether inversion symmetry is present. Through polarization resolution, rotational anisotropy, microscopy, and higher-order multipole analysis, SHG can distinguish crystallographic, magnetic, polar, chiral, ferro-rotational, and multipolar electronic orders. In this sense, SHG provides a bridge between optical spectroscopy and Landau-type symmetry analysis, domain imaging, and multipole-sensitive tensor classification. This role is particularly important for quantum materials, where the relevant broken symmetry is often subtle, spatially inhomogeneous, intertwined with other degrees of freedom, or difficult to access using conventional diffraction, transport, or linear optical probes. Several opportunities and challenges remain for applying SHG to quantum materials. Below, we provide an outlook of SHG for a quantitative probe of order-parameter dynamics, discuss the connection between SHG and electronic and magnetic states based on theoretical calculations, and also briefly comment on the future role of SHG in condensed matter physics.

### 5.1. Resonant SHG spectroscopy

Photon-energy-dependent resonant SHG provides a route for adding spectroscopic information to the symmetry sensitivity of RA-SHG. Conventional fixed-wavelength RA-SHG primarily constrains point-group and order-parameter symmetry, whereas tuning the fundamental or second-harmonic photon energy through electronic resonances can selectively enhance nonlinear pathways involving particular electronic or collective states. Near resonance, selected nonlinear optical pathways are strongly enhanced, and their spectral dependence can reflect the orbital, spin, parity, and symmetry character of the participating states. Resonant SHG can therefore combine the symmetry sensitivity of RA-SHG with spectroscopic information, providing access to electronic and magnetic states [13,94], coupled quasiparticles [148], critical behavior near phase transitions [76], and potentially quantum-geometric properties of electronic wave functions [149].

At the same time, resonant SHG introduces additional challenges compared with fixed-wavelength symmetry measurements. Near resonance, the magnitudes and complex phases of different susceptibility components can vary rapidly with photon energy. Consequently, an RA-SHG pattern may change substantially even when the underlying point-group symmetry remains unchanged. Such changes can arise simply from variations in the relative amplitudes and interference of symmetry-allowed tensor components. Multiple intermediate states and ED, MD, and EQ pathways may also overlap spectrally, making resonant enhancement alone insufficient to identify a microscopic origin. Resonant SHG should therefore be analyzed through the combined dependence on photon energy, polarization, rotational anisotropy, temperature, and magnetic field, rather than from intensity spectra alone. First-principle calculations of the complex nonlinear susceptibility will be particularly important for identifying the participating intermediate states and distinguishing genuine symmetry changes from resonance-induced modifications of the SHG response. With these developments, resonant SHG could provide a powerful bridge between symmetry-resolved characterization and microscopic spectroscopy of quantum phases.

### 5.2. Dynamics, fluctuations, and nonequilibrium order

Most SHG studies reviewed here focus on static or quasi-static symmetry breaking. Yet many of the most important questions in quantum materials concern fluctuations and dynamics. These include how an order parameter forms, how competing phases exchange spectral weight, and how ultrafast light fields can selectively and coherently control order parameters. Time-resolved SHG provides a natural route to these questions because it can track the symmetry of a transient state, providing a unique advantage over conventional transient spectroscopy [150–154].

One future direction is ultrafast order-parameter tracking. In pump-probe SHG, the pump pulse with resonant frequencies can be chosen to drive phonons, magnons, excitons, collective electronic modes, and other quasiparticles, while the SHG probe reads out the instantaneous symmetry. This approach can reveal whether a light-induced phase is truly distinct in symmetry or merely a thermalization of the equilibrium phase. For example, coherent phonon excitation may transiently break inversion symmetry, rotate a polar axis, induce a ferro-rotational distortion, or couple to an electronic nematic order. THz excitation may drive soft polar modes, magnetic resonances, or nonlinear currents connected to Berry curvature. In these cases, time-resolved

RA-SHG can determine whether the driven response has the symmetry expected for the target order parameter.

The second direction is fluctuation-sensitive SHG. Near a phase transition, SHG can contain contributions from order-parameter fluctuations even above the static transition temperature [12,76]. This capability is especially promising for low-dimensional and strongly correlated systems, where fluctuations may dominate over long-range order. However, fluctuation SHG is difficult to interpret because it may arise from short-range domains, surface ordering, precursor lattice distortions, or nonlinear optical resonance effects. Solving this problem will require temperature-dependent tensor analysis, statistical imaging of domains, comparison with scattering measurements, and theoretical modeling of fluctuation-induced nonlinear susceptibility.

## 5.3. Quantitative microscopic interpretation of SHG susceptibilities

One significant conceptual challenge for SHG studies of quantum materials is to connect experimentally measured tensor components with their microscopic origins. Group theory determines which nonlinear susceptibility elements are symmetry-allowed, but it does not, by itself, identify the microscopic mechanisms responsible for their magnitude, phase, or resonant behavior. This limitation is especially important in quantum materials, where strong correlations, spin-orbit coupling, multiband electronic structures, and electron–lattice interactions can all contribute to the same symmetry-allowed tensor element.

A practical route forward is to adopt a three-level hierarchy of interpretation. The first level is symmetry. SHG directly constrains the allowed tensor components and, in turn, narrows the possible crystallographic or magnetic point groups. The second level is phenomenology. The temperature, field, strain, fluence, and polarization dependence of individual tensor elements can reveal the relevant order parameters, their couplings, and their evolution under external control. A recent study of bilayer CrSBr provides a representative example [78]. Field-dependent RA-SHG established a direct phenomenological relation between the nonlinear susceptibilities and the spin canting angle $\theta$ between the magnetic moments in the two layers. By relating the extracted susceptibility component to $\theta$, the measured SHG response could be quantitatively mapped onto the field-dependent spin configurations and the associated toroidal and polar order parameters. Such phenomenological analysis therefore allows RA-SHG to go beyond identifying broken symmetry and to track quantitatively how multiple order parameters evolve and couple to one another. The third level is microscopic theory. First-principles or model-based calculations of nonlinear susceptibility, incorporating electronic structure, spin-orbit coupling, excitonic effects, and electronic correlations, can then be guided by the symmetry and phenomenological constraints obtained from the first two levels. This hierarchy is essential because purely microscopic calculations may be challenging without symmetry constraints, whereas purely symmetry-based analysis can remain underdetermined without microscopic input.

Future theoretical efforts should also include expanding practical tools for magnetic and multipolar SHG analysis. Magnetic point groups, time-reversal-even and time-reversal-odd susceptibilities, spin point groups, and altermagnetic symmetries should be incorporated into accessible tensor-analysis software. Such tools should be able to generate allowed ED, MD, and EQ tensor forms and simulate RA-SHG patterns for realistic experimental geometries. This capability would make SHG analysis more robust and reduce ambiguity when multiple candidate symmetries produce similar angular patterns.

### 5.4. Future role of SHG in condensed matter physics

In the realm of condensed matter physics, the enduring value of SHG lies in its unique capability to provide symmetry information that is otherwise challenging to obtain. In particular, RA-SHG constrains point-group and order-parameter symmetries through the symmetry-allowed nonlinear susceptibility tensors, whereas many conventional probes are more directly sensitive to translational order, microscopic magnetic structure, local electronic structure and density of states, or elementary excitations. These techniques therefore provide complementary rather than redundant information. Table 8 summarizes the principal information obtained from RA-SHG and several commonly used probes of ordered states.

Table 8. Comparison of RA-SHG with complementary probes of ordered states

| Technique | Main information | Complementary strength |
|---|---|---|
| RA-SHG | Point-group/order-parameter symmetry | Sensitive to subtle symmetry breaking |
| Raman spectroscopy | Phonon, magnon, electronic excitations | Symmetry-resolved collective dynamics |
| Neutron scattering | Magnetic structure; ordering wavevector | Microscopic magnetic structure |
| Resonant x-ray scattering | Charge, orbital, magnetic, multipolar order | Element- and momentum-resolved order |
| X-ray/electron diffraction | Crystal structure; atomic displacements | Translational and space-group information |
| Scanning tunneling microscopy | Local density of states | Atomic-scale real-space imaging of structural and charge orders |
| Kerr/Faraday effect | Magneto-optical response | Sensitive to magnetic/TR symmetry breaking responses |

Because these probes access complementary aspects of ordered states, integrating SHG with these measurements under identical tuning conditions can provide a more thorough understanding of the studied materials. For instance, pairing SHG with X-ray diffraction can effectively decouple electronic symmetry breaking from underlying lattice distortions. Similarly, combining SHG with resonant X-ray or neutron scattering can correlate point-group symmetry with specific ordering wavevectors and microscopic magnetic or multipolar structures. When integrated with transport measurements, SHG can directly link broken symmetries to nonlinear Hall, magnetoelectric, or superconducting behaviors. Furthermore, utilizing SHG alongside ultrafast spectroscopy will be critical for determining whether light-induced phases represent genuine, fundamental alterations in symmetry.

Viewed through this broader lens, SHG offers far more than a simple diagnostic measurement; it establishes a comprehensive "symmetry language" for quantum materials. By analyzing which tensor elements emerge, how they transform, interfere, and respond to external perturbations, SHG can unveil the intricate hierarchy of order parameters in complex systems. Future progress in the field will hinge on making this technique more quantitative, spatially resolved, dynamically sensitive, and tightly integrated with microscopic theory. If these challenges are successfully met, SHG is poised to become an essential, transformative probe for discovering and controlling hidden, intertwined, and nonequilibrium phases in quantum materials.

**CRediT authorship contribution statement**

Xiaoyu Guo: Visualization, Writing – original draft, Writing – review & editing. Chang Jae Roh: Visualization, Writing – original draft, Writing – review & editing. Youngjun Ahn: Conceptualization, Funding acquisition, Visualization, Writing – original draft, Writing – review & editing.

**Declaration of competing interest**

The authors declare that they have no known competing financial interests or personal relationships that could have appeared to influence the work reported in this paper.

**Acknowledgements**

This work was supported by the U.S. Department of Energy (DOE), Office of Science (SC), Basic Energy Sciences, Materials Science and Engineering Division.